\documentclass[12pt]{article}

\usepackage{graphicx}

\usepackage{setspace}
\usepackage{subcaption}
\usepackage{latexsym}
\usepackage{amsmath}
\usepackage{natbib}
\usepackage{epsfig}
\usepackage{multirow}
\usepackage{fancybox}
\usepackage{color}
\usepackage{xcolor,colortbl}
\usepackage{tabu}
\usepackage{amssymb,amsmath}
\usepackage{epsfig, graphicx}
\usepackage{mathrsfs}
\usepackage{bbm}
\usepackage{hyperref}
\usepackage{enumitem}
\usepackage{booktabs}
\usepackage[figuresright]{rotating}
\usepackage{pdflscape}

\def \bZ {{\bf Z}}

\def \bb {{\bf b}}
\def \Sfrak{\mathfrak{S}}

\def \Cbbm{\mathbbm{C}}

\def \onebb{\mathbbm{1}}
\def \Rbb{\mathbb{R}}
\def \Pbb{\mathbb{P}}
\def \Ebb{\mathbb{E}}
\def \Var{\mathrm{Var}}
\def \rmd{\mathrm{d}}

\def \bbeta{\boldsymbol{\beta}}
\def \bbetahat{\widehat{\bbeta}}

\def \transpose{{\sf \scriptscriptstyle{T}}}
\def \trans{^{\transpose}}
\def \btheta{\boldsymbol{\theta}}
\def \bthetahat{\widehat{\btheta}}

\def \alphahat{\widehat{\alpha}}

\def \bSigma{\boldsymbol{\Sigma}}

\newcommand{\bgamma}{\boldsymbol{\gamma}}
\newcommand{\bW}{{\bf W}}

\def \Psihat{\widehat{\Psi}}

\def \ellhat{\widehat{\ell}}

\def \Bcal{\mathcal B}
\def \Ncal{\mathcal N}
\def \Ical{\mathcal I}
\def \Vcal{\mathcal V}

\usepackage[figuresright]{rotating}

\newtheorem{theorem}{Theorem}

\newtheorem{remark}{Remark}

\title{Two-Stage Pseudo Maximum Likelihood Estimation of Semiparametric Copula-based Regression Models for Semi-Competing Risks Data}

\author{Sakie J. Arachchige \\
	   Department of Mathematics and Statistics, \\
	   Mississippi State University, MS 39762, USA
          \and
        Xinyuan Chen \\
	   Department of Mathematics and Statistics, \\
	   Mississippi State University, MS 39762, USA
	   \and 
	  Qian M. Zhou$^*$\\
	   Department of Mathematics and Statistics, \\
	   Mississippi State University, MS 39762, USA\\
	   $*${\it email}: qz70@msstate.edu
	   }
\date{}

\begin{document}
\maketitle

\abstract{We propose a two-stage estimation procedure for a copula-based model with semi-competing risks data, where the non-terminal event is subject to dependent censoring by the terminal event, and both events are subject to independent censoring. With a copula-based model, the marginal survival functions of individual event times are specified by semiparametric transformation models, and the dependence between the bivariate event times is specified by a parametric copula function. For the estimation procedure, in the first stage, the parameters associated with the marginal of the terminal event are estimated using only the corresponding observed outcomes, and in the second stage, the marginal parameters for the non-terminal event time and the copula parameter are estimated together via maximizing a pseudo-likelihood function based on the joint distribution of the bivariate event times. We derived the asymptotic properties of the proposed estimator and provided an analytic variance estimator for inference. Through simulation studies, we showed that our approach leads to consistent estimates with less computational cost and more robustness than the one-stage procedure developed in \citet{chen2012maximum}, where all parameters were estimated simultaneously. In addition, our approach demonstrates more desirable finite-sample performances over another existing two-stage estimation method proposed in \citet{zhu2021semiparametric}. An \textbf{R} package \textbf{PMLE4SCR} is developed to implement our proposed method.}\\


\section{Introduction}

Medical studies oftentimes involve situations where multiple time-to-event outcomes exist simultaneously. These events can be divided into two categories: \emph{non-terminal}, e.g., cancer relapse, and \emph{terminal}, e.g., death. The occurrence of the terminal event precludes that of the non-terminal, but not vice versa; the resulting data are commonly referred to as the \emph{semi-competing risks} \citep{fine2001semi}. Let $T$ and $D$ denote times to a non-terminal and terminal event, respectively, and they are assumed to be subject to independent administrative censoring. The distribution of $T$ and the dependence between $T$ and $D$ are often of interest.

Copula-based models have been used to model the between-event dependency for semi-competing risks data that arise from studies of aging \citep{varadhan2014semicompeting, sun2023semiparametric,sun2024penalised} and disease progression among cancer patients \citep{zhou2016semiparametric}. In such models, the joint survival function of $(T,D)$, $S(t,d)\equiv\Pbb(T>t,D>d)$, is directly specified as $S(t,d)=\Cbbm(S_T(t),S_D(d))$. Here, $\Cbbm(\cdot,\cdot):[0,1]^2\rightarrow[0,1]$ is a bivariate copula function \citep{sklar1959functions}, and $S_T(t) \equiv \Pbb(T>t)$ and $S_D(d)\equiv \Pbb(D>d)$ are marginal survival functions of $T$ and $D$, respectively. This copula-based approach enjoys the flexibility from the separate modeling of (i) the marginals and (ii) the dependence structure between $T$ and $D$. \citet{shih1995inferences} proposed a two-stage estimation procedure for multivariate survival outcome under independent censoring, where, in the first stage, the marginals were estimated separately for each event without accounting for the between-event dependencies, and, in the second stage, the parameter associated with the copula was estimated via maximizing the log-likelihood with the plug-in estimates of the marginals. The resulting estimator for the copula parameter is referred to as the pseudo maximum likelihood estimator (PMLE). 

Accurately estimating the copula parameter relies on the consistent estimation of the marginals. The validity of the above stage-wise estimation procedure requires that individual event times are only subject to independent censoring. Thus, such an approach has been mostly used on clustered survival data. However, this assumption does not hold for semi-competing risks data because $T$ is dependently censored by $D$. In other words, standard tools such as the Kaplan-Meier (KM) estimator are not consistent for $S_T(\cdot)$, which poses challenges in estimating the copula parameter via the above two-stage procedure. As a result, earlier estimators were obtained from estimating equations based on a concordance indicator \citep{fine2001semi}, the Doob–Meyer (DM) decomposition of a counting process \citep{wang2003estimating}, or conditional Kendall's $\tau$ \citep{lakhal2008estimating}. Later, these procedures were extended to incorporate a discrete covariate for the marginal distribution via a regression model and for the between-event dependence parameter \citep{ghosh2006semiparametric,hsieh2008regression}. For estimating the marginal of $T$, they predominantly considered a \emph{plug-in} estimator, first proposed in \citet{fine2001semi}, through the relationship of $S_T(\cdot)$ with $S_D(\cdot)$ and the survival function $S_{T^*}(\cdot)$ of $T^* \equiv T\wedge D$, the time to the first event (non-terminal or terminal). However, this method is limited to the Archimedean copula family, which provides an explicit expression for such a relationship. Different from the above methods, \citet{peng2007regression} considered a class of time-varying coefficient semiparametric regression models for the marginals and copulas with time-varying parameters. They simultaneously estimated the copula parameter and the marginal of $T$ using non-linear estimating equations. 

\citet{chen2012maximum} considered the maximum likelihood estimation (MLE) approach, i.e., simultaneous estimation of all parameters associated with the marginals and the copula via maximizing the log-likelihood function. \citet{zhou2016semiparametric} proposed a PMLE for the copula parameter based on the log-likelihood function expressed in terms of $S_D(\cdot)$ and $S_{T^*}(\cdot)$ without covariates. \citet{zhu2021semiparametric} extended the procedure in \citet{zhou2016semiparametric} to incorporate covariates into the marginals specified by time-varying coefficient semiparametric regression models. They proposed a two-stage estimation procedure: in the first stage, the PMLE of the copula parameter was obtained following the idea in \citet{zhou2016semiparametric}, and, in the second stage, the regression coefficients associated with $S_T(\cdot)$ were estimated from non-linear estimating equations developed in \citet{peng2007regression}.

In this article, we propose a new estimation procedure for copula-based models to analyze semi-competing risks data, with the primary goals of investigating (i) the association between $T$ and $D$ and (ii) the direct covariate effects on their marginals. Along the line of \citet{chen2012maximum}, we consider a class of models where a parametric copula is employed to model the dependence between $T$ and $D$, and the marginals are specified by semi-parametric transformation models. We estimate the marginals and copula parameters via a two-stage PMLE approach. Specifically, in the first stage, we obtain a consistent estimate of the marginal of $D$ since it is only subject to independent censoring, and in the second stage, we estimate the marginal of $T$ and the copula parameter simultaneously by maximizing the pseudo-likelihood function with the plug-in estimates of $S_D(\cdot)$.

Our two-stage estimation approach adapts from the one-stage MLE method of \citet{chen2012maximum}, which is expected to be {asymptotically} more efficient but likely computationally challenging in practice. In contrast, our PMLE approach reduces the dimension of the parameter space in each stage while yielding consistent estimates. Despite not utilizing the association between $T$ and $D$ in the first stage, which might result in efficiency loss, we observed through simulation studies that, compared to the MLE, the PMLE approach gives more desirable finite-sample performances, including smaller biases for copula parameters. In addition, our approach enjoys robustness against copula misspecification when estimating the marginals. When the copula function is misspecified, our estimation of $S_D(\cdot)$ in the first stage is guaranteed to be consistent. Furthermore, although the PMLE of $S_T(\cdot)$ is not consistent under copula misspecification, with a consistent estimate of $S_D(\cdot)$, it leads to a smaller bias compared to the one-stage MLE, which is additionally affected by the biased estimates of $S_D(\cdot)$. 

{Our method exhibits three advantages compared to the stage-wise estimation procedure developed by \citet{zhu2021semiparametric}.} First, our method does not involve the first-event time $T^*$, and consequently, besides Archimedean copulas, it can also be applied to the elliptical copula family, e.g., Gaussian or $t$ copulas. Second, given the same estimator of the marginal of $D$ from both methods, our estimators for the marginal of $T$ and the copula parameter can be regarded as the MLE, but the estimator in \citet{zhu2021semiparametric} is still the PMLE based on plug-in estimates. Consequently, our estimators are inherently more efficient. Third, we provide variance estimators based on analytic expressions for inference. Compared to the non-parametric bootstrap employed in \citet{zhu2021semiparametric} and \citet{zhou2016semiparametric}, our inference procedure is computationally more efficient. To implement our method, an \textbf{R} package \textbf{PMLE4SCR} is available in the GitHub repository \url{https://github.com/michellezhou2009/PMLE4SCR}.

The remainder of the article is organized as follows. In Section \ref{sec:method}, we propose the two-stage PMLE estimation procedure and discuss its asymptotic properties. Section \ref{sec:numerical} presents simulation studies evaluating the finite-sample performance of the PMLE and the analysis of Bone Marrow Transplant (BMT) data as an illustration. We provide concluding remarks in Section \ref{sec:conclusion}. {Technical details and additional numerical results} are given in the Online Supplement of the Supporting Information.

\section{Method} \label{sec:method}

\subsection{Copula-based model specification}

Let $A$ denote the time to independent administrative censoring. The observed event times are denoted by $C\equiv D\wedge A$ and $X\equiv T\wedge C$, with respective censoring status defined as $\delta_D=\onebb(D\leq A)$ and $\delta_T=\onebb(T\leq C)$. In addition, let $\bZ\in\Rbb^p$ denote a vector of baseline covariates, and given $\bZ$, $A$ is independent of both $T$ and $D$, but $T$ and $D$ are associated. The observed data from $n$ independent subjects is denoted by $\Sfrak = \{(X_i, C_i, \delta_{T,i}, \delta_{D,i}, \bZ_i), i = 1, \ldots, n\}$. We have the following commonly adopted assumptions regarding the study setting and covariates:
\begin{itemize}[leftmargin=0.8cm]
    \item[(A1)] There exists a maximum follow-up time $\xi<\infty$ in the study such that $\Pbb(X>\xi|\bZ)>0$, $\Pbb(\delta_T=0,\delta_D=0, X = \xi|\bZ)>0$, $\Pbb(C>\xi|\bZ)>0$, and $\Pbb(\delta_T=1, \delta_D=0, C=\xi|\bZ)>0$, with probability 1;

    \item[(A2)] The covariates $\bZ$ is bounded with probability 1. 
\end{itemize}

{Assumptions (A1) - (A2) are regulatory conditions which have seen frequent adoption in the semi-competing risks literature, see \citet[\S 1.2]{fine2001semi}, \citet[Conditions 2 - 3]{zeng2006efficient}, \citet[Assumptions (b) - (c)]{Chen2010}, and \citet[Assumptions (II) - (III)]{chen2012maximum}. These two conditions are required to obtain the uniform convergence of the estimators for $S_T(t|\bZ)$ and $S_D(d|\bZ)$ on $[0,\xi]$, which will be given in the later sections. Assumption (A1) is plausible in practice because, in general, most studies terminate after a finite observation time due to various constraints, i.e., time and/or budget, and subjects who have not experienced any events throughout the entire follow-up are regarded as administratively censored. For example, in the critical care trial of acute lung injury patients \citep{ARDS2000} studied in \citet{Chen2024}, the administrative censoring (maximum follow-up) time is 180 days from when a patient is admitted to the intensive care unit.} 

Given covariates $\bZ$, the copula-based approach models the joint survival function of $(T,D)$, $\Pbb(T>t, D>d|\bZ)$, as a function of their respective marginals $S_T(t|\bZ)=\Pbb(T>t|\bZ)$ and $S_D(d|\bZ)=\Pbb(D > d|\bZ)$, i.e.,
\begin{equation}\label{equ:survival_copula}
	\Pbb(T>t, D>d|\bZ) = \Cbbm\left(S_T(t|\bZ), S_D(d|\bZ);\alpha\right),
\end{equation}
where $\Cbbm(\cdot,\cdot;\alpha): [0,1]^2 \rightarrow [0,1]$ is a copula function parametrically specified with parameter $\alpha$. In the remainder of this article, 
we define $\dot \Cbbm_{u_j} \equiv \partial\Cbbm(u_T,u_D;\alpha)/\partial u_j$, $j=T,D$, as the first-order partial derivatives of the copula function, and $\ddot \Cbbm_{u_Tu_D} \equiv \partial^2\Cbbm(u_T,u_D;\alpha)/\partial u_Tu_D$ as the second-order partial derivative. 
\begin{remark} \label{rmk:copula}
    The Archimedean and elliptical copula families are often selected for specifying $\Cbbm(\cdot,\cdot;\alpha)$. Archimedean copulas are defined via a generator function $\zeta(u;\alpha)$ for $u\in[0,1]$, which is continuous, convex, and strictly decreasing with $\zeta(1;\alpha)=0$. A bivariate Archimedean copula can be constructed as $\Cbbm(u_1,u_2;\alpha) = \zeta^{-1}\{\zeta(u_1;\alpha)+\zeta(u_2;\alpha);\alpha\}$. Below, we present some commonly used generator functions and corresponding bivariate copulas and refer readers to \citet{Nelsen2006} for a more detailed introduction.
    \begin{enumerate}
        \item[(i)] \textit{Clayton copula}, $\zeta(u;\alpha)=(u^{-\alpha}-1)/\alpha$ for $\alpha\in(-1,\infty)\backslash\{0\}$, and
        \begin{align} \label{eq:Clayton_copula}
            \Cbbm(u_1,u_2;\alpha)=\left\{\max\left(u_1^{-\alpha}+u_2^{-\alpha}-1,0\right)\right\}^{-1/\alpha};
        \end{align}
        \item[(ii)] \textit{Frank copula}, $\zeta(u;\alpha)=-\log\{(e^{-\alpha u}-1)/(e^{-\alpha}-1)\}$ for $\alpha\in\Rbb\backslash\{0\}$, and 
        \begin{align} \label{eq:Frank_copula}
            \Cbbm(u_1,u_2;\alpha)=-\alpha^{-1}\log\left\{1+\frac{(e^{-\alpha u_1}-1)(e^{-\alpha u_2}-1)}{e^{-\alpha}-1}\right\};
        \end{align}
        \item[(iii)] \textit{Gumbel copula}, $\zeta(u;\alpha)=\{-\log(u)\}^\alpha$ for $\alpha\in[1,\infty)$, and
        \begin{align}\label{eq:Gumbel_copula}
            \Cbbm(u_1,u_2;\alpha)=\exp\left(-\left[\{-\log(u_1)\}^\alpha +\{-\log(u_2)\}^\alpha\right]^{1/\alpha}\right).
        \end{align}
    \end{enumerate}
    Elliptical copulas join marginals via an elliptical distribution, e.g., a Gaussian distribution. A bivariate Gaussian copula with parameter $\alpha$ is defined as 
    \begin{align} \label{eq:Gaussian_copula}
        \Cbbm(u_1,u_2;\alpha) = \Phi_2\{\Phi^{-1}(u_1),\Phi^{-1}(u_2);\alpha\},
    \end{align}
    where $\Phi(\cdot)$ is the cumulative distribution function of the standard univariate Gaussian distribution, and $\Phi_2(\cdot,\cdot;\alpha)$ is the standard bivariate Gaussian cumulative distribution function with $\alpha$ being the correlation parameter.
\end{remark}

We specify the marginal survival functions of $T$ and $D$ via semi-parametric transformation models. Specifically, for $j=T,D$,
\begin{equation}\label{equ:marginal_semitrans}
	S_j(t|\bZ) = \exp\left[-G_j\left\{R_j(t) e^{\bbeta_j\trans\bZ} \right\}\right],
\end{equation}
where $\bbeta_j$ is a $p$-dimensional vector of regression coefficients, the unspecified baseline function $R_j(\cdot)$ is assumed to be increasing, and $G_j(\cdot)$ is a specified non-negative, strictly increasing, and continuously differentiable function. For example, if $G_j(t) = t$, \eqref{equ:marginal_semitrans} becomes a Cox proportional hazards (PH) model; if $G_j(t)=\log(1+t)$, \eqref{equ:marginal_semitrans} becomes a proportional odds model. The marginal survival function in \eqref{equ:marginal_semitrans} can be easily extended to accommodate time-varying covariates $\bZ(t)$, where $S_j(t|\bZ) = \exp[-G_j\{\int_0^ t e^{\bbeta_j\trans\bZ(s)}\text{d} R_j(s) \}]$.

\subsection{The two-stage estimation procedure}

Let $\btheta_D = \{\bbeta_D, R_D(\cdot)\}$ and $\btheta_T = \{\bbeta_T, R_T(\cdot)\}$ denote the parameters associated with the marginals of $D$ and $T$, respectively. We propose a two-stage estimation procedure for $\btheta=\{\btheta_D,\btheta_T,\alpha\}$, where, in the first stage, $\btheta_D$ is estimated by itself via an MLE approach with observed data $\Sfrak_D = \{(C_i, \delta_{D,i}, \bZ_i), i = 1,\ldots,n\}$, and, in the second stage, $\btheta_1\equiv\{\btheta_T,\alpha\}$ is estimated via maximizing a pseudo-log-likelihood depending on the estimate $\bthetahat_D$ obtained from the first stage. We hereafter refer to the proposed estimation procedure as the two-stage PMLE for brevity.

\subsubsection{The first stage estimation of $\btheta_D$} \label{subsec:first-stage}

In the semi-competing risks setting, $\btheta_D$ can be consistently estimated by an MLE with observed data $\Sfrak_D$, since $D$ is only subject to independent censoring by $A$ \citep{zeng2006efficient}. The baseline function $R_D(\cdot)$ is regarded as a non-decreasing step function that jumps only at the observed event times. Let $0<d_1 < d_2 < \cdots < d_{\kappa_D}\leq\xi$ denote the \emph{ordered} observed death times, and let $\rmd R_{D,l}>0$ denote the jump size of $R_D(\cdot)$ at time $d_l$, for $l=1,\ldots,\kappa_D$. Thus, $\btheta_D$ can be re-expressed as $\btheta_D = (\bbeta_D\trans, \rmd R_{D,1},\ldots, \rmd R_{D,\kappa_D})\trans$. The log-likelihood function of $\btheta_D$ given $\Sfrak_D$ is $\ell_D(\btheta_D)=n^{-1}\sum_{i=1}^n \ell_{D,i}(\btheta_D)$, where
\begin{equation}\label{equ:ll_D} 
    \begin{aligned}
	   \ell_{D,i}(\btheta_D) =& \int_0^\xi \left[\log \rmd R_D(t) + \bbeta_D\trans \bZ_i + \log \dot{G}_D\left\{R_D(t) e^{\bbeta_D\trans\bZ_i}\right\}\right]\rmd N_{D,i}(t)\\
    &-G_D\left\{R_D(C_i)e^{\bbeta_D\trans\bZ_i}\right\},
    \end{aligned}
\end{equation}
with $N_{D,i}(t) = \onebb(C_i\leq t)\delta_{D,i}$. The MLE $\bthetahat_D=(\bbetahat_D\trans,\widehat{\rmd R}_{D,1},\ldots,\widehat{\rmd R}_{D,\kappa_D})\trans$ is the solution to score equations $\Psi_{D}(\btheta_D) = n^{-1}\sum_{i=1}^n \Psi_{D,i}(\btheta_D) = 0$ with $\Psi_{D,i}(\btheta_D) = \partial\ell_{D,i}(\btheta_D)/\partial \btheta_D$. 

\subsubsection{The second stage estimation of $\btheta_1$} \label{subsec:second-stage}

Different from $\btheta_D$, $\btheta_T$ cannot be consistently estimated using only the observed data $\{(X_i, \delta_{T,i},\bZ_i),\allowbreak i=1, \ldots,n\}$, since $T$ is subject to both dependent and independent censoring by $D$ and $A$, respectively. A consistent estimator must be obtained from a full likelihood function based on the joint survival function of $(T,D)$ with $\Sfrak$ \citep{fine2001semi}. Similar to $R_D(\cdot)$, $R_T(\cdot)$ is also regarded as a non-decreasing step function that jumps only at the observed non-terminal event times $0<t_1 < t_2 <\cdots<t_{\kappa_T}\leq\xi$, where $\rmd R_{T,l}>0$ denotes the jump size of $R_T(\cdot)$ at time $t_l$, for $l=1,\ldots,\kappa_T$. The parameter $\btheta_T$ can be re-expressed as $\btheta_T=(\bbeta_T\trans, \rmd R_{T,1},\ldots, \rmd R_{T,\kappa_T})\trans$. The full log-likelihood function given $\Sfrak$ is $\ell(\btheta) = n^{-1}\sum_{i=1}^n \ell_i(\btheta)$, where
\begin{equation} \label{equ:full_ll}
\begin{aligned}
	\ell_i(\btheta) =&~ \delta_{T,i}\delta_{D,i}\log \ddot \Cbbm_{u_Tu_D} +   \delta_{T,i}(1 - \delta_{D,i})\log \dot \Cbbm_{u_T} \\
	&~+ (1-\delta_{T,i}) \delta_{D,i} \log \dot \Cbbm_{u_D} +  (1-\delta_{T,i}) (1-\delta_{D,i})\log \Cbbm\\
	&~+ \delta_{T,i} \left[-G_T\left\{R_T(X_i) e^{\bbeta_T\trans\bZ_i} \right\}  + \log \dot G_T\left\{R_T(X_i) e^{\bbeta_T\trans\bZ_i} \right\}\right. +\log \rmd R_T(X_i) + \bbeta_T\trans \bZ_i \Big] \\
	&~+ \delta_{D,i} \left[-G_D\left\{R_D(C_i) e^{\bbeta_D\trans\bZ_i} \right\}+ \log \dot G_D\left\{R_D(C_i) e^{\bbeta_D\trans\bZ_i} \right\}\right.+ \log \rmd R_D(C_i) + \bbeta_D\trans \bZ_i \Big].
\end{aligned}
\end{equation}
The PMLE of $\btheta_1$ is obtained by maximizing the following pseudo-log-likelihood function: 
\begin{align*}
   \ellhat(\btheta_1)\equiv\ell(\btheta_1,\bthetahat_D) = n^{-1} \sum_{i=1}^n \ell_i(\btheta_1, \bthetahat_D), 
\end{align*}
where $\ell_i$ is given in \eqref{equ:full_ll}. The PMLE $\bthetahat_1=(\alphahat,\bbetahat_T\trans,\widehat{\rmd R}_{T,1},\ldots,\widehat{\rmd R}_{T,\kappa_T})\trans$ is the solution to pseudo score equations $\Psihat_1(\btheta_1) = n^{-1}\sum_{i=1}^n \Psihat_{1,i}(\btheta_1) = 0$ with $\Psihat_{1,i}(\btheta_1)=\dot \ell_{\btheta_1,i}(\btheta_1,\bthetahat_D) = \partial\ell_i(\btheta_1, \bthetahat_D)/\partial \btheta_1$. The score function $\ell_{\btheta_1,i}(\btheta_1,\btheta_D)=(\Psi_{\alpha,i},\Psi_{\bbeta_T,i}\trans, \Psi_{\rmd R_{T1},i},\ldots, \Psi_{\rmd R_{T\kappa_T},i})\trans$, with the expressions given in Section S1 of the Online Supplement, is a function of $\btheta_D$ only through $U_{D,i}=S_D(C_i|\bZ_i;\btheta_D)$.


\begin{remark} \label{rmk:copula-cov}
    The copula in \eqref{equ:survival_copula} can be extended by allowing the copula parameter to depend on some baseline covariates. {{In the data analysis of \citet{chen2012maximum}, the copula parameter varies across different levels of a categorical variable. \cite{nikoloulopoulos2008multivariate} regressed the copula parameter on covariates for nonsurvival outcomes. Following these works,}} we can assume $\alpha = \phi(\bgamma\trans \bW)$, where $\bW$ is a subset of the covariates $\bZ$, and $\phi(\cdot)$ is a specified link function. The copula parameter thus becomes $\bgamma$, and its corresponding score function is modified as $\Psi_{\bgamma,i} = \Psi_{\alpha,i}\dot \phi(\bgamma\trans\bW_i) \bW_i$, where $\dot f(\cdot)$ denotes the first-order derivative of a univariate function $f(\cdot)$.
\end{remark}
\begin{remark}\label{rmk:ll}
    The log-likelihood function in \eqref{equ:full_ll} is equivalent to the log-likelihood in \citet{chen2012maximum} derived based on three observed counting processes: the first event being non-terminal, the first event being terminal, and the second event occurring subsequently to the uncensored first event. Therefore, the one-stage MLE obtained via maximizing \eqref{equ:full_ll} shares the same asymptotic properties given in Theorems 1 and 2 of \citet{chen2012maximum}.
\end{remark}

\subsection{Asymptotic properties of the two-stage PMLE}\label{sec:asymptotic}

We present the asymptotic properties of the two-stage PMLE. Define the following quantities:
\begin{align*}
    \Upsilon_1(g_T, g_D) = -\log \Cbbm\left(e^{-G_T(g_T)}, e^{-G_D(g_D)};\alpha\right)~~\text{and}~~\eta_{1j}(g_T,g_D) = \frac{\partial}{\partial g_j} \Upsilon_1(g_T, g_D),
\end{align*}
for $j=T,D$; also
\begin{align*}
    \Upsilon_2(g_T,g_D) = - \log \dot \Cbbm_{u_T}\left(e^{-G_T(g_T)}, e^{-G_D(g_D)};\alpha\right)~~\text{and}~~\eta_2(g_T,g_D) =  \frac{\partial}{\partial g_D} \Upsilon_2(g_T, g_D).
\end{align*}
Let $\btheta^0 = \{\alpha^0, \bbeta_T^0, R_T^0(\cdot), \bbeta_D^0, R_D^0(\cdot)\}$ be the true values of $\btheta$, and $\boldsymbol{\Theta}$ denote the total parameter space. The following conditions are required to establish the asymptotic properties of $\bthetahat=(\bthetahat_D\trans,\bthetahat_1\trans)\trans$.

\begin{itemize}[leftmargin=0.8cm]
	
    \item[(C1)] The true baseline functions $R_T^0(\cdot)$ and $R_D^0(\cdot)$ are strictly increasing and continuously differentiable, and the true values $\alpha^0$, $\bbeta_T^0$ and $\bbeta_D^0$, fall in the interior of a compact parameter space. 
	
	

    \item[(C2)] For any sequence $0<h_1<\cdots<h_r\leq g$,
    \begin{equation*}
        \prod_{s=1}^r\{(1+h_s)\dot G_D(h_s)\}\exp\{-G_D(g)\}\leq \mu_0^r(1+g)^{-\nu_0},
    \end{equation*}
    where $\mu_0$ and $\nu_0$ are positive constants.

	
    \item[(C3)] The function $G_j(\cdot)$ satisfies the conditions: $G_j(0)=0$, $G_j(\infty) = \infty$, and $\dot G_j(0)>0$ for $j=T,D$. The copula function $\Cbbm(\cdot,\cdot;\alpha)$ satisfies the conditions: $\Cbbm(0,0;\alpha) = 0$ and $\eta_{1j}(0,0)>0$ for $j=T,D$. Also, for positive constants $c_1$, $c_2$, and $c_3$,
	\begin{align*}
		& \lim_{g_1,g_2\rightarrow \infty}\frac{\sum_{j=1}^2 \log\{g_{j}\times\sup_{s_1\leq g_1, s_2\leq g_2}\eta_{1j}(s_1,s_2)\}}{\Upsilon_1(c_1g_1, c_2g_2)} = 0,\displaybreak[0]\\
		& \lim_{g_1,g_2\rightarrow \infty}  \frac{ \log\{g_2\times\sup_{s_1\leq g_1, s_2\leq g_2}\eta_2(s_1,s_2)\}}{\inf_{w_1\geq 0, w_2\geq 0}\{\Upsilon_2(c_1w_1, c_3g_2)-\Upsilon_2(c_1w_1, c_2w_2)\}} = 0.
	\end{align*} 
	

    \item[(C4)] $\sum_{i=1}^n \ell_i(\btheta)$, $\sum_{i=1}^n \dot \ell_{\btheta_1,i}(\btheta)$, $\sum_{i=1}^n \ddot \ell_{\btheta_1\btheta_1,i}(\btheta)$, and $\sum_{i=1}^n \ddot \ell_{\btheta_1u_D,i}(\btheta)$ are continuous and bounded on $[0,\xi]\times \boldsymbol{\Theta}$, where $\dot \ell_{\btheta_1,i}(\btheta) = \partial \ell_i(\btheta) /\partial \btheta_1$, $\ddot \ell_{\btheta_1\btheta_1,i}(\btheta) = \partial^2 \ell_i(\btheta)/\partial \btheta_1\partial \btheta_1\trans$, and $\ddot \ell_{\btheta_1 u_D,i}(\btheta) = \partial^2 \ell_i(\btheta)/\partial \btheta_1\partial u_D$.
	
    \item[(C5)] Information matrices $\Ical_D(\btheta_D)=\Ebb\{\Ical_{D,i}(\btheta_D)\}$ with $\Ical_{D,i}(\btheta_D)=-\partial^2\ell_{D,i}(\btheta_D)/\partial\btheta_D\partial \btheta_D\trans$ and $\Ical_1(\btheta) =\Ebb\{\Ical_{1,i}(\btheta)\}$ with $\Ical_{1,i}(\btheta)=-\partial ^2 \ell_i(\btheta)/\partial \btheta_1\partial \btheta_1\trans$ evaluated at the true value of $\btheta_D^0$ and $\btheta^0$ are positive definite, and their eigenvalues are bounded below and above by some finite positive constants. The explicit expressions of $\Ical_D(\btheta_D)$ and $\Ical_1(\btheta)$ are given in Section S2 of the Online Supplement.
 
    \item[(C6)] The variance-covariance matrix $\Var\{\dot \ell_{\btheta_1,i}(\btheta) + n^{-1}\sum_{k=1}^n \ddot \ell_{\btheta_1u_D,k}(\btheta) \psi_{u_{D,k},i}(\btheta_D)\}$ evaluated at $\btheta^0$ is positive definite, and its eigenvalues are bounded below and above by some finite positive constants, where $\psi_{u_{D,k},i}(\btheta_D)$ is given in Section S3 of the Online Supplement.
	
\end{itemize}

{Assumptions (A1) - (A2) and Conditions (C1) - (C2)} are from the regulatory conditions listed in \citet{zeng2006efficient} for establishing the consistency and asymptotic distribution of $\bthetahat_D$. Specifically, with probability 1, $\bbetahat_D$ converges to $\bbeta_D^0$ and $\widehat R_D(\cdot)\equiv\{\widehat{\rmd R}_{D,l},l=1,\ldots,\kappa_D\}$ converges to $R_D^0(\cdot)$ uniformly in the interval $[0,\xi]$. In addition, $\sqrt{n}(\bthetahat_D-\btheta_D^0)$ converges weakly to a zero-mean Gaussian process. Let $\Ical_D^0= \Ebb\{\Ical_{D,i}(\btheta_D ^ 0)\}$. For a Hamadard differentiable functional $F_D(\btheta_D)$ of $\btheta_D$ \citep{chen2012maximum}, the functional delta method suggests that $\sqrt{n}\{F_D(\bthetahat_D)-F_D(\btheta_D^0)\}$ converges weakly to a zero-mean Gaussian process with the variance-covariance matrix $\dot F_D(\btheta_D^0)\trans(\Ical_D ^ 0)^{-1}\dot F_D(\btheta_D^0)$, where $\dot F_D(\btheta_D)$ is the derivative of $F_D(\btheta_D)$ with respect to $\btheta_D$. The matrix $(\Ical_D^0)^{-1}$ is estimated by $\bar\Ical_{D,n}^{-1}(\bthetahat_D)$ where $\bar\Ical_{D,n}(\bthetahat_D)=n^{-1}\sum_{i=1}^n\Ical_{D,i}(\bthetahat_D)$. Alternatively, one can estimate $(\Ical_D ^ 0)^{-1}$ via a robust estimator, given by $\widehat\bSigma_{D,n} \equiv n^{-1}\sum_{i=1}^n \psi_{\btheta_D,i}(\bthetahat_D)\psi_{\btheta_D,i}(\bthetahat_D)\trans$ with $\psi_{\btheta_D,i}(\bthetahat_D) = \bar\Ical_{D,n}^{-1}(\bthetahat_D)\Psi_{D,i}(\bthetahat_D)$, where $\Psi_{D,i}(\btheta_D)$ is the score function for the log-likelihood function $\ell_{D,i}$ discussed in Section \ref{subsec:first-stage}.

{Condition (C3)} was adopted in \citet{chen2012maximum}, which regulates the tail behavior of the estimated baseline functions, and together with {Assumption (A1)}, the boundedness of $\widehat R_D(\cdot)$ and $\widehat R_T(\cdot)$ is guaranteed. {Conditions (C4) - (C5)} are standard conditions guaranteeing the existence of the score functions, information matrices, and certain components for the variance-covariance matrix of $\bthetahat_1$. { Condition (C6)} assumes the variance-covariance matrix of $\bthetahat_1$ is well-defined. Additionally, since, for the Gumbel copula, $\eta_{1j}(0,0)>0$ {(in Condition (C3))} does not hold, we adopt the same modification as \citet{chen2012maximum} by adding a small positive value, e.g., $n^{-1}$, to the cumulative hazard function, $G_j\{R_j(t)e^{\bbeta_j\trans\bZ}\}$, at time 0 to make $\eta_{1j}(0,0)$ positive for $j=T,D$.

The consistency and asymptotic distribution of $\bthetahat_1$ are established in the following Theorems \ref{thm:PMLE-consistency} and \ref{thm:PMLE-asymptotic}, respectively.
\begin{theorem}\label{thm:PMLE-consistency}
    {Assume Assumptions (A1) - (A2) and Conditions (C1) - (C6) hold.} With probability 1, $\alphahat$ converges to $\alpha^0$, $\bbetahat_T$ converges to $\bbeta_T^0$, and $\widehat R_T(\cdot)\equiv\{\widehat{\rmd R}_{T,l},l=1,\ldots,\kappa_T\}$ converges to $R_T^0(\cdot)$ uniformly in the interval $[0,\xi]$.
\end{theorem}

\begin{theorem}\label{thm:PMLE-asymptotic}
    {Assume Assumptions (A1) - (A2) and Conditions (C1) - (C6) hold,} and then $\sqrt{n}(\bthetahat_1 - \btheta_1^0)$ converges weakly to a zero-mean Gaussian process. Let $\Ical_1^0=\Ebb\{\Ical_{1,i}(\btheta^0)\}$. For a Hamadard differentiable functional $F_1(\btheta_1)$ of $\btheta_1$, $\sqrt{n}\{F_1(\bthetahat_1)-F_1(\btheta_1^0)\}$ converges weakly to a zero-mean Gaussian process with the variance-covariance matrix $\dot F_1(\btheta_1^0)\trans\bSigma_1^0\dot F(\btheta_1^0)$, where $\dot F_1(\btheta_1)$ is the derivative of $F_1(\btheta_1)$ with respect to $\btheta_1$, and $\bSigma_1^0 \equiv \Var\{\psi_{\btheta_1,i}(\btheta^0)\}$ with 
    \begin{align*}
        \psi_{\btheta_1,i}(\btheta^0) = \left(\Ical_1^0\right) ^{-1} \left\{\dot \ell_{\btheta_1,i}(\btheta^0) + n^{-1}\sum_{k=1}^n \ddot \ell_{\btheta_1u_D,k}(\btheta^0) \psi_{u_{D,k},i}(\btheta_D^0) \right\}.
    \end{align*}
	The expression of $\psi_{u_{D,k},i}(\btheta_D^0)$ is given in Section S3 of the Online Supplement.
\end{theorem}

An application of Theorem \ref{thm:PMLE-asymptotic} is to consider linear functionals in the form of { $ \varphi = \bb_{\beta_T}\trans\bbeta_T+\int_0^\xi b_{R_T}(t)\rmd R_T(t)+b_\alpha\alphahat$ for $F_1(\btheta_1)$, where $b_\alpha$ and $\bb_{\beta_T}$ are a known scalar and vector, respectively, and $\bb_{R_T}\equiv(b_{R_T,1},\ldots,b_{R_T,\kappa_T})\trans$ is the known vector of the values of $b_{R_T}(t)$ evaluated at $0< t_1< t_2<\cdots<t_{\kappa_T}\leq\xi$. The asymptotic variance of $\varphi$ is given by $\bb_1\trans\bSigma_1^0\bb_1$, where $\bb_1=(\bb_{\beta_T}\trans,\bb_{R_T}\trans,b_\alpha)\trans$}. The variance-covariance matrix $\bSigma_1^0$ can be consistently estimated by
\begin{equation}\label{eq:Sigmahat}
    \widehat\bSigma_{1,n} = n^{-1}\sum_{i=1}^n \psi_{\btheta_1,i}(\bthetahat)\psi_{\btheta_1,i}(\bthetahat)\trans,
\end{equation}
where 
\begin{align*}
    \psi_{\btheta_1,i}(\bthetahat) = \bar\Ical_{1,n}^{-1}(\bthetahat) \left\{\dot \ell_{\btheta_1,i}(\bthetahat) + n^{-1}\sum_{k=1}^n \ddot \ell_{\btheta_1u_D,k}(\bthetahat) \psi_{u_{D,k},i}(\bthetahat_D)\right\}
\end{align*}
with $\bar\Ical_{1,n}(\bthetahat) = n^{-1}\sum_{i=1}^n \Ical_{1,i}(\bthetahat)$. This estimator can be re-written in the form of a robust sandwich variance estimator: $ \widehat\bSigma_{1,n}  = \bar\Ical_{1,n}^{-1}(\bthetahat)\bar\Vcal_{1,n}(\bthetahat)\bar\Ical_{1,n}^{-1}(\bthetahat)$ with $\bar\Vcal_{1,n}(\bthetahat)=n^{-1}\sum_{i=1}^n \{\dot \ell_{\btheta_1,i}(\bthetahat) + n^{-1}\sum_{k=1}^n \ddot \ell_{\btheta_1u_D,k}(\bthetahat) \psi_{u_{D,k},i}(\bthetahat_D)\}\{\dot \ell_{\btheta_1,i}(\bthetahat) + n^{-1}\sum_{k=1}^n \ddot \ell_{\btheta_1u_D,k}(\bthetahat) \psi_{u_{D,k},i}(\bthetahat_D)\}\trans$. 

{ With such an analytic variance estimator, we can construct a 95\% confidence interval (CI) for $\alpha$: $\alphahat \pm 1.96 \times s_{\alphahat}$, where the standard error (SE) $s_{\alphahat}$ is the square root of the last diagonal element of $\widehat\bSigma_{1,n}$ (i.e., $\bb_{\beta_T}={\bf 0}$, $\bb_{R_T}={\bf 0}$, and $b_\alpha=1$); similarly, a 95\% CI for $\beta_{T,k}$, the $k$-th element of $\bbeta_{T}$: $\widehat \beta_{T,k} \pm 1.96 \times s_{\widehat\beta_{T,k}}$, where $\widehat \beta_{T,k}$ is the $k$-th element of $\widehat\bbeta_T$, and the SE $s_{\widehat\beta_{T,k}}$ is the square root of the $k$-th diagonal element of $\widehat\bSigma_{1,n}$ (i.e., $\bb_{\beta_T}$ is a vector with the $k$-th element being one and other elements being zero, $\bb_{R_T}={\bf 0}$, and $b_\alpha = 0$). For a given $t_0$, let $k_{t_0}=\sum_{k=1}^{\kappa_T}\onebb (t_k \leq t_0)$. A point estimate of the baseline survival probabilities $S_{T,0}(t_0)=\exp[-G_T\{R_T(t_0)\}]$ is $\widehat S_T(t_0) = \exp[-G_T\{\widehat R_T(t_0)\}]$ with $\widehat R_T(t_0) = \bb_{R_T(t_0)}\trans \widehat{\rmd R}_T$, where $\widehat{\rmd R}_T=(\widehat{\rmd R}_{T,1}, \cdots, \widehat{\rmd R}_{T,\kappa_T})\trans$, and   $\bb_{R_T(t_0)}$ is a $\kappa_T$-dimensional vector with the first $k_{t_0}$ elements being one, and the rest being zero. In addition, a 95\% CI for $S_{T,0}(t_0)$ is $[e^{-G_T(R_L)}, e^{-G_T(R_U)}]$ with $[R_L, R_U] = \widehat R_T(t_0) \pm 1.96 \times s_{\widehat R_T(t_0)}$, where the SE $s_{\widehat R_T(t_0)}$ is the square root of $\bb_1\trans \widehat\bSigma_{1,n}\bb_1$ with $\bb_1 = ({\bf 0}\trans, \bb_{R_T(t_0)}\trans,0)\trans$ (i.e., $\bb_{\beta_T}={\bf 0}$ and $b_\alpha = 0$).}

\section{Numerical studies} \label{sec:numerical}


\subsection{Simulation studies}

\subsubsection{Data generation}

We generated baseline covariates $Z_{1,i}\sim\Ncal(1, 0.5)\times\onebb[0,2]$, a truncated normal distribution, and $Z_{2,i}\sim \Bcal(0.8)$, a Bernoulli distribution. Given $\bZ_i=(Z_{1,i},Z_{2,i})\trans$, the outcomes $(T,D)$ were generated from $\log(T_i/3) = -\beta_{T,1} Z_{1,i} - \beta_{T,2} Z_{2,i} + e_{T,i}$ and $\log(D_i/3) = - \beta_{D,1} Z_{1,i} - \beta_{D,2} Z_{2,i} + e_{D,i}$, where $(e_{T,i}, e_{D,i})$ followed a bivariate distribution that corresponds to a certain parametric copula function. Here, we considered two commonly used copula families: Gumbel and Clayton, with different strengths of association specified by Kendall's $\tau$, defined as $\tau=4\int_{[0,1]^2}\Cbbm(u_1,u_2)\rmd\Cbbm(u_1,u_2) - 1$. Kendall's $\tau$ has a one-to-one correspondence with $\alpha$: for example, $\tau \equiv \tau(\alpha) = (\alpha-1)/\alpha$ for Gumbel and $\tau(\alpha)=\alpha/(\alpha+2)$ for Clayton. We considered three dependence levels $\tau=0.4,0.6,0.8$. As an example, to generate $(e_{T,i}, e_{D,i})$ from the Gumbel copula with $\tau = 0.6$, we first generated $(u_{T,i},u_{D,i})$ with parameter $\alpha = 2.5$ corresponding to $\tau = 0.6$, and then obtained {{$e_{j,i} = \log\{-\log(u_{j,i})\}$}} for $j=T,D$. We set $\bbeta_D = (0.2, 0)\trans$, which leads to about 18\% of death times censored, and considered two sets of $\bbeta_T = (1,1)\trans$ and $(1,0.5)\trans$, which leads to different censoring rates, 3\%-12\%, and 5\%-18\%, receptively, for the non-terminal event. 

\subsubsection{Simulation study I}

The first simulation study focused on evaluating the estimation accuracy and efficiency of the proposed PMLE approach and comparing its finite-sample performances with the one-stage MLE method from \citet{chen2012maximum} under each combination of $(\tau,\bbeta_T)$. The sample size $n$ was set to 200 and 400, with $1,000$ replications of simulated datasets. We specified $G_T(t) = G_D(t) = t$ (i.e., PH model) for the marginal distributions and the copula function as the true copula family. Both the PMLE and one-stage MLE were obtained {{based on}} the same log-likelihood function in \eqref{equ:full_ll} using the {\bf R} package {\bf trust} \citep{greyer2020trust}, which implements the trust region optimization algorithm \citep{fletcher2000practical}. { The \textbf{trust} package requires users to provide the starting values of the parameters. To achieve faster convergence, we used the following estimates instead of random values. For $\btheta_D$ and $\btheta_T$, we fit marginal models with the data $(C_i, \delta_{D,i}, \bZ_i)$ and $(X_i, \delta_{T,i}, \bZ_i)$, correspondingly. These two estimates were plugged into \eqref{equ:full_ll}, and the resulting pseudo-log-likelihood function was maximized to obtain the starting value of the copula parameter $\alpha$.} For most combinations of the copula family and $(n, \tau, \bbeta_T)$, the optimization converged for all $1,000$ replications. The exceptions are under Clayton with Kendall's $\tau = 0.8$, there were seven and four non-converged replications for $\bbeta_T = (1,1)\trans$ and $(1,0.5)\trans$ respectively with $n = 200$, and three and one non-converged replications for $\bbeta_T = (1,1)\trans$ and $(1,0.5)\trans$ respectively with $n = 400$.

{ Tables \ref{tab:est_gumbel} and \ref{tab:est_clayton} present the summary statistics of the MLE and PMLE of $(\bbeta_T,\alpha)$ for sample size $n=200$ under the Gumbel and Clayton copula, respectively. In addition, we examined the estimation of the baseline survival function of $T$, where, specifically, three time points, $t_0=0.863$, $2.079$, and $4.159$, were selected such that the corresponding baseline survival probability $S_{T,0}(t_0) = 0.75$, $0.5$, and $0.25$. Tables \ref{tab:BHaz_gumbel200} and \ref{tab:BHaz_clayton200} present the summary statistics for the MLE and PMLE of these three baseline survival probabilities. The summary statistics include the relative bias (BIAS), empirical standard deviation (ESD), average standard error (ASE), and the root mean square error (rMSE), as well as the empirical coverage percentage of the 95\% CI using the analytic standard error estimator. The PMLE's SEs were calculated via the procedure described in Section \ref{sec:asymptotic}. The MLE's SEs were calculated by a robust variance-covariance matrix estimator: $\bar\Ical_{n}^{-1}(\bthetahat_{\mbox{\tiny MLE}})\bar\Vcal_n(\bthetahat_{\mbox{\tiny MLE}})\bar\Ical_{n}^{-1}(\bthetahat_{\mbox{\tiny MLE}})$ with $\bar\Ical_{n}(\btheta)=-n^{-1}\sum_{i=1}^n \ddot \ell_{\btheta\btheta,i}(\btheta)$ and $\bar\Vcal_{n}(\btheta)=n^{-1}\sum_{i=1}^n \dot \ell_{\btheta,i}(\btheta)\allowbreak \dot \ell_{\btheta,i}(\btheta)\trans$, where  $\dot  \ell_{\btheta,i}(\btheta) = \partial \ell_i(\btheta)/\partial \btheta$ and $\ddot \ell_{\btheta\btheta,i}(\btheta) = \partial ^2 \ell_i(\btheta)/\partial \btheta\partial \btheta\trans$ are the first-order and second-order derivatives of the log-likelihood function $\ell_i(\btheta)$ in \eqref{equ:full_ll} with respect to $\btheta$. The results for sample size $n=400$ are reported in Web Tables 1 - 4 of the Online Supplement.

For both copula families, the PMLE performs well with a small bias relative to the estimation variability. As the sample size $n$ increases, the bias decreases, demonstrating its consistency, and the ESD and, consequently, rMSE also decrease. The censoring rate also affects PMLE's performance. Compared to $\bbeta_T = (1,1)\trans$, $\bbeta_T = (1,0.5)\trans$, resulting in a higher censoring rate, leads to a larger BIAS, ESD, and rMSE. The same pattern is also observed for the MLE. In addition, for most scenarios, the PMLE's ASE is close to its ESD, and the empirical coverage percentages of the CI are close to the nominal level, 95\%. These results confirm the validity of our analytic SE estimator. The only exception is for the baseline survival probability $S_{T,0}(t_0)=0.25$, where the empirical coverage percentages for both MLE and PMLE are significantly lower than 95\%. This is because fewer subjects have observed events at a larger time point due to censoring. 

Compared with the MLE, the PMLE's ESD for the marginal distribution parameters is generally larger, which is expected since the MLE is asymptotically more efficient. The PMLE of the copula parameter $\alpha$ has a smaller ESD in some scenarios, especially for sample size $n=200$. When the sample size increases to $400$, the lead by the PMLE diminishes, and for some scenarios, the MLE exhibits better efficiency. To confirm the relationship of asymptotic efficiency between the PMLE and the MLE, we conducted additional simulations for sample size $n=1,000$ under the scenario of Kendall's $\tau=0.8$ and $\bbeta_T=(1,1)\trans$; the results are reported in Web Table 5 of the Online Supplement. When the sample size increases to $1,000$, the MLE has a smaller ESD than the PMLE, confirming that the MLE is asymptotically more efficient.

We also compared the computational time between the PMLE and MLE with sample sizes $n=200$, $400$, and $1,000$ for Kendall's $\tau=0.8$ and $\bbeta_T=(1,1)\trans$. When the copula family is Gumbel, using a Mac Air with Apple M2 chip and 24 GB memory, the average running times of 10 replications for the PMLE are 0.53 secs, 2.14 secs, and 20.5 secs, for sample sizes $n=200$, $400$, and $1,000$, respectively, and the average running times for MLE are 1.85 secs, 10.5 secs, and 200 secs. When the copula family is Clayton, the average running times for the three sample sizes are 0.33 secs, 1.16 secs, 13.8 secs for the PMLE, 1.02 secs, 5.6 secs, and 171 secs for MLE. This, together with the estimation results, shows that the PMLE is computationally more efficient than the MLE while maintaining comparable performance. Furthermore, we compared the MLE of the marginal distribution parameter for $D$ with our first-stage estimator $\widehat \btheta_D$ using its own survival outcome without accounting for the dependence between $T$ and $D$. The results are reported in Web Tables 6 - 13 of the Online Supplement. Note that the three time points $t_0=0.863, 2.079, 4.159$, selected for the baseline probabilities of $T$, also leads to the baseline probabilities of D being $S_{D,0}(t_0)=0.75, 0.5, 0.25$, respectively. Both estimators have a small bias, and their rMSE is dominated by the variance. The MLE has a smaller ESD because it utilizes extra information on the joint distribution of $(T,D)$.
}


\begin{table}[htbp]
    \centering
    \begin{tabular}{lll rccc>{}c}
        \toprule
        $\tau$ & $\bbeta_{T}$ & Method & \multicolumn{1}{c}{BIAS} & ESD & ASE & rMSE & CP  \\
        \midrule
        \multicolumn{3}{c}{} & \multicolumn{5}{c}{$\widehat\beta_{T,1}$} \\
        \cmidrule{4-8}
        \multirow{4}{*}{0.4} & \multirow{2}{*}{(1,1)}   & MLE & 0.005 & 0.160 & 0.160 & 0.160 & 94.0\\
        &  & PMLE & 0.006 & 0.162 & 0.163 & 0.162 & 94.7\\      
        \cmidrule{4-8} 
        & \multirow{2}{*}{(1,0.5)}  & MLE & 0.006 & 0.163 & 0.163 & 0.163 & 94.6\\
        &  & PMLE & 0.007 & 0.165 & 0.166 & 0.166 & 94.7\\
        \midrule
        \multirow{4}{*}{0.6} & \multirow{2}{*}{(1,1)}  & MLE & 0.001 & 0.156 & 0.157 & 0.156 & 94.3\\
        &  & PMLE & 0.002 & 0.163 & 0.165 & 0.163 & 94.7\\
        \cmidrule{4-8} 
        & \multirow{2}{*}{(1,0.5)}  & MLE & 0.000 & 0.158 & 0.158 & 0.158 & 94.6\\
        &  & PMLE & 0.003 & 0.164 & 0.166 & 0.164 & 94.4\\
        \midrule
        \multirow{4}{*}{0.8} & \multirow{2}{*}{(1,1)}  & MLE & 0.007 & 0.156 & 0.156 & 0.156 & 94.4\\
        &  & PMLE & 0.005 & 0.168 & 0.171 & 0.168 & 95.3\\
        \cmidrule{4-8} 
        & \multirow{2}{*}{(1,0.5)}  & MLE & 0.005 & 0.157 & 0.157 & 0.157 & 94.4\\
        &  & PMLE & 0.006 & 0.169 & 0.171 & 0.169 & 94.8\\
        \midrule 
        \multicolumn{3}{c}{} & \multicolumn{5}{c}{$\widehat\beta_{T,2}$}\\
        \cline{4-8}
        \multirow{4}{*}{0.4} & \multirow{2}{*}{(1,1)}   & MLE & 0.012 & 0.203 & 0.197 & 0.203 & 94.2\\
        &  & PMLE & 0.009 & 0.204 & 0.201 & 0.205 & 95.4\\
        \cmidrule{4-8} 
        & \multirow{2}{*}{(1,0.5)}   & MLE & 0.024 & 0.393 & 0.378 & 0.394 & 93.8\\
        &  & PMLE & 0.020 & 0.396 & 0.385 & 0.397 & 93.8\\
        
        \midrule
        \multirow{4}{*}{0.6} & \multirow{2}{*}{(1,1)}   & MLE & 0.016 & 0.198 & 0.191 & 0.198 & 93.6\\
        &  & PMLE & 0.009 & 0.204 & 0.200 & 0.204 & 95.2\\
        \cmidrule{4-8} 
        & \multirow{2}{*}{(1,0.5)}   & MLE & 0.025 & 0.383 & 0.366 & 0.383 & 93.3\\
        &  & PMLE & 0.015 & 0.394 & 0.384 & 0.394 & 93.9\\
        \midrule 
        \multirow{4}{*}{0.8} & \multirow{2}{*}{(1,1)}  & MLE & 0.024 & 0.196 & 0.189 & 0.197 & 93.4\\
        &  & PMLE & 0.006 & 0.211 & 0.205 & 0.211 & 94.6\\
        \cmidrule{4-8} 
        & \multirow{2}{*}{(1,0.5)}  & MLE & 0.033 & 0.382 & 0.363 & 0.383 & 93.2\\
        &  & PMLE & 0.011 & 0.409 & 0.394 & 0.409 & 94.5\\
        \midrule 
        \multicolumn{3}{c}{} &\multicolumn{5}{c}{$\widehat\alpha$} \\
        \cmidrule{4-8}
        \multirow{4}{*}{0.4} & \multirow{2}{*}{(1,1)}  & MLE & 0.036 & 0.075 & 0.078 & 0.083 & 96.8\\
        &  & PMLE & 0.023 & 0.073 & 0.075 & 0.077 & 96.4\\
        \cmidrule{4-8} 
        & \multirow{2}{*}{(1,0.5)}   & MLE & 0.037 & 0.078 & 0.080 & 0.086 & 96.1\\
        &  & PMLE & 0.024 & 0.075 & 0.076 & 0.079 & 96.3\\
        \midrule
        \multirow{4}{*}{0.6} & \multirow{2}{*}{(1,1)} & MLE & 0.054 & 0.087 & 0.088 & 0.103 & 93.6\\
        &  & PMLE & 0.026 & 0.083 & 0.084 & 0.087 & 95.6\\
        
        \cmidrule{4-8} 
        & \multirow{2}{*}{(1,0.5)} & MLE & 0.055 & 0.089 & 0.090 & 0.105 & 94.1\\
        &  & PMLE & 0.027 & 0.084 & 0.085 & 0.088 & 95.7\\
        \midrule
        \multirow{4}{*}{0.8} & \multirow{2}{*}{(1,1)}  & MLE & 0.096 & 0.105 & 0.104 & 0.142 & 89.7\\
        &  & PMLE & 0.020 & 0.093 & 0.095 & 0.095 & 95.3\\
        \cmidrule{4-8} 
        & \multirow{2}{*}{(1,0.5)}   & MLE & 0.094 & 0.106 & 0.105 & 0.142 & 89.5\\
        &  & PMLE & 0.020 & 0.093 & 0.095 & 0.096 & 95.3\\
        \bottomrule 
    \end{tabular}
    \vspace{0.4cm}
    \caption{Simulation results for the estimates of $(\bbeta_T,\alpha)$ under the Gumbel copula with sample size $n=200$. BIAS: relative bias; ESD: relative empirical standard deviation; ASE: relative average standard error; rMSE: relative root mean square error; CP: empirical coverage percentage of 95\% confidence interval.}
    \label{tab:est_gumbel}
\end{table}

\begin{table}[htbp]
	\centering
\begin{tabular}{lll rccc>{}c}
	\toprule
	$\tau$ & $\bbeta_{T}$ & Method & \multicolumn{1}{c}{BIAS} & ESD & ASE & rMSE & CP  \\
	\midrule
	\multicolumn{3}{c}{} & \multicolumn{5}{c}{$\widehat\beta_{T,1}$} \\
	\cmidrule{4-8}
	\multirow{4}{*}{0.4} & \multirow{2}{*}{(1,1)}  & MLE & 0.014 & 0.154 & 0.152 & 0.155 & 94.6\\
	&  & PMLE & 0.014 & 0.160 & 0.160 & 0.161 & 94.7\\      
	\cmidrule{4-8} 
	& \multirow{2}{*}{(1,0.5)}  & MLE & 0.016 & 0.160 & 0.157 & 0.161 & 93.5\\
	&  & PMLE & 0.016 & 0.166 & 0.165 & 0.167 & 93.8\\
	\midrule
	\multirow{4}{*}{0.6} & \multirow{2}{*}{(1,1)} & MLE & 0.018 & 0.143 & 0.139 & 0.144 & 93.5\\
	&  & PMLE & 0.018 & 0.161 & 0.163 & 0.162 & 95.6\\
	\cmidrule{4-8} 
	& \multirow{2}{*}{(1,0.5)} & MLE & 0.022 & 0.147 & 0.142 & 0.149 & 93.6\\
	&  & PMLE & 0.021 & 0.165 & 0.166 & 0.167 & 95.3\\
	\midrule
	\multirow{4}{*}{0.8} & \multirow{2}{*}{(1,1)} & MLE & 0.034 & 0.133 & 0.133 & 0.137 & 94.4\\
	&  & PMLE & 0.027 & 0.173 & 0.174 & 0.175 & 96.0\\
	\cmidrule{4-8} 
	& \multirow{2}{*}{(1,0.5)} & MLE & 0.035 & 0.135 & 0.135 & 0.140 & 94.9\\
	&  & PMLE & 0.029 & 0.174 & 0.175 & 0.177 & 95.7\\
	\midrule 
	\multicolumn{3}{c}{} & \multicolumn{5}{c}{$\widehat\beta_{T,2}$}\\
	\cline{4-8}
	\multirow{4}{*}{0.4} & \multirow{2}{*}{(1,1)} & MLE & 0.009 & 0.197 & 0.191 & 0.197 & 93.6\\
	&  & PMLE & 0.003 & 0.207 & 0.201 & 0.207 & 93.8\\
	\cmidrule{4-8} 
	& \multirow{2}{*}{(1,0.5)} & MLE & 0.026 & 0.372 & 0.359 & 0.373 & 93.4\\
	&  & PMLE & 0.008 & 0.392 & 0.381 & 0.393 & 93.2\\
	\midrule
	\multirow{4}{*}{0.6} & \multirow{2}{*}{(1,1)} & MLE & 0.008 & 0.177 & 0.171 & 0.177 & 93.4\\
	&  & PMLE & 0.012 & 0.210 & 0.201 & 0.210 & 93.9\\
	\cmidrule{4-8} 
	& \multirow{2}{*}{(1,0.5)} & MLE & 0.023 & 0.332 & 0.317 & 0.333 & 93.6\\
	&  & PMLE & 0.005 & 0.392 & 0.377 & 0.392 & 92.8\\
	\midrule 
	\multirow{4}{*}{0.8} & \multirow{2}{*}{(1,1)} & MLE & 0.001 & 0.161 & 0.158 & 0.161 & 94.2\\
	&  & PMLE & 0.025 & 0.223 & 0.210 & 0.225 & 93.6\\
	\cmidrule{4-8} 
	& \multirow{2}{*}{(1,0.5)} & MLE & 0.011 & 0.303 & 0.295 & 0.304 & 94.1\\
	&  & PMLE & 0.019 & 0.413 & 0.393 & 0.414 & 93.5\\
	\midrule 
	\multicolumn{3}{c}{} &\multicolumn{5}{c}{$\widehat\alpha$} \\
	\cmidrule{4-8}
	\multirow{4}{*}{0.4} & \multirow{2}{*}{(1,1)} & MLE & 0.063 & 0.186 & 0.186 & 0.196 & 94.8\\
	&  & PMLE & 0.026 & 0.184 & 0.183 & 0.186 & 94.4\\
	
	\cmidrule{4-8} 
	& \multirow{2}{*}{(1,0.5)} & MLE & 0.061 & 0.191 & 0.190 & 0.201 & 94.7\\
	&  & PMLE & 0.025 & 0.189 & 0.186 & 0.190 & 94.6\\
	\midrule
	\multirow{4}{*}{0.6} & \multirow{2}{*}{(1,1)} & MLE & 0.067 & 0.148 & 0.146 & 0.162 & 94.8\\
	&  & PMLE & 0.011 & 0.146 & 0.145 & 0.146 & 94.5\\
	\cmidrule{4-8} 
	& \multirow{2}{*}{(1,0.5)} & MLE & 0.065 & 0.151 & 0.148 & 0.165 & 94.1\\
	&  & PMLE & 0.010 & 0.149 & 0.146 & 0.150 & 93.6\\
	\midrule
	\multirow{4}{*}{0.8} & \multirow{2}{*}{(1,1)} & MLE & 0.101 & 0.138 & 0.136 & 0.171 & 92.1\\
	&  & PMLE & 0.011 & 0.125 & 0.128 & 0.126 & 95.1\\
	\cmidrule{4-8} 
	& \multirow{2}{*}{(1,0.5)}  & MLE & 0.098 & 0.139 & 0.136 & 0.170 & 92.7\\
	&  & PMLE & 0.013 & 0.127 & 0.128 & 0.127 & 93.7\\
	\bottomrule 
\end{tabular}
    \vspace{0.4cm}
	\caption{Simulation results for the estimates of $(\bbeta_T,\alpha)$ under the Clayton copula with $n=200$. BIAS: relative bias; ESD: relative empirical standard deviation; ASE: relative average standard error; rMSE: relative root mean square error; CP: empirical coverage percentage of 95\% confidence interval.}
	\label{tab:est_clayton}
\end{table}

\begin{table}[htbp]
	\centering
	\begin{tabular}{ll cc l rcccc}
		\toprule
		$\tau$ & $\bbeta_{T}$ & Time & Prob. & Method & \multicolumn{1}{c}{BIAS} & ESD & ASE & rMSE & CP  \\
		\midrule
		\multirow{12}{*}{0.4} & \multirow{6}{*}{(1,1)}  & \multirow{2}{*}{0.863} & \multirow{2}{*}{0.750} & MLE &  0.007 & 0.077 & 0.076 & 0.078 & 94.3\\
		&  & &  & PMLE & 0.005 & 0.078 & 0.077 & 0.078 & 94.2\\		
		\cmidrule{3-10} 
		&  & \multirow{2}{*}{2.079} & \multirow{2}{*}{0.500} & MLE & 0.009 & 0.172 & 0.170 & 0.173 & 93.4\\
		&  &  &  & PMLE & 0.006 & 0.174 & 0.172 & 0.174 & 93.5\\
		\cmidrule{3-10} 
		&  & \multirow{2}{*}{4.159} & \multirow{2}{*}{0.250} & MLE & 0.162 & 0.397 & 0.334 & 0.428 & 82.4\\
		&  &  &  & PMLE & 0.168 & 0.397 & 0.338 & 0.432 & 81.7\\  
		\cmidrule{3-10} 
		& \multirow{6}{*}{(1,0.5)} & \multirow{2}{*}{0.863} & \multirow{2}{*}{0.750} & MLE & 0.007 & 0.077 & 0.075 & 0.077 & 94.0\\
		&  &  &  & PMLE & 0.005 & 0.077 & 0.076 & 0.078 & 93.6\\
		\cmidrule{3-10} 
		&  & \multirow{2}{*}{2.079} & \multirow{2}{*}{0.500} & MLE & 0.008 & 0.169 & 0.169 & 0.170 & 93.3\\
		&  & &  & PMLE & 0.005 & 0.170 & 0.171 & 0.171 & 93.8\\
		\cmidrule{3-10} 
		&  & \multirow{2}{*}{4.159} & \multirow{2}{*}{0.250} & MLE & 0.066 & 0.373 & 0.339 & 0.379 & 89.0\\
		&  &  &  & PMLE & 0.073 & 0.375 & 0.343 & 0.382 & 89.0\\
		\midrule
		\multirow{12}{*}{0.6} & \multirow{6}{*}{(1,1)} & \multirow{2}{*}{0.863} & \multirow{2}{*}{0.750} & MLE & 0.009 & 0.075 & 0.074 & 0.075 & 95.2\\
		&  &  &  & PMLE & 0.006 & 0.077 & 0.076 & 0.077 & 94.8\\
		\cmidrule{3-10} 
		& & \multirow{2}{*}{2.079} & \multirow{2}{*}{0.500} & MLE & 0.009 & 0.166 & 0.165 & 0.166 & 94.2\\
		&  &  &  & PMLE & 0.004 & 0.169 & 0.170 & 0.169 & 94.0\\
		\cmidrule{3-10} 
		&  & \multirow{2}{*}{4.159} & \multirow{2}{*}{0.250} & MLE & 0.133 & 0.379 & 0.327 & 0.401 & 84.1\\
		&  &  &  & PMLE & 0.142 & 0.382 & 0.337 & 0.408 & 84.7\\
		\cmidrule{3-10}  
		& \multirow{6}{*}{(1,0.5)} & \multirow{2}{*}{0.863} & \multirow{2}{*}{0.750} & MLE & 0.008 & 0.074 & 0.073 & 0.075 & 94.5\\
		&  &  &  & PMLE & 0.005 & 0.076 & 0.075 & 0.076 & 94.7\\
		\cmidrule{3-10} 
		&  & \multirow{2}{*}{2.079} & \multirow{2}{*}{0.500} & MLE & 0.008 & 0.162 & 0.163 & 0.162 & 93.8\\
		&  &  &  & PMLE & 0.003 & 0.166 & 0.168 & 0.166 & 93.9\\
		\cmidrule{3-10} 
		&  & \multirow{2}{*}{4.159} & \multirow{2}{*}{0.250} & MLE & 0.052 & 0.357 & 0.328 & 0.360 & 89.2\\
		&  &  &  & PMLE & 0.060 & 0.362 & 0.339 & 0.367 & 89.6\\
		\midrule
		\multirow{12}{*}{0.8} & \multirow{6}{*}{(1,1)} & \multirow{2}{*}{0.863} & \multirow{2}{*}{0.750} & MLE & 0.011 & 0.073 & 0.072 & 0.074 & 95.7\\
		&  &  &  & PMLE & 0.005 & 0.077 & 0.076 & 0.077 & 94.4\\
		\cmidrule{3-10}
		&  & \multirow{2}{*}{2.079} & \multirow{2}{*}{0.500} & MLE & 0.008 & 0.163 & 0.162 & 0.163 & 93.9\\
		&  &  &  & PMLE & 0.001 & 0.171 & 0.171 & 0.171 & 93.5\\
		\cmidrule{3-10} 
		&  & \multirow{2}{*}{4.159} & \multirow{2}{*}{0.250} & MLE & 0.116 & 0.359 & 0.321 & 0.378 & 86.3\\
		&  &  &  & PMLE & 0.130 & 0.369 & 0.340 & 0.391 & 86.8\\
		\cmidrule{3-10} 
		& \multirow{6}{*}{(1,0.5)} & \multirow{2}{*}{0.863} & \multirow{2}{*}{0.750} & MLE & 0.011 & 0.073 & 0.072 & 0.074 & 95.2\\
		&  &  &  & PMLE & 0.004 & 0.077 & 0.075 & 0.077 & 94.2\\
		\cmidrule{3-10}
		&  & \multirow{2}{*}{2.079} & \multirow{2}{*}{0.500} & MLE & 0.007 & 0.160 & 0.160 & 0.160 & 93.4\\
		&  &  &  & PMLE & 0.000 & 0.169 & 0.169 & 0.169 & 93.5\\
		\cmidrule{3-10}
		&  & \multirow{2}{*}{4.159} & \multirow{2}{*}{0.250} & MLE & 0.050 & 0.338 & 0.323 & 0.341 & 90.7\\
		&  &  &  & PMLE & 0.062 & 0.349 & 0.342 & 0.355 & 91.2\\
		\bottomrule 
	\end{tabular}
	\vspace{0.4cm}
	\caption{Simulation results for the baseline survival function of $T$ under the Gumbel copula with sample size $n=200$. BIAS: relative bias; ESD:  relative empirical standard deviation; ASE:  relative average standard error; rMSE: relative root mean square error; CP: empirical coverage percentage of 95\% confidence interval.}
	\label{tab:BHaz_gumbel200}
\end{table}

\begin{table}[htbp]
	\centering
	\begin{tabular}{ll cc l rcccc}
		\toprule
		$\tau$ & $\bbeta_{T}$ & Time & Prob. & Method & \multicolumn{1}{c}{BIAS} & ESD & ASE & rMSE & CP  \\
		\midrule
		\multirow{12}{*}{0.4} & \multirow{6}{*}{(1,1)}  & \multirow{2}{*}{0.863} & \multirow{2}{*}{0.750} & MLE & 0.003 & 0.077 & 0.075 & 0.077 & 93.2\\
		&  &  &  & PMLE & 0.001 & 0.078 & 0.078 & 0.078 & 93.5\\		
		\cmidrule{3-10} 
		&  & \multirow{2}{*}{2.079} & \multirow{2}{*}{0.500}& MLE & 0.002 & 0.168 & 0.163 & 0.168 & 93.2\\
		&  &  &  & PMLE & 0.005 & 0.172 & 0.170 & 0.172 & 93.1\\
		\cmidrule{3-10} 
		&  & \multirow{2}{*}{4.159} & \multirow{2}{*}{0.250} & MLE & 0.134 & 0.380 & 0.313 & 0.403 & 81.6\\
		&  &  &  & PMLE & 0.151 & 0.381 & 0.326 & 0.410 & 82.5\\ 
		\cmidrule{3-10} 
		& \multirow{6}{*}{(1,0.5)} & \multirow{2}{*}{0.863} & \multirow{2}{*}{0.750} & MLE & 0.002 & 0.076 & 0.075 & 0.076 & 92.9\\
		&  &  &  & PMLE & 0.000 & 0.077 & 0.077 & 0.077 & 93.5\\
		\cmidrule{3-10} 
		&  & \multirow{2}{*}{2.079} & \multirow{2}{*}{0.500} & MLE & 0.001 & 0.164 & 0.164 & 0.164 & 92.9\\
		&  &  &  & PMLE & 0.005 & 0.168 & 0.171 & 0.169 & 93.3\\
		\cmidrule{3-10} 
		&  & \multirow{2}{*}{4.159} & \multirow{2}{*}{0.250} & MLE & 0.056 & 0.350 & 0.319 & 0.355 & 89.1\\
		&  &  &  & PMLE & 0.073 & 0.358 & 0.333 & 0.365 & 89.2\\
		\midrule
		\multirow{12}{*}{0.6} & \multirow{6}{*}{(1,1)} & \multirow{2}{*}{0.863} & \multirow{2}{*}{0.750} & MLE & 0.000 & 0.069 & 0.068 & 0.069 & 92.8\\
		&  &  &  & PMLE & 0.001 & 0.077 & 0.076 & 0.077 & 92.6\\
		\cmidrule{3-10} 
		& & \multirow{2}{*}{2.079} & \multirow{2}{*}{0.500} & MLE & 0.000 & 0.151 & 0.146 & 0.151 & 92.9\\
		&  &  &  & PMLE & 0.009 & 0.167 & 0.165 & 0.167 & 93.1\\
		\cmidrule{3-10} 
		&  & \multirow{2}{*}{4.159} & \multirow{2}{*}{0.250} & MLE & 0.120 & 0.341 & 0.286 & 0.362 & 83.1\\
		&  &  &  & PMLE & 0.145 & 0.358 & 0.321 & 0.386 & 85.0\\
		\cmidrule{3-10}  
		& \multirow{6}{*}{(1,0.5)} & \multirow{2}{*}{0.863} & \multirow{2}{*}{0.750} & MLE & 0.002 & 0.068 & 0.067 & 0.068 & 92.8\\
		&  &  &  & PMLE & 0.002 & 0.076 & 0.075 & 0.076 & 93.1\\
		\cmidrule{3-10} 
		&  & \multirow{2}{*}{2.079} & \multirow{2}{*}{0.500} & MLE & 0.003 & 0.148 & 0.146 & 0.148 & 93.2\\
		&  &  &  & PMLE & 0.007 & 0.165 & 0.165 & 0.165 & 92.9\\
		\cmidrule{3-10} 
		&  & \multirow{2}{*}{4.159} & \multirow{2}{*}{0.250}& MLE & 0.048 & 0.309 & 0.290 & 0.312 & 90.5\\
		&  &  &  & PMLE & 0.070 & 0.334 & 0.325 & 0.341 & 91.0\\
		\midrule
		\multirow{12}{*}{0.8} & \multirow{6}{*}{(1,1)} & \multirow{2}{*}{0.863} & \multirow{2}{*}{0.750} & MLE & 0.006 & 0.063 & 0.062 & 0.063 & 93.2\\
		&  &  &  & PMLE & 0.005 & 0.078 & 0.076 & 0.079 & 92.6\\
		\cmidrule{3-10}
		&  & \multirow{2}{*}{2.079} & \multirow{2}{*}{0.500}& MLE & 0.008 & 0.139 & 0.137 & 0.139 & 93.5\\
		&  &  &  & PMLE & 0.014 & 0.169 & 0.168 & 0.170 & 92.0\\
		\cmidrule{3-10} 
		&  & \multirow{2}{*}{4.159} & \multirow{2}{*}{0.250} & MLE & 0.141 & 0.320 & 0.272 & 0.350 & 83.7\\
		&  &  &  & PMLE & 0.171 & 0.362 & 0.330 & 0.400 & 85.5\\
		\cmidrule{3-10} 
		& \multirow{6}{*}{(1,0.5)} & \multirow{2}{*}{0.863} & \multirow{2}{*}{0.750} & MLE & 0.009 & 0.062 & 0.062 & 0.063 & 92.5\\
		&  &  &  & PMLE & 0.005 & 0.078 & 0.077 & 0.078 & 92.9\\
		\cmidrule{3-10}
		&  & \multirow{2}{*}{2.079} & \multirow{2}{*}{0.500} & MLE & 0.009 & 0.135 & 0.136 & 0.135 & 93.4\\
		&  &  &  & PMLE & 0.010 & 0.167 & 0.168 & 0.167 & 92.7\\
		\cmidrule{3-10}
		&  & \multirow{2}{*}{4.159} & \multirow{2}{*}{0.250} & MLE & 0.058 & 0.288 & 0.275 & 0.293 & 90.9\\
		&  &  &  & PMLE & 0.079 & 0.338 & 0.335 & 0.347 & 90.7\\
		\bottomrule 
	\end{tabular}
	\vspace{0.4cm}
	\caption{Simulation results for the baseline survival function of $T$ under the Clayton copula with sample size $n=200$. BIAS: relative bias; ESD:  relative empirical standard deviation; ASE:  relative average standard error; rMSE: relative root mean square error; CP: empirical coverage percentage of 95\% confidence interval.}
	\label{tab:BHaz_clayton200}
\end{table}

\subsubsection{Simulation study II}

We conducted a second simulation study to assess the robustness of the PMLE and MLE under the misspecification of the copula function. Specifically, we considered two settings: (i) we generated $(T,D)$ from the Gumbel copula but fit the Clayton copula, and (ii) vice versa. The sample size was $n=400$, Kendall's $\tau=0.6$, and $\bbeta_T=(1,0.5)\trans$. When the copula function is misspecified, the MLE or PMLE estimator $\widehat \alpha$ will not converge to the true value of the copula parameter under the true copula family, and the plug-in estimate of Kendall's $\tau$, $\widehat\tau = \tau(\widehat \alpha)$, will not converge to the true Kendall's $\tau$ either. Thus, it is more meaningful to focus on estimating the marginals in this setting. Table \ref{tab:misspeci} presents the BIAS, ESD, and rMSE for $\bbeta_D$ and $\bbeta_T$ with or without the copula misspecification. {When the copula is correctly specified, the bias of the PMLE and MLE for $\bbeta_D$ are similar; when the copula is misspecified, the PMLE's bias remains unchanged. This is because the PMLE estimate of $\bbeta_D$ is obtained without the information of the copula function and, consequently, is still a consistent estimator under copula misspecification. In contrast, the MLE's bias increases substantially under the copula misspecification.} For the estimation of $\bbeta_T$, both the MLE and PMLE perform less ideally due to the misspecification. Still, when the true copula is Gumbel, but the Clayton copula is fitted, the PMLE gives a much smaller bias, demonstrating better finite-sample robustness. When the true copula is Clayton but the Gumbel copula is fitted, the biases for the PMLE and MLE are similar.

\begin{table}[htbp]
	\centering
	\begin{tabular}{ll rr cc cc}
		\toprule
		Parameter & Method & \multicolumn{2}{c}{BIAS} & \multicolumn{2}{c}{ESD} & \multicolumn{2}{c}{MSE}  \\
		\midrule
		\multicolumn{2}{l}{True: Gumbel} & \multicolumn{1}{c}{GC} & \multicolumn{1}{c}{GG} & GC & GG & GC & GG  \\
		\midrule
		\multirow{2}{*}{$\beta_{D,1}$} & MLE & 0.070 & 0.003 & 0.108 & 0.111 & 0.129 & 0.111 \\
		& PMLE & 0.002 & 0.002 & 0.118 & 0.118 & 0.118 & 0.118 \\
		\multirow{2}{*}{$\beta_{D,2}$} & MLE & 0.019 & 0.002 & 0.122 & 0.134 & 0.124 & 0.134 \\
		& PMLE & 0.001 & 0.001 & 0.144 & 0.144 & 0.144 & 0.144 \\
		\multirow{2}{*}{$\beta_{T,1}$} & MLE & 0.195 & 0.001 & 0.116 & 0.113 & 0.226 & 0.113 \\
		& PMLE & 0.078 & 0.002 & 0.115 & 0.118 & 0.139 & 0.118 \\
		\multirow{2}{*}{$\beta_{T,2}$} & MLE & 0.104 & 0.008 & 0.130 & 0.137 & 0.166 & 0.137 \\
		& PMLE & 0.045 & 0.006 & 0.137 & 0.143 & 0.144 & 0.143 \\
		\midrule
		\multicolumn{2}{l}{True: Clayton} & \multicolumn{1}{c}{CG} & \multicolumn{1}{c}{CC} & CG & CC & CG & CC \\
		\midrule
		\multirow{2}{*}{$\beta_{D,1}$} & MLE & 0.053 & 0.005 & 0.123 & 0.095 & 0.133 & 0.095 \\
		& PMLE & 0.003 & 0.003 & 0.117 & 0.117 & 0.117 & 0.117 \\
		\multirow{2}{*}{$\beta_{D,2}$} & MLE & 0.021 & 0.013 & 0.149 & 0.112 & 0.151 & 0.113 \\
		& PMLE & 0.001 & 0.001 & 0.145 & 0.145 & 0.145 & 0.145 \\
		\multirow{2}{*}{$\beta_{T,1}$} & MLE & 0.041 & 0.011 & 0.123 & 0.103 & 0.129 & 0.103 \\
		& PMLE & 0.041 & 0.013 & 0.125 & 0.116 & 0.131 & 0.117 \\
		\multirow{2}{*}{$\beta_{T,2}$} & MLE & 0.013 & 0.006 & 0.152 & 0.118 & 0.152 & 0.118 \\
		& PMLE & 0.018 & 0.003 & 0.154 & 0.140 & 0.155 & 0.140 \\
		\bottomrule
	\end{tabular}
    \vspace{0.4cm}
	\caption{Simulation results for copula misspecification. GC: true Gumbel - fit Clayton; GG: Gumbel - Gumbel; GC: Gumbel - Clayton; CG: Clayton - Gumbel; CC: Clayton - Clayton; CG: Clayton - Gumbel. BIAS: bias; ESD: empirical standard deviation; rMSE: root mean square error.}
	\label{tab:misspeci}
\end{table}

\subsubsection{Simulation study III}

We also compared our PMLE with the two-stage estimator developed in \citet{zhu2021semiparametric} under the Gumbel copula. Specifically, \citet{zhu2021semiparametric}, in the first stage, estimated the regression coefficient $\bbeta_D$ first, which produced $\widehat U_{D,i}=S_D(X_{D,i} | \bZ_i;\bbetahat_D)$, and with the estimated marginal of $D$ and a non-parametric estimator of the marginal of $T^*$ (time to the first event, which is only subject to independent censoring), a PMLE of the copula parameter was obtained. In the second stage, the time-varying regression coefficient, $\bbeta_T(t)$, was estimated via a separate set of non-linear estimating equations. For comparison purposes, we followed the same simulation setting of \citet{zhu2021semiparametric}. Since the marginal specifications are different, we focused on the comparison of the copula parameter estimation in terms of the BIAS, ESD, and rMSE, presented in Table \ref{tab:copula_zhu}, where the results of \citet{zhu2021semiparametric}'s method were extracted from Table 1 of their paper directly. It needs to be pointed out that the definition of the Gumbel copula parameter in \citet{zhu2021semiparametric} is different from ours, and their Gumbel copula function is defined as
\begin{equation*}
    \Cbbm(u_1,u_2;\alpha^*)=\exp\left(-\left[\{-\log(u_1)\}^{1/\alpha^*} +\{-\log(u_2)\}^{1 / \alpha ^ *}\right]^{\alpha^*}\right),
\end{equation*}
which indicates that our Gumbel copula parameter in \eqref{eq:Gumbel_copula} $\alpha = 1 / \alpha^*$. Based on this relationship, we obtained the PMLE of $\alpha^*$ by $\alphahat ^ * = 1/\alphahat$, and with this estimator, we calculated the summary statistics presented in Table \ref{tab:copula_zhu}. Across all the combinations $(n, \tau,\bbeta_T)$, our PMLE has given more desirable performances in all three metrics. 

\begin{table}[htbp]
    \centering
    \begin{tabular}{llll rcc rcc}
        \toprule
        \multicolumn{4}{c}{} & \multicolumn{3}{c}{$n=200$} & \multicolumn{3}{c}{$n=400$} \\
        \cmidrule{5-10} 
        $\tau$ & $\alpha ^ *$ & $\bbeta_T$ & Method & \multicolumn{1}{c}{BIAS} & ESD & rMSE & \multicolumn{1}{c}{BIAS} & ESD & rMSE \\
        \midrule
        \multirow{4}{*}{0.4} &  \multirow{4}{*}{0.6} & \multirow{2}{*}{(1,1)} & PMLE & 0.011 & 0.042 & 0.043 & 0.005 & 0.030 & 0.030 \\
        &  & & \citet{zhu2021semiparametric} & 0.034 & 0.059 & 0.068 & 0.028 & 0.058 & 0.064 \\
        \cmidrule{5-10} 
        & & \multirow{2}{*}{(1,0.5)} & PMLE & 0.011 & 0.043 & 0.044 & 0.006 & 0.031 & 0.031 \\
        &  & & \citet{zhu2021semiparametric} & 0.032 & 0.061 & 0.069 & 0.029 & 0.059 & 0.066 \\
        \midrule 
        \multirow{4}{*}{0.6} &  \multirow{4}{*}{0.4} & \multirow{2}{*}{(1,1)} & PMLE & 0.008 & 0.031 & 0.032 & 0.004 & 0.022 & 0.023 \\
        &  & & \citet{zhu2021semiparametric} & 0.058 & 0.064 & 0.086 & 0.041 & 0.057 & 0.070 \\
        \cmidrule{5-10} 
        & & \multirow{2}{*}{(1,0.5)} & PMLE & 0.008 & 0.032 & 0.033 & 0.004 & 0.023 & 0.023 \\
        &  & & \citet{zhu2021semiparametric} & 0.051 & 0.060 & 0.079 & 0.039 & 0.056 & 0.068 \\
        \midrule 
        \multirow{4}{*}{0.8} &  \multirow{4}{*}{0.2} & \multirow{2}{*}{(1,1)} & PMLE & 0.002 & 0.018 & 0.018 & 0.001 & 0.012 & 0.012 \\
        &  & & \citet{zhu2021semiparametric} & 0.078 & 0.056 & 0.096 & 0.057 & 0.042 & 0.071 \\
        \cmidrule{5-10} 
        & & \multirow{2}{*}{(1,0.5)} & PMLE & 0.002 & 0.018 & 0.018 & 0.001 & 0.012 & 0.012 \\
        &  & & \citet{zhu2021semiparametric} & 0.068 & 0.048 & 0.083 & 0.055 & 0.042 & 0.069 \\
        \bottomrule 
    \end{tabular}
    \vspace{0.4cm}
    \caption{Simulation results comparing the proposed PMLE and the method in \citet{zhu2021semiparametric}. BIAS: bias; ESD: empirical standard deviation; rMSE: root mean square error. $\alpha^*$ is the copula parameter of Gumbel copula given by $\Cbbm(u_1,u_2;\alpha^*)=\exp(-[\{-\log(u_1)\}^{1/\alpha^*} +\{-\log(u_2)\}^{1 / \alpha ^ *}]^{\alpha^*})$.}
    \label{tab:copula_zhu}
\end{table}

\subsection{Analysis of the BMT data}

We applied the proposed method to the BMT data, available in the \textbf{R} package \textbf{SemiCompRisks}, which consists of 137 patients with acute myelocytic leukemia (AML) or acute lymphoblastic leukemia (ALL) aged 7 to 52 \citep{klein2003survival}. This data set contains time to death (terminal event, $D$), subject to the administrative censoring of study time, and time to relapse (non-terminal event, $T$), which is subject to censoring by both death and study time. We were interested in investigating the dependence between times to relapse and death and the effects of disease groups, AML-low, AML-high, and ALL, on the time to relapse and death. Here, AML-low was set as the baseline group. 

The PH models were assumed for both marginal distributions, and three copula families, Clayton, Gumbel, and Frank, were considered. In addition, we assumed that under each copula family, the copula parameter may vary with the disease group. Both the MLE and PMLE approaches were implemented to estimate the parameters. {{For inference, the standard errors of PMLE and MLE were calculated by the same procedure used for the simulation study I.}} 

The parameter estimates with their standard errors (in the parentheses) are shown in  Table \ref{tab:data_ana}. For the dependence between the bivariate event times, we calculated the plug-in estimate of Kendall's $\tau$: $\widehat \tau = \tau(\widehat \alpha)$ under each copula family, where $\widehat \alpha$ is the MLE or PMLE of $\alpha$. As described earlier, $\tau(\alpha) = (\alpha - 1)/\alpha$ for Gumbel, and $\tau(\alpha) =\alpha/(\alpha+2)$ for Clayton. In addition, for Frank, $\tau(\alpha)= 1 - \frac{4}{\alpha}+\frac{4}{\alpha}\int_0^\alpha \frac{t}{e^t - 1}\rmd t$. The standard error of $\widehat \tau$ was obtained via the delta method: $\widehat \sigma_{\tau} = |\dot \tau(\alphahat)|\widehat\sigma_{\alpha}$, where $\widehat \sigma_{\tau}$ and $\widehat \sigma_{\alpha}$ denote the standard error for $\widehat\tau$ and $\alphahat$, respectively. 

\begin{table}[htbp]
	\centering
	\begin{tabu}{c cc cc cc}
		\toprule
		Covariates & \multicolumn{2}{c}{Gumbel} & \multicolumn{2}{c}{Clayton}  & \multicolumn{2}{c}{Frank}\\
        \cmidrule{2-7}
        & PMLE & MLE & PMLE & MLE & PMLE & MLE \\
        \midrule
        \multicolumn{7}{l}{Regression coefficients for leukemia relapse}\\
        AML high & 1.239 (0.317) & 1.147 (0.313) & 1.168 (0.311) & 1.116 (0.306) & 1.137 (0.302) & 1.032 (0.305) \\
        ALL & 0.854 (0.345) & 0.764 (0.325) & 0.710 (0.324) & 0.669 (0.320) & 0.716 (0.318) & 0.645 (0.318) \\
        \midrule
        \multicolumn{7}{l}{Regression coefficients for death}\\
        AML high & 1.022 (0.276) & 0.953 (0.281) & 1.022 (0.276) & 0.977 (0.271) & 1.022 (0.276) & 0.905 (0.272) \\
        ALL & 0.611 (0.285) & 0.553 (0.284) & 0.611 (0.285) & 0.577 (0.280) & 0.611 (0.285) & 0.535 (0.279) \\
        \midrule
        \multicolumn{7}{l}{Kendall's $\tau$ between event times}\\
        AML low & 0.679 (0.140) & 0.709 (0.122) & 0.814 (0.084) & 0.821 (0.079) & 0.737 (0.105) & 0.738 (0.106) \\
        AML high & 0.726 (0.090) & 0.769 (0.089) & 0.769 (0.108) & 0.777 (0.102) & 0.760 (0.094) & 0.770 (0.095) \\
        ALL & 0.686 (0.090) & 0.726 (0.079) & 0.767 (0.100) & 0.773 (0.093) & 0.721 (0.094) & 0.730 (0.095) \\
        \midrule
        \rowfont{}Log-Likelihood & -4.474 & -4.456 & -4.436 & -4.432 & -4.447 & -4.443 \\
        \bottomrule
	\end{tabu}
    \vspace{0.4cm}
	\caption{Estimation results on the BMT data using the PMLE and MLE methods. Standard errors are given in parentheses.}
	\label{tab:data_ana}
\end{table}

\begin{figure}[htbp]
    \centering
    \includegraphics[width=\textwidth]{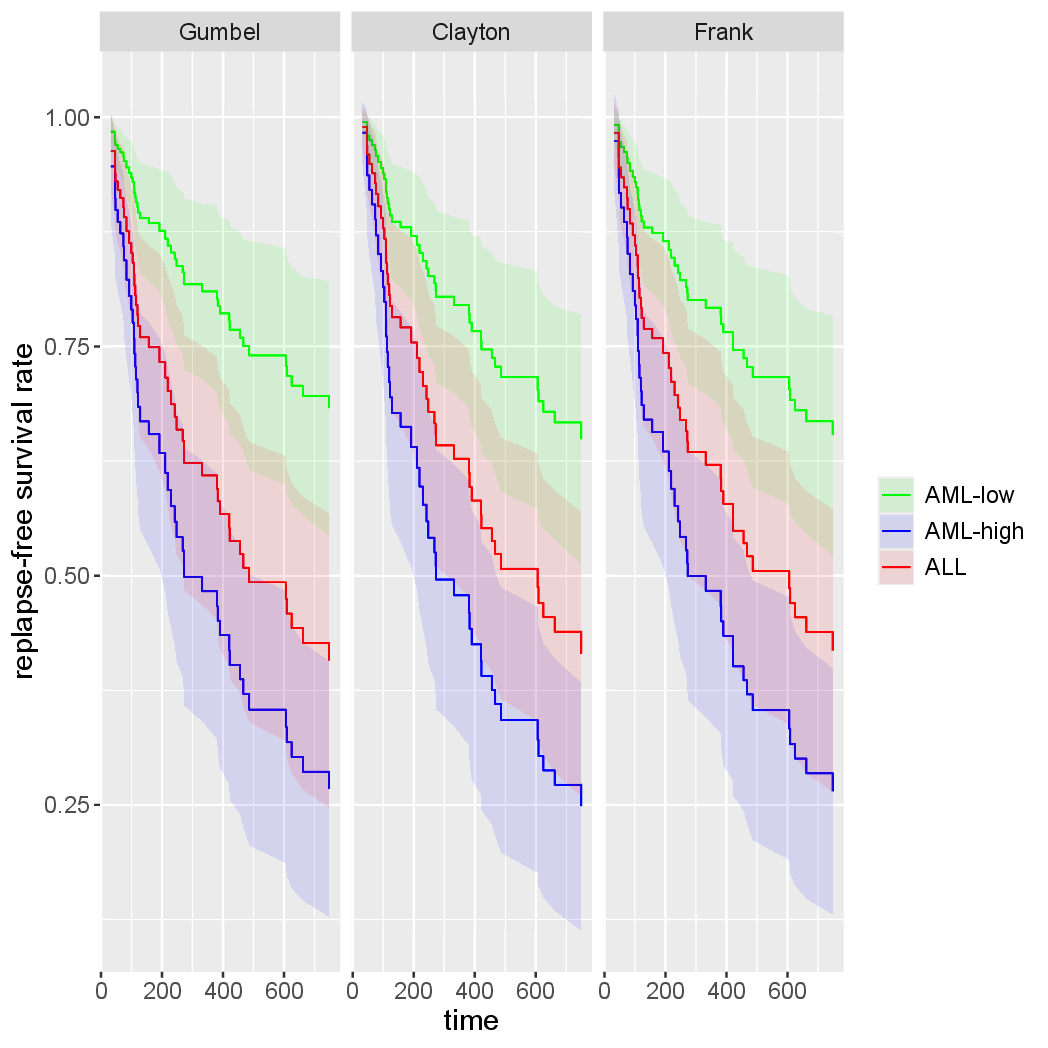}
    \caption{{{PMLE}} estimates of relapse-free survival probabilities $\Pbb(T>t)$ for each disease group. The shaded areas indicate the 95\% pointwise confidence intervals.}\label{fig:data_ana}
\end{figure}

{{Comparing the MLE and PMLE under each copula family, the two approaches produce similar point estimates and standard errors. For marginal distributions, compared to the AML-low group, both the AML-high and ALL groups have significantly higher risks of death and relapse of leukemia under all copula families. Figure \ref{fig:data_ana} plots the estimated survival functions for the time to relapse for each disease group with 95\% confidence intervals from the PMLE. The estimates from the MLE are given in Web Figure 1 of the Online Supplement. The AML-low group has the highest survival rates, and the AML-high has the lowest rates. 

Regarding the dependence between times to relapse and death, under each copula family, a strong and similar association is observed for the three groups. A difference across the three copula families is that under Clayton, the association is strongest for the AML-low group, but the AML-high group has the strongest association under Frank and Gumbel. We compared the log-likelihood function evaluated at the PMLE and MLE, and the Clayton copula family has the highest value using both methods. Thus, Clayton is the most suitable of these three families, using the log-likelihood as the selection criterion (equivalent to using the Akaike information criterion or Bayesian information criterion because all alternatives have the same number of parameters). Furthermore, under Clayton, the difference between the PMLE and MLE for $\bbeta_D$ is the smallest. As shown in simulation study II, the PMLE and MLE for the marginal distribution parameter of $D$ are similar when the copula family is correctly specified, but under copula misspecification, they diverge. Thus, the smallest gap between these two estimators under Clayton is another piece of evidence suggesting that Clayton is the best family among the three candidates.
}} 

\section{Conclusions} \label{sec:conclusion}

This article proposed a two-stage likelihood-based estimation procedure for analyzing semi-competing risks data. The dependence between the non-terminal and terminal event times is modeled by a parametric copula function, and the marginals are specified via semi-parametric transformation models. One motivation for such a stage-wise estimation is to alleviate the computational challenge arising from the one-stage MLE approach. Through simulation studies, our proposed PMLE method has given a more desirable finite-sample performance than the MLE for the copula parameter. {{On the other hand, despite being less efficient than the MLE for the regression coefficients of the marginal for $D$, the proposed PMLE method is robust against copula misspecification. This is because the estimation of $S_D(\cdot)$ in the first stage is guaranteed to be consistent. As a result, the estimation of $S_T(\cdot)$ might lead to a smaller bias than the one-stage MLE. Despite this robustness, the PMLE of $S_T(\cdot)$ is inconsistent if the copula function is misspecified. Thus, it is ideal to pair the method with a goodness-of-fit test for the copula specification, where the biggest challenge is to distinguish the misspecification of the copula function from the misspecification of $S_T(\cdot)$, which is a future research problem.}}

Although our log-likelihood function in \eqref{equ:full_ll} is equivalent to the counting-process-based log-likelihood in \citet{chen2012maximum} (Remark \ref{rmk:ll}), the respective one-stage MLEs would be different for finite samples due to the discretization of $R_j(\cdot)$ for $j=T,D$. Thus, we observe close but not exactly the same results between Table \ref{tab:data_ana} of this article and Table 5 of \citet{chen2012maximum}. However, we expect this difference will decrease with a larger sample size.

The popular strategy for addressing the dependent censoring in estimating the survival function of $T$ was using the first-event time $T^*$. However, its implementation is limited to the Archimedean copula family, and when $T$ is heavily censored by $D$, the estimation for the tail of $S_T(\cdot)$ is challenging \citep{fine2001semi,lakhal2008estimating}. In addition, $S_{T^*}(\cdot)$ is often estimated nonparametrically even with the covariates, contributing to the less ideal estimation efficiency. Though this could be improved by using a semi-parametric model, it is likely to violate the model assumptions imposed on both $T$ and $D$. For example, if one assumes that $T$ and $D$ both follow PH models, then the survival function of $T^*$ does not follow a PH model and might not even have an explicit expression. 

{{In the semiparametric copula model, the copula parameter could depend on the baseline covariates via a parametric specification. In the nonsurvival context, some works focused on nonparametrically modeling the functional relationship between the copula parameter and covariates, e.g., \citet{acar2011dependence} and \citet{abegaz2012semiparametric}. Specifically, let $W$ be a continuous covariate, and it is connected with the copula parameter $\alpha$ via $\eta(w) = \phi^{-1}(\alpha)$, where $\phi^{-1}$ is the inverse of the link function $\phi$ introduced in Remark \ref{rmk:copula-cov}. In these papers, $\eta(w)$ was approximated by a Taylor expansion of polynomial of degree $q$ with the coefficient parameters $\bgamma=(\gamma_0, \cdots, \gamma_{q})\trans$. \citet{geerdens2018conditional} extended this model to right-censored clustered survival data. In these articles, the function $\eta(w)$ was estimated by maximizing a kernel-smoothed pseudo-log-likelihood function: $\ell(\bgamma)=\sum_{i=1}^n \ell_i(\bgamma)K_h(W_i-w)$, where $\ell_i$ is the local pseudo log-likelihood contribution from subject $i$, and $K_h(\cdot)$ is a kernel function. Since our estimation procedure is also likelihood-based, it can potentially accommodate such a nonparametric specification $\eta(w)$ for the copula parameter. In the above existing works, the marginal survival functions were estimated separately. However, for semi-competing risks data, $\eta(w)$ would be estimated along with the parameters associated with the marginal distribution of $T$.
}}

\section*{acknowledgements}
This work is supported by the grant DMS-2210481 from the National Science Foundation.

%
\section*{Conflict of interest}

The authors declare that they have no conflict of interest.

\bibliographystyle{apalike}      
\bibliography{PMLE_SCR}   

\end{document}


\maketitle

\onehalfspacing

Section \hyperref[appendix:score]{S1} gives the expressions of score functions.

Section \hyperref[appendix:im]{S2} gives the expressions of the information matrices

Section \hyperref[appendix:pf]{S3} gives the proof of Theorems 1 and 2.

Section \hyperref[appendix:sim-res]{S4} gives additional results from numerical studies, including Web Tables 1 - 13 and Web Figure 1.

\section{Expressions of score functions}\label{appendix:score}

The score function $\Psi(\btheta)$ is a $(2p+\kappa_T+\kappa_D+1)$-dimensional vector, given by $\Psi = \sum_{i=1}^n \Psi_i$, where 
\begin{align*}
    \Psi_i = \frac{\partial}{\partial \btheta} \ell_i = \left(
    \begin{array}{ccccc}
    	\Psi_{\alpha,i} & \Psi_{\bbeta_T,i}\trans & \Psi_{\rmd R_T,i}\trans & \Psi_{\bbeta_D,i}\trans & \Psi_{\rmd R_D,i}\trans
    \end{array}
    \right)\trans.
\end{align*}
The elements $\Psi_{\alpha,i}$ is scalar, $\Psi_{\bbeta_T,i}$ and $\Psi_{\bbeta_D,i}$ are $p$-dimensional vectors, $\Psi_{\rmd R_{T},i}= (\Psi_{\rmd R_{T1},i}, \ldots,\allowbreak \Psi_{\rmd R_{T\kappa_T},i})\trans$  is an $\kappa_T$-dimensional vector, and $\Psi_{\rmd R_{D},i}=(\Psi_{\rmd R_{D1},i}, \ldots, \Psi_{\rmd R_{D\kappa_D},i})\trans$ is a $\kappa_D$-dimensional vector. 

To derive their expressions, we define the following notation: let $\dot f(x) \equiv \rmd f(x)/\rmd x$ and $\ddot f(x) \equiv \rmd^2 f(x)/\rmd x^2$ denote the first-order and second-order derivatives of a univariate function $f$. In addition, $\dot \Cbbm_\alpha = \partial\Cbbm/\partial \alpha$, $\ddot\Cbbm_{u_j\alpha} = \partial^2\Cbbm/\partial u_j\partial \alpha$, and $\ddot \Cbbm_{u_j^2} = \partial^2\Cbbm\partial u_j^2$ for $j=T,D$, and $\dddot\Cbbm_{u_T^{\sdone} u_D ^{\sdtwo}\alpha^{\sdthree}} = \partial^3\Cbbm/\partial u_T ^{\sdone} \partial u_D ^{\sdtwo}\partial \alpha ^ {\sdthree}$ for integers $d_1, d_2, d_3 \geq 0$ with $d_1 + d_2 + d_3 = 3$. In the remainder of the Appendix, for simplicity, we omit $(U_{T,i}, U_{D,i};\alpha)$ from the copula function $\Cbbm$ as well as its partial derivatives. 

The expression of $\Psi_{\alpha,i}$ is given as
\begin{equation*}
	\Psi_{\alpha,i} = \delta_{T,i}\delta_{D,i}\frac{\dddot \Cbbm_{u_Tu_D\alpha}}{\ddot \Cbbm_{u_Tu_D}} + \delta_{T,i}(1 - \delta_{D,i}) \frac{\ddot \Cbbm_{u_T\alpha}}{\dot \Cbbm_{u_T}} + (1-\delta_{T,i}) \delta_{D,i} \frac{\ddot \Cbbm_{u_D\alpha}}{\dot \Cbbm_{u_D}} +  (1-\delta_{T,i}) (1-\delta_{D,i})\frac{\dot \Cbbm_\alpha}{\Cbbm}.
\end{equation*}
Define
\begin{align*}
	\Psi_{u_T,i} & = \delta_{T,i}\delta_{D,i}\frac{\dddot \Cbbm_{u_T^2 u_D}}{\ddot \Cbbm_{u_Tu_D}} + \delta_{T,i}(1 - \delta_{D,i}) \frac{\ddot \Cbbm_{u_T^2}}{\dot \Cbbm_{u_T}} + (1-\delta_{T,i}) \delta_{D,i} \frac{\ddot \Cbbm_{u_Tu_D}}{\dot \Cbbm_{u_D}} +  (1-\delta_{T,i}) (1-\delta_{D,i})\frac{\dot \Cbbm_{u_T}}{\Cbbm},
\end{align*}
and
\begin{align*}
	\Psi_{u_D,i} & = \delta_{T,i}\delta_{D,i}\frac{\dddot \Cbbm_{u_T u_D^2}}{\ddot \Cbbm_{u_Tu_D}} + \delta_{T,i}(1 - \delta_{D,i}) \frac{\ddot \Cbbm_{u_Tu_D}}{\dot \Cbbm_{u_T}}+(1-\delta_{T,i}) \delta_{D,i} \frac{\ddot \Cbbm_{u_D^2}}{\dot \Cbbm_{u_D}} + (1-\delta_{T,i}) (1-\delta_{D,i})\frac{\dot \Cbbm_{u_D}}{\Cbbm}.
\end{align*}
Let $R_{T,i} = R_T(X_i)$ and $\Lambda_{T,i}=R_{T,i}e^{\bbeta_T'\bZ_i}$. In addition, let $R_{D,i} = R_D(C_i)$ and $\Lambda_{iD}=R_{D,i}e^{\bbeta_D'\bZ_i}$. The expressions of $\Psi_{\bbeta_T,i}$ and $\Psi_{\bbeta_D,i}$ are, for $j=T,D$,
\begin{equation*}
	\Psi_{\bbeta_j,i} = \bZ_i \left\{ - \Psi_{u_j,i}U_{j,i}\dot G_j\left(\Lambda_{j,i} \right)R_{j,i} e^{\bbeta_j\trans\bZ_i} - \delta_{j,i}  \dot G_j\left(\Lambda_{j,i} \right)R_{j,i} e^{\bbeta_j\trans\bZ_i} + \delta_{j,i}   \frac{\ddot G_j\left(\Lambda_{j,i}\right)}{\dot G_j\left(\Lambda_{j,i} \right)}R_{j,i} e^{\bbeta_j\trans\bZ_i} +  \delta_{j,i} \right\}.
\end{equation*}

Let $Y_{T,l,i} = \onebb(X_i\geq t_l)$ and $I_{T,l,i}=\onebb(X_i=t_l)$. In addition, let $Y_{D,l,i}=\onebb(C_i\geq d_l)$ and $I_{D,l,i}=\onebb(C_i=d_l)$. The expressions of $\Psi_{\rmd R_{T,l},i}$ and $\Psi_{\rmd R_{D,l},i}$ follow a similar form. For $j=T,D$,
\begin{equation*}
	\Psi_{\rmd R_{j,l},i} =  Y_{j,l,i} \left\{ - \Psi_{u_j,i}U_{j,i}\dot G_j\left(\Lambda_{j,i} \right)e^{\bbeta_j\trans\bZ_i} - \delta_{j,i}  \dot G_j\left(\Lambda_{j,i} \right)e^{\bbeta_j\trans\bZ_i} + \delta_{j,i} \frac{\ddot G_j\left(\Lambda_{j,i}\right)}{\dot G_j\left(\Lambda_{j,i} \right)}e^{\bbeta_j\trans\bZ_i} \right\}+ \delta_{j,i} I_{j,l,i} \rmd R_{j,l}^{-1}.
\end{equation*}

\section{Expressions of information matrices}\label{appendix:im}

The full information matrix $\bar\Ical_n(\btheta) = n^{-1}\sum_{i=1}^n \Ical_i(\btheta)$ is a $(2p+\kappa_D+\kappa_T+1)$-dimensional square matrix, where 
\begin{align*}
\Ical_i(\btheta) = -\frac{\partial^2}{\partial \btheta \partial \btheta\trans}\ell_i(\btheta) = 
\left[
\begin{array}{ccccc}
	\Ical_{\alpha\alpha,i} & \Ical_{\alpha\bbeta_T,i}\trans & \Ical_{\alpha \rmd R_T,i}\trans & \Ical_{\alpha\bbeta_D,i}\trans & \Ical_{\alpha \rmd R_D,i}\trans\\
	\Ical_{\alpha\bbeta_T,i} & \Ical_{\bbeta_T\bbeta_T,i} & \Ical_{\bbeta_T \rmd R_T,i}\trans & \Ical_{\bbeta_T \bbeta_D,i}\trans & \Ical_{\bbeta_T \rmd R_D,i}\trans\\
	\Ical_{\alpha \rmd R_T,i} & \Ical_{\rmd R_T \bbeta_T,i} & \Ical_{\rmd R_T \rmd R_T,i} & \Ical_{\rmd R_T\bbeta_D,i}\trans & \Ical_{\rmd R_T \rmd R_D,i}\trans\\
	\Ical_{\alpha \bbeta_D} & \Ical_{\bbeta_D \bbeta_T,i} & \Ical_{\bbeta_D \rmd R_T,i} & \Ical_{\bbeta_D \bbeta_D,i} & \Ical_{\bbeta_D \rmd R_D,i}\trans\\
	\Ical_{\alpha \rmd R_D} & \Ical_{\rmd R_D \bbeta_T,i} & \Ical_{\rmd R_D \rmd R_T,i} & \Ical_{\rmd R_D \bbeta_D,i} & \Ical_{\rmd R_D \rmd R_D,i}
\end{array}
\right].
\end{align*}
To derive the expression of the elements in $\Ical_i(\btheta)$, we need to define let $\ddddot\Cbbm_{u_T^{\sdone} u_D^{\sdtwo}\alpha^{\sdthree}} = \partial^4 \Cbbm/\partial u_T^{\sdone} \partial u_D^{\sdtwo}\partial \alpha^{\sdthree}$ for integers $d_1, d_2, d_3 \geq 0$ with $d_1 + d_2 + d_3 = 4$. We have $\Ical_{\alpha\alpha,i}$, a scalar, given by
\begin{align*}
	\Ical_{\alpha\alpha,i} & = \delta_{T,i}\delta_{D,i}\frac{\ddddot \Cbbm_{u_Tu_D\alpha^2} \ddot \Cbbm_{u_Tu_D} - \dddot \Cbbm_{u_Tu_D\alpha} ^ 2 }{\ddot \Cbbm_{u_Tu_D} ^ 2} +   \delta_{T,i}(1 - \delta_{D,i}) \frac{\dddot \Cbbm_{u_T\alpha^2} \dot \Cbbm_{u_T}- \ddot \Cbbm_{u_T\alpha} ^2}{\dot \Cbbm_{u_T} ^ 2} \displaybreak[0]\\
	& \quad+  (1-\delta_{T,i}) \delta_{D,i} \frac{\dddot \Cbbm_{u_D\alpha^2}\dot \Cbbm_{u_D}-\ddot \Cbbm_{u_D\alpha} ^ 2}{\dot \Cbbm_{u_D} ^ 2} +  (1-\delta_{T,i}) (1-\delta_{D,i})\frac{\ddot \Cbbm_{\alpha ^ 2}\Cbbm - \dot \Cbbm_\alpha ^ 2}{\Cbbm^2}.
\end{align*}
Define the following
\begin{align*}
	\Ical_{u_T\alpha,i} & = \delta_{T,i}\delta_{D,i}\frac{\ddddot \Cbbm_{u_T^2 u_D \alpha}\ddot \Cbbm_{u_Tu_D} - \dddot \Cbbm_{u_T^2 u_D}  \dddot \Cbbm_{u_T u_D \alpha}}{\ddot \Cbbm_{u_Tu_D}^2} + \delta_{T,i}(1 - \delta_{D,i}) \frac{\dddot \Cbbm_{u_T^2\alpha}\dot \Cbbm_{u_T} - \ddot \Cbbm_{u_T^2} \ddot \Cbbm_{u_T\alpha}}{\dot \Cbbm_{u_T}^2} \displaybreak[0]\\
	& \quad + (1-\delta_{T,i}) \delta_{D,i} \frac{\dddot \Cbbm_{u_T u_D\alpha}\dot \Cbbm_{u_D}-\ddot \Cbbm_{u_Tu_D} \ddot \Cbbm_{u_D\alpha}}{\dot \Cbbm_{u_D}^2} + (1-\delta_{T,i}) (1-\delta_{D,i})\frac{\ddot\Cbbm_{u_T\alpha}\Cbbm - \dot \Cbbm_{u_T} \dot \Cbbm_\alpha}{\Cbbm^2},
\end{align*}
\begin{align*}
	\Ical_{u_D\alpha,i} & = \delta_{T,i}\delta_{D,i}\frac{\ddddot \Cbbm_{u_T u_D^2 \alpha} \ddot \Cbbm_{u_Tu_D}- \dddot \Cbbm_{u_T u_D^2} \dddot \Cbbm_{u_T u_D \alpha}}{\ddot \Cbbm_{u_Tu_D}^2} + \delta_{T,i}(1 - \delta_{D,i}) \frac{\dddot \Cbbm_{u_Tu_D\alpha}\dot \Cbbm_{u_T} - \ddot \Cbbm_{u_Tu_D} \ddot \Cbbm_{u_T\alpha}}{\dot \Cbbm_{u_T}^2}  \displaybreak[0]\\
	& \quad + (1-\delta_{T,i}) \delta_{D,i} \frac{\dddot \Cbbm_{u_D^2\alpha}\dot \Cbbm_{u_D}-\ddot \Cbbm_{u_D^2} \ddot \Cbbm_{u_D\alpha}}{\dot \Cbbm_{u_D}^2} + (1-\delta_{T,i}) (1-\delta_{D,i})\frac{\ddot\Cbbm_{u_D\alpha}\Cbbm - \dot \Cbbm_{u_D} \dot \Cbbm_\alpha}{\Cbbm^2},
\end{align*}
\begin{align*}
	\Ical_{u_Tu_T,i} & = \delta_{T,i}\delta_{D,i}\frac{\ddddot \Cbbm_{u_T^3 u_D} \ddot \Cbbm_{u_Tu_D} - \dddot \Cbbm_{u_T^2 u_D}^2}{\ddot \Cbbm_{u_Tu_D}^2} + \delta_{T,i}(1 - \delta_{D,i}) \frac{\dddot \Cbbm_{u_T^3}\dot \Cbbm_{u_T} - \ddot \Cbbm_{u_T^2} ^2}{\dot \Cbbm_{u_T} ^ 2}  \displaybreak[0]\\
	& \quad + (1-\delta_{T,i}) \delta_{D,i} \frac{\dddot \Cbbm_{u_T^2 u_D}\dot \Cbbm_{u_D}-\ddot \Cbbm_{u_Tu_D} ^ 2}{\dot \Cbbm_{u_D} ^ 2} +  (1-\delta_{T,i}) (1-\delta_{D,i})\frac{\ddot\Cbbm_{u_T^ 2}\Cbbm - \dot \Cbbm_{u_T} ^ 2}{\Cbbm^2},
\end{align*}
\begin{align*}
	\Ical_{u_Du_D,i} & = \delta_{T,i}\delta_{D,i}\frac{\ddddot \Cbbm_{u_T u_D^3} \ddot \Cbbm_{u_Tu_D} - \dddot \Cbbm_{u_T u_D^2}^2}{\ddot \Cbbm_{u_Tu_D}^2} + \delta_{T,i}(1 - \delta_{D,i}) \frac{\dddot \Cbbm_{u_Tu_D^2} \dot \Cbbm_{u_T} - \ddot \Cbbm_{u_Tu_D}^2}{\dot \Cbbm_{u_T}^2} \displaybreak[0]\\
	& \quad + (1-\delta_{T,i}) \delta_{D,i} \frac{\dddot \Cbbm_{u_D^3}\dot \Cbbm_{u_D}-\ddot \Cbbm_{u_D^2}^2}{\dot \Cbbm_{u_D}^2} + (1-\delta_{T,i}) (1-\delta_{D,i})\frac{\ddot\Cbbm_{u_D^2} \Cbbm - \dot \Cbbm_{u_D}^2}{\Cbbm^2},
\end{align*}
and
\begin{align*}
	\Ical_{u_Tu_D,i} & = \delta_{T,i}\delta_{D,i}\frac{\ddddot \Cbbm_{u_T^2 u_D^2} \ddot \Cbbm_{u_Tu_D}- \dddot \Cbbm_{u_T u_D^2}  \dddot \Cbbm_{u_T^2 u_D}}{\ddot \Cbbm_{u_Tu_D} ^ 2} +   \delta_{T,i}(1 - \delta_{D,i}) \frac{\dddot \Cbbm_{u_T ^ 2 u_D} \dot \Cbbm_{u_T}- \ddot \Cbbm_{u_Tu_D} \ddot \Cbbm_{u_T^2}}{\dot \Cbbm_{u_T} ^ 2}  \displaybreak[0]\\
	& \quad + (1-\delta_{T,i}) \delta_{D,i} \frac{\dddot \Cbbm_{u_Tu_D^2}\dot \Cbbm_{u_D}-\ddot \Cbbm_{u_D^2} \ddot\Cbbm_{u_Tu_D}}{\dot \Cbbm_{u_D} ^ 2} +  (1-\delta_{T,i}) (1-\delta_{D,i})\frac{\ddot\Cbbm_{u_Tu_D}\Cbbm - \dot \Cbbm_{u_D} \dot\Cbbm_{u_T}}{\Cbbm^2}.
\end{align*}
Both $\Ical_{\bbeta_T\bbeta_T,i}$ and $\Ical_{\bbeta_D\bbeta_D,i}$ are $p$-dimensional square matrices, and their expression follow similar forms. For $j=T,D$,
\begin{align*}
	\Ical_{\bbeta_j\bbeta_j,i} = &~ \bZ_i\bZ_i\trans \Bigg\{\Ical_{u_ju_j,i}U_{j,i}^2 \dot G_j\left(\Lambda_{j,i} \right) ^ 2 R_{j,i} ^ 2 e^{2\bbeta_j\trans\bZ_i} + \Psi_{u_j,i}U_{j,i}\dot G_j\left(\Lambda_{j,i} \right)^2 R_{j,i} ^ 2  e^{2\bbeta_j\trans\bZ_i} \displaybreak[0]\\
	&\qquad\quad  - \Psi_{u_j,i}U_{j,i} \ddot G_j\left(\Lambda_{j,i} \right)R_{j,i} ^ 2 e^{2\bbeta_j\trans\bZ_i} - \Psi_{u_j,i}U_{j,i} \dot G_j\left(\Lambda_{j,i} \right)R_{j,i} e^{\bbeta_j\trans\bZ_i} \displaybreak[0]\\
	&\qquad\quad - \delta_{j,i} \ddot G_j\left(\Lambda_{j,i} \right)R_{j,i}^2 e^{2\bbeta_j\trans\bZ_i} - \delta_{j,i}  \dot G_j\left(\Lambda_{j,i} \right)R_{j,i} e^{\bbeta_j\trans\bZ_i} \displaybreak[0]\\
	&\qquad\quad + \delta_{j,i} \frac{\dddot G_j\left(\Lambda_{j,i}\right)\dot G_j\left(\Lambda_{j,i} \right) - \ddot G_j\left(\Lambda_{j,i} \right)^2 }{\dot G_j\left(\Lambda_{j,i} \right)}R_{j,i} ^ 2 e^{2\bbeta_j\trans\bZ_i} + \delta_{j,i}   \frac{\ddot G_j\left(\Lambda_{j,i}\right)}{\dot G_j\left(\Lambda_{j,i} \right)}R_{j,i} e^{\bbeta_j\trans\bZ_i} \Bigg\}.
\end{align*}

The expressions of $\Ical_{\rmd R_T\rmd R_T,i}$, a $\kappa_T\times \kappa_T$-dimensional matrix, and $\Ical_{\rmd R_D\rmd R_D,i}$, a $\kappa_D\times\kappa_D$-dimensional matrix, follow similar forms. For $j=T,D$ and $l,l'=1,\cdots,\kappa_j$, the $(l,l')$ element of the matrix $\Ical_{\rmd R_j\rmd R_j,i}$ is given by
\begin{align*}
	\Ical_{\rmd R_{j,l}\rmd R_{j,l'},i} = &~ Y_{j,l,i} Y_{j,l',i}\Bigg\{\Ical_{u_ju_j,i}U_{j,i} ^ 2 \dot G_j\left(\Lambda_{j,i} \right)^2 e^{2\bbeta_j\trans\bZ_i} + \Psi_{u_j,i}U_{j,i}\dot G_j\left(\Lambda_{j,i} \right) ^ 2 e^{2\bbeta_j\trans\bZ_i} - \Psi_{u_j,i}U_{j,i} \ddot G_j\left(\Lambda_{j,i} \right) e^{2\bbeta_j\trans\bZ_i}  \displaybreak[0]\\
	& \qquad\qquad\quad- \delta_{j,i} \ddot G_j\left(\Lambda_{j,i} \right)e^{2\bbeta_j\trans\bZ_i} + \delta_{j,i} \frac{\dddot G_j\left(\Lambda_{j,i}\right)\dot G_j\left(\Lambda_{j,i} \right) - \ddot G_j\left(\Lambda_{j,i} \right)^2 }{\dot G_j\left(\Lambda_{j,i} \right)} e^{2\bbeta_j\trans\bZ_i} - \delta_{j,i} I_{j,l,i} \onebb(l=l') \rmd R_{j,l}^{-2} \Bigg\}.
\end{align*}
Both $\Ical_{\alpha\bbeta_T,i}$ and $\Ical_{\alpha\bbeta_D,i}$ are $p$-dimensional vectors, and their expressions follow similar forms. For $j=T,D$,
\begin{equation*}
	\Ical_{\alpha\bbeta_j,i} = - \Ical_{u_j\alpha,i}U_{j,i}\dot G_j\left(\Lambda_{j,i}\right)R_{j,i}e^{\bbeta_j\trans\bZ_i}\bZ_i.
\end{equation*}
The expressions of $\Ical_{\alpha \rmd R_T,i}$, a $\kappa_T$-dimensional vector, and $\Ical_{\alpha \rmd R_D,i}$, a $\kappa_D$-dimensional vector, follow similar forms. For $j=T,D$ and $l=1,\ldots,\kappa_j$, the $l$-th element of the vector $\Ical_{\alpha \rmd R_j,i}$ is given as
\begin{equation*}
	\Ical_{\alpha \rmd R_{j,l},i} = - Y_{j,l,i}\Ical_{u_j\alpha,i}U_{j,i}\dot G_j\left(\Lambda_{j,i} \right)e^{\bbeta_j\trans\bZ_i}.
\end{equation*}
$\Ical_{\bbeta_T\bbeta_D,i}$ is a $p\times p$-dimensional matrix, given by
\begin{equation*}
    \Ical_{\bbeta_T\bbeta_D,i} = \bZ_i\bZ_i\trans \Ical_{u_Tu_D,i}U_{T,i}U_{D,i}e^{(\bbeta_T+\bbeta_D)\trans\bZ_i}R_{T,i}R_{D,i}\dot G_T\left(\Lambda_{T,i} \right)\dot G_D\left(\Lambda_{D,i} \right).
\end{equation*}
$\Ical_{\rmd R_T \rmd R_D,i}$ is a $\kappa_T \times \kappa_D$-dimensional matrix with the $(l,l')$ element, given by
\begin{equation*}
    \Ical_{\rmd R_{T,l}\rmd R_{D,l'},i} = Y_{T,l,i}Y_{D,l',i}\Ical_{u_Tu_D,i}U_{T,i}U_{D,i}e^{(\bbeta_T+\bbeta_D)\trans\bZ_i}\dot G_T\left(\Lambda_{T,i} \right)\dot G_D\left(\Lambda_{D,i} \right).
\end{equation*}
The expressions of $\Ical_{\bbeta_T\rmd R_T,i}$, a $p\times \kappa_T$-dimensional matrix, and $\Ical_{\bbeta_D\rmd R_D,i}$, a $p\times \kappa_D$-dimensional matrix, follow a similar form. For $j=T,D$ and $l=1,\cdots \kappa_j$, the $l$-th column of the matrix $\Ical_{\bbeta_j\rmd R_j,i}$ is given by
\begin{align*}
	\Ical_{\bbeta_j \rmd R_{j,l},i} = &~ \bZ_iY_{j,l,i} \Bigg\{\Ical_{u_ju_j,i}U_{j,i} ^ 2 \dot G_j\left(\Lambda_{j,i} \right) ^ 2 R_{j,i} e^{2\bbeta_j\trans\bZ_i} + \Psi_{u_j,i}U_{j,i}\dot G_j\left(\Lambda_{j,i} \right) ^ 2 R_{j,i} e^{2\bbeta_j\trans\bZ_i} \displaybreak[0]\\
	& \qquad \qquad  - \Psi_{u_j,i}U_{j,i} \ddot G_j\left(\Lambda_{j,i} \right)R_{j,i} e^{2\bbeta_j\trans\bZ_i} - \Psi_{u_j,i}U_{j,i} \dot G_j\left(\Lambda_{j,i} \right)e^{\bbeta_j\trans\bZ_i} \displaybreak[0]\\
	& \qquad\qquad - \delta_{j,i} \ddot G_j\left(\Lambda_{j,i} \right)R_{j,i}^2 e^{2\bbeta_j\trans\bZ_i} - \delta_{j,i}  \dot G_j\left(\Lambda_{j,i} \right) e^{\bbeta_j\trans\bZ_i} \displaybreak[0]\\
	& \qquad\qquad + \delta_{j,i} \frac{\dddot G_j\left(\Lambda_{j,i}\right)\dot G_j\left(\Lambda_{j,i} \right) - \ddot G_j\left(\Lambda_{j,i} \right)^2 }{\dot G_j\left(\Lambda_{j,i} \right)}R_{j,i} e^{2\bbeta_j\trans\bZ_i} + \delta_{j,i} \frac{\ddot G_j\left(\Lambda_{j,i}\right)}{\dot G_j\left(\Lambda_{j,i} \right)}e^{\bbeta_j\trans\bZ_i} \Bigg\}.
\end{align*}
The expressions of $\Ical_{\bbeta_T\rmd R_D,i}$, a $p\times \kappa_D$-dimensional matrix, and $\Ical_{\bbeta_D\rmd R_T,i}$, a $p\times \kappa_T$-dimensional matrix, follow similar forms. For $j,j'=T,D$ with $j\neq j'$ and $l=1,\cdots \kappa_{j'}$, the $l$-th column of the matrix $\Ical_{\bbeta_j\rmd R_{j'},i}$ is given by
\begin{equation*}
    \Ical_{\bbeta_j \rmd R_{j',k},i} = \bZ_i Y_{j',k,i} U_{T,i}U_{D,i}e^{(\bbeta_T+\bbeta_D)'\bZ_i}\dot G_T(\Lambda_{T,i})\dot G_T(\Lambda_{D,i})R_{j,i}.
\end{equation*}

\section{Proof of Theorems 1 and 2}\label{appendix:pf}%

The proof of Theorems 1 and 2 requires the asymptotic properties of $\bthetahat_D$ obtained in the first stage, which was established in \citet{zeng2006efficient}, where it was shown that $\bbetahat_D$ converges to $\bbeta_D^0$ and $\Rhat_D(\cdot)$ converges to $R_D^0(\cdot)$ uniformly in the interval $[0,\xi]$. In addition, by Theorem 3.3.1 of \citet{vanderVaart1996}, $\sqrt{n}(\bthetahat_D-\btheta_D^0)$ can be expressed as $\sqrt{n}(\bthetahat_D-\btheta_D^0) = n^{-1/2}\sum_{i=1}^n \psi_{\btheta_D,i}(\btheta_D^0) + o_p(1)$ with $\psi_{\btheta_D,i}(\btheta_D^0) = \left(\Ical_D^0\right)^{-1} \Psi_{D,i}(\btheta_D^0)$, where $\Ical_D^0 = \Ebb\{\Ical_{D,i}(\btheta_D^0)\}$. 
Since the variance-covariance matrix of $n^{-1/2}\sum_{i=1}^n\Psi_{D,i}(\btheta_D^0)$ is equal to $\Ical_D^0$, by the functional delta method, $\sqrt{n}\{F_D(\bthetahat_D)-F_D(\btheta_D^0)\}$ converges to a zero-mean Gaussian process with the variance-covariance matrix $\dot F_D(\btheta_D^0)\trans(\Ical_D^0)^{-1}\dot F_D(\btheta_D^0)$.

Define $\Hfrak(\btheta_D;t,\bz) = \exp[-G_D\{R_D(t) e^{\bbeta_D\trans\bz} \}]$ and let
\begin{equation*}
    \hfrak(\btheta_D;t,\bz) = \frac{\partial \Hfrak(\btheta_D;t,\bz)}{\partial \btheta_D} = \left(\hfrak_{\bbeta_D}\trans,\hfrak_{\rmd R_{D,1}}, \ldots, \hfrak_{\rmd R_{D,\kappa_D}}\right)\trans,
\end{equation*}
where
\begin{align*}
	\hfrak_{\bbeta_D} & = -\hfrak(\btheta_D;t,\bz)\dot G_D\left\{R_D(t)e^{\bbeta_D\trans \bz}\right\}R_D(t)\bz, \displaybreak[0]\\
	\hfrak_{\rmd R_{D,l}} & = -\hfrak(\btheta_D;t,\bz)\dot G_D\left\{R_D(t)e^{\bbeta_D\trans \bz}\right\}e^{\bbeta_D\trans \bz}\onebb(t\geq d_l), \, l = 1,\ldots,\kappa_D.
\end{align*}
Let $\Uhat_{D,i} = \Hfrak(\bthetahat_D;C_i,\bZ_i)$ and $U_{D,i}^0 = \Hfrak(\btheta_D^0;C_i,\bZ_i)$. By the functional delta method, we have $\sqrt{n}(\Uhat_{D,i}-U_{D,i}^0) = \hfrak(\btheta_D^0;C_i,\bZ_i)\trans \{\sqrt{n}(\bthetahat_D-\btheta_D^0)\} + o_p(1)$. And, by the expansion of $\sqrt{n}(\bthetahat_D-\btheta_D^0)$ above, we have 
\begin{equation}\label{equ:U_exp}
	\sqrt{n}(\Uhat_{D,i}-U_{D,i}^0) = n^{-1/2}\sum_{k=1}^n \psi_{u_{D,i},k}(\btheta_D^0) + o_p(1), 
\end{equation}
where $\psi_{u_{D,i},k}(\btheta_D^0) = \hfrak(\btheta_D^0;C_i,\bZ_i)\psi_{\btheta_D,k}(\btheta_D^0)$.

\subsection{Proof of Theorem 1}

Let $\breve \btheta_1 = \arg\max_{\btheta_1}\ell(\btheta_1,\btheta_D^0)$. By similar arguments in \citet{chen2012maximum}, we have $\breve \alpha$ converges $\alpha^0$ and $\breve \bbeta_T$ converges $\bbeta_T^0$, and $\breve R_T(\cdot)\equiv\{\breve{\rmd R}_{T,l},l=1,\ldots,\kappa_T\}$ converges to $R_T^0(\cdot)$ uniformly in the interval $[0,\xi]$. By C4 and the consistency of $\bthetahat_D$, we have, almost surely, $|\ell(\btheta_1, \bthetahat_D) - \ell(\btheta_1, \btheta_D^0)|_{\infty} = o_p(1)$. Thus, $|\alphahat - \breve \alpha| = o_p(1)$, $\|\bbetahat_T - \breve \bbeta_T\|=o_p(1)$, and $|\Rhat_T - \breve R_T|_{\infty}=o_p(1)$. With probability 1, $\alphahat$ converges $\alpha^0$, $\bbetahat_T$ converges $\bbeta_T^0$, and $\Rhat_T(\cdot)$ converges to $R_T^0(\cdot)$ uniformly on $[0,\xi]$.

\subsection{Proof of Theorem 2}

By Theorem 3.3.1 of \citet{vanderVaart1996}, $\sqrt{n}(\bthetahat_1-\btheta_1^0)$ can be expressed as 
\begin{equation*}
    \sqrt{n}(\bthetahat_1-\btheta_1^0) = \bar\Ical_{1,n}^{-1}(\btheta_1^0, \bthetahat_D)\left\{n^{-1/2}\sum_{i=1}^n \dot \ell_{\btheta_1,i}(\btheta_1^0,\bthetahat_D)\right\} + o_p(1),
\end{equation*}
where 
\begin{equation}\label{equ:Ical_11n}
	\bar\Ical_{1,n}(\btheta_1^0, \bthetahat_D) = - n^{-1}\sum_{i=1}^n \ddot \ell_{\btheta_1\btheta_1,i}(\btheta_1^0, \bthetahat_D)
\end{equation} 
with $\ddot \ell_{\btheta_1\btheta_1,i}(\btheta_1^0,\bthetahat_D) = \partial^2\ell_i(\btheta)/\partial\btheta_1\partial \btheta_1\trans$. By C6 and the consistency of $\bthetahat_D$, 
\begin{equation*}
    \left\|\bar\Ical_{1,n}(\btheta_1^0, \bthetahat_D) - \Ebb\left\{-\ddot \ell_{\btheta_1\btheta_1,i}(\btheta^0)\right\}\right\|=o_p(1).
\end{equation*}
Thus, 
\begin{equation*}
    \sqrt{n}(\bthetahat_1-\btheta_1^0) = \left(\Ical_1^0\right) ^{-1}\left\{n^{-1/2}\sum_{i=1}^n \dot \ell_{\btheta_1,i}(\btheta_1^0, \bthetahat_D)\right\} + o_p(1),
\end{equation*}
where $\Ical_1^0 = \Ebb\{- \ddot \ell_{\btheta_1\btheta_1,i}(\btheta^0)\}$. Since $\dot \ell_{\btheta_1,i}(\btheta_1^0, \bthetahat_D)$ is a function of $\bthetahat_D$ only through $\Uhat_{D,i}$, we can write $\dot \ell_{\btheta_1,i}(\btheta_1^0, \bthetahat_D) = \dot \ell_{\btheta_1,i}(\btheta_1^0, \Uhat_{D,i})$. By the mean value theorem, we have 
\begin{equation*}
    \dot \ell_{\btheta_1,i}(\btheta_1^0, \Uhat_{D,i}) = \dot \ell_{\btheta_1,i}(\btheta_1^0, U_{D,i}^0) + \ddot \ell_{\btheta_1u_D,i}(\btheta_1^0, U_{D,i}^0)(\Uhat_{D,i}-U_{D,i}^0) + o_p(1),
\end{equation*}
where $\ddot \ell_{\btheta_1u_D,i}(\btheta_1^0, U_{D,i}^0)=\partial\Psi_{1,i}(\btheta_1^0)/\partial u_D = \partial^2\ell_i(\btheta^0)/\partial u_D\partial \btheta_1$. Thus, we can express 
\begin{align*}
	\sqrt{n}(\bthetahat_1-\btheta_1^0)=\left(\Ical_1^0\right)^{-1} n^{-1/2}\sum_{i=1}^n \left\{\dot \ell_{\btheta_1,i}(\btheta_1^0, U_{D,i}^0) + \ddot \ell_{\btheta_1u_D,i}(\btheta_1^0, U_{D,i}^0)(\Uhat_{D,i}-U_{D,i}^0)\right\}+ o_p(1).
\end{align*}
By the expansion of $\sqrt{n}(\Uhat_{D,i}-U_{D,i}^0)$ in \eqref{equ:U_exp}, we can express $\sqrt{n}(\bthetahat_1 - \btheta_1^0) = n^{-1/2}\sum_{i=1}^n \psi_{\btheta_1,i}(\btheta^0)\allowbreak+o_p(1)$, where 
\begin{equation*}
    \psi_{\btheta_1,i}(\btheta^0) = \left(\Ical_1^0\right) ^{-1} \left\{\dot \ell_{\btheta_1,i}(\btheta^0) + n^{-1}\sum_{k=1}^n \ddot \ell_{\btheta_1u_D,k}(\btheta^0) \psi_{u_{D,k},i}(\btheta_D^0) \right\}.
\end{equation*}
Thus, $\sqrt{n}\{F_1(\bthetahat_1) - F_1(\btheta_1^0)\}$ converges to a zero-mean Gaussian process with variance-covariance matrix $\dot F_1(\btheta_1^0)\trans\bSigma_1^0 \dot F_1(\btheta_1^0)$, where $\bSigma_1^0= \Var\{\psi_{\btheta_1,i}(\btheta^0)\}$.

\section{Additional results from numerical studies}
\label{appendix:sim-res}

\begin{table}[H]
	\centering
	\begin{tabular}{lll rcccc}
		\toprule
		$\tau$ & $\bbeta_{T}$ & Method & \multicolumn{1}{c}{BIAS} & ESD & ASE & rMSE & CP  \\
		\midrule
		\multicolumn{3}{c}{} & \multicolumn{5}{c}{$\widehat\beta_{T,1}$} \\
		\cmidrule{4-8}
		\multirow{4}{*}{0.4} & \multirow{2}{*}{(1,1)} & MLE & 0.001 & 0.113 & 0.113 & 0.113 & 94.8\\
		&  & PMLE & 0.002 & 0.115 & 0.115 & 0.115 & 94.9\\      
		\cmidrule{4-8} 
		& \multirow{2}{*}{(1,0.5)}  & MLE & 0.001 & 0.114 & 0.115 & 0.114 & 95.2\\
		&  & PMLE & 0.002 & 0.116 & 0.116 & 0.116 & 95.3\\
		\midrule
		\multirow{4}{*}{0.6} & \multirow{2}{*}{(1,1)}   & MLE & 0.002 & 0.111 & 0.111 & 0.111 & 94.1\\
		&  & PMLE & 0.001 & 0.117 & 0.116 & 0.117 & 94.2\\
		\cmidrule{4-8} 
		& \multirow{2}{*}{(1,0.5)} & MLE & 0.001 & 0.113 & 0.112 & 0.113 & 93.7\\
		&  & PMLE & 0.002 & 0.118 & 0.116 & 0.118 & 94.1\\
		\midrule
		\multirow{4}{*}{0.8} & \multirow{2}{*}{(1,1)}   & MLE & 0.006 & 0.111 & 0.110 & 0.112 & 94.5\\
		&  & PMLE & 0.002 & 0.121 & 0.120 & 0.121 & 94.5\\
		\cmidrule{4-8} 
		& \multirow{2}{*}{(1,0.5)}  & MLE & 0.004 & 0.112 & 0.111 & 0.113 & 94.1\\
		&  & PMLE & 0.003 & 0.121 & 0.120 & 0.121 & 94.7\\
		\midrule 
		\multicolumn{3}{c}{} & \multicolumn{5}{c}{$\widehat\beta_{T,2}$}\\
		\cline{4-8}
		\multirow{4}{*}{0.4} & \multirow{2}{*}{(1,1)}    & MLE & 0.003 & 0.148 & 0.139 & 0.148 & 93.7\\
		&  & PMLE & 0.002 & 0.150 & 0.140 & 0.150 & 93.0\\
		\cmidrule{4-8} 
		& \multirow{2}{*}{(1,0.5)} & MLE & 0.009 & 0.285 & 0.267 & 0.285 & 92.7\\
		&  & PMLE & 0.007 & 0.289 & 0.270 & 0.289 & 92.7\\
		\midrule
		\multirow{4}{*}{0.6} & \multirow{2}{*}{(1,1)} & MLE & 0.008 & 0.142 & 0.134 & 0.142 & 94.1\\
		&  & PMLE & 0.005 & 0.148 & 0.140 & 0.148 & 94.3\\
		\cmidrule{4-8} 
		& \multirow{2}{*}{(1,0.5)}  & MLE & 0.017 & 0.274 & 0.258 & 0.274 & 93.8\\
		&  & PMLE & 0.012 & 0.285 & 0.268 & 0.285 & 93.3\\
		\midrule 
		\multirow{4}{*}{0.8} & \multirow{2}{*}{(1,1)}  & MLE & 0.011 & 0.139 & 0.133 & 0.139 & 93.9\\
		&  & PMLE & 0.003 & 0.151 & 0.143 & 0.151 & 94.5\\
		\cmidrule{4-8} 
		& \multirow{2}{*}{(1,0.5)}  & MLE & 0.018 & 0.269 & 0.257 & 0.270 & 93.7\\
		&  & PMLE & 0.009 & 0.291 & 0.275 & 0.291 & 94.0\\
		\midrule 
		\multicolumn{3}{c}{} &\multicolumn{5}{c}{$\widehat\alpha$} \\
		\cmidrule{4-8}
		\multirow{4}{*}{0.4} & \multirow{2}{*}{(1,1)}  & MLE & 0.018 & 0.051 & 0.052 & 0.054 & 95.5\\
		&  & PMLE & 0.011 & 0.051 & 0.052 & 0.052 & 95.2\\
		\cmidrule{4-8} 
		& \multirow{2}{*}{(1,0.5)} & MLE & 0.019 & 0.053 & 0.054 & 0.056 & 94.9\\
		&  & PMLE & 0.012 & 0.052 & 0.053 & 0.054 & 95.0\\
		\midrule
		\multirow{4}{*}{0.6} & \multirow{2}{*}{(1,1)}  & MLE & 0.027 & 0.057 & 0.059 & 0.064 & 94.7\\
		&  & PMLE & 0.013 & 0.057 & 0.058 & 0.059 & 95.2\\
		\cmidrule{4-8} 
		& \multirow{2}{*}{(1,0.5)}  & MLE & 0.028 & 0.059 & 0.060 & 0.065 & 94.5\\
		&  & PMLE & 0.013 & 0.058 & 0.059 & 0.060 & 94.9\\
		\midrule
		\multirow{4}{*}{0.8} & \multirow{2}{*}{(1,1)} & MLE & 0.049 & 0.065 & 0.066 & 0.081 & 91.9\\
		&  & PMLE & 0.010 & 0.062 & 0.064 & 0.063 & 95.4\\
		\cmidrule{4-8} 
		& \multirow{2}{*}{(1,0.5)}  & MLE & 0.049 & 0.065 & 0.067 & 0.082 & 92.0\\
		&  & PMLE & 0.010 & 0.063 & 0.064 & 0.064 & 95.4\\
		\bottomrule 
	\end{tabular}
	\vspace{0.4cm}
	\caption{Simulation results for the estimates of $(\bbeta_T,\alpha)$ under the Gumbel copula with sample size $n=400$. BIAS: relative bias; ESD: relative empirical standard deviation; ASE: relative average standard error; rMSE: relative root mean square error; CP: empirical coverage percentage of 95\% confidence interval.}
	\label{tab:est_gumbel400}
\end{table}

\begin{table}[H]
	\centering
	\begin{tabular}{lll rcccc}
            \toprule
		$\tau$ & $\bbeta_{T}$ & Method & \multicolumn{1}{c}{BIAS} & ESD & ASE & rMSE & CP  \\
		
		\midrule
		\multicolumn{3}{c}{} & \multicolumn{5}{c}{$\widehat\beta_{T,1}$} \\
		\cmidrule{4-8}
		\multirow{4}{*}{0.4} & \multirow{2}{*}{(1,1)} & MLE & 0.008 & 0.108 & 0.107 & 0.108 & 94.1\\
		&  & PMLE & 0.008 & 0.112 & 0.112 & 0.112 & 94.8\\      
		\cmidrule{4-8} 
		& \multirow{2}{*}{(1,0.5)} & MLE & 0.008 & 0.112 & 0.110 & 0.112 & 94.4\\
		&  & PMLE & 0.008 & 0.115 & 0.116 & 0.115 & 94.8\\
		\midrule
		\multirow{4}{*}{0.6} & \multirow{2}{*}{(1,1)} & MLE & 0.011 & 0.101 & 0.097 & 0.101 & 93.2\\
		&  & PMLE & 0.011 & 0.114 & 0.114 & 0.115 & 94.8\\
		\cmidrule{4-8} 
		& \multirow{2}{*}{(1,0.5)} & MLE & 0.011 & 0.103 & 0.099 & 0.103 & 94.1\\
		&  & PMLE & 0.013 & 0.116 & 0.117 & 0.117 & 95.4\\
		\midrule
		\multirow{4}{*}{0.8} & \multirow{2}{*}{(1,1)} & MLE & 0.016 & 0.091 & 0.089 & 0.093 & 93.6\\
		&  & PMLE & 0.016 & 0.122 & 0.123 & 0.123 & 95.0\\
		\cmidrule{4-8} 
		& \multirow{2}{*}{(1,0.5)} & MLE & 0.016 & 0.092 & 0.090 & 0.094 & 94.2\\
		&  & PMLE & 0.018 & 0.123 & 0.124 & 0.124 & 94.9\\
		\midrule 
		\multicolumn{3}{c}{} & \multicolumn{5}{c}{$\widehat\beta_{T,2}$}\\
		\cline{4-8}
		\multirow{4}{*}{0.4} & \multirow{2}{*}{(1,1)} & MLE & 0.002 & 0.145 & 0.133 & 0.145 & 92.5\\
		&  & PMLE & 0.008 & 0.150 & 0.140 & 0.150 & 93.8\\
		\cmidrule{4-8} 
		& \multirow{2}{*}{(1,0.5)} & MLE & 0.007 & 0.275 & 0.251 & 0.276 & 92.3\\
		&  & PMLE & 0.002 & 0.284 & 0.265 & 0.284 & 93.1\\
		\midrule
		\multirow{4}{*}{0.6} & \multirow{2}{*}{(1,1)} & MLE & 0.001 & 0.125 & 0.118 & 0.125 & 94.3\\
		&  & PMLE & 0.012 & 0.147 & 0.140 & 0.148 & 94.3\\
		\cmidrule{4-8} 
		& \multirow{2}{*}{(1,0.5)} & MLE & 0.012 & 0.236 & 0.220 & 0.236 & 93.8\\
		&  & PMLE & 0.007 & 0.279 & 0.262 & 0.279 & 93.2\\
		\midrule 
		\multirow{4}{*}{0.8} & \multirow{2}{*}{(1,1)} & MLE & 0.002 & 0.111 & 0.106 & 0.111 & 94.6\\
		&  & PMLE & 0.018 & 0.156 & 0.148 & 0.157 & 93.7\\
		\cmidrule{4-8} 
		& \multirow{2}{*}{(1,0.5)} & MLE & 0.012 & 0.208 & 0.199 & 0.209 & 94.0\\
		&  & PMLE & 0.014 & 0.291 & 0.275 & 0.292 & 94.1\\
		\midrule 
		\multicolumn{3}{c}{} &\multicolumn{5}{c}{$\widehat\alpha$} \\
		\cmidrule{4-8}
		\multirow{4}{*}{0.4} & \multirow{2}{*}{(1,1)} & MLE & 0.025 & 0.124 & 0.126 & 0.127 & 96.0\\
		&  & PMLE & 0.007 & 0.125 & 0.126 & 0.125 & 95.2\\
		\cmidrule{4-8} 
		& \multirow{2}{*}{(1,0.5)} & MLE & 0.026 & 0.127 & 0.129 & 0.129 & 96.0\\
		&  & PMLE & 0.007 & 0.127 & 0.129 & 0.127 & 95.6\\
		\midrule
		\multirow{4}{*}{0.6} & \multirow{2}{*}{(1,1)} & MLE & 0.028 & 0.094 & 0.097 & 0.099 & 95.1\\
		&  & PMLE & 0.000 & 0.098 & 0.100 & 0.098 & 94.5\\
		
		\cmidrule{4-8} 
		& \multirow{2}{*}{(1,0.5)} & MLE & 0.027 & 0.095 & 0.099 & 0.099 & 95.9\\
		&  & PMLE & 0.001 & 0.098 & 0.101 & 0.098 & 95.1\\
		\midrule
		\multirow{4}{*}{0.8} & \multirow{2}{*}{(1,1)} & MLE & 0.046 & 0.080 & 0.086 & 0.092 & 94.8\\
		&  & PMLE & 0.011 & 0.080 & 0.087 & 0.081 & 94.6\\
		\cmidrule{4-8} 
		& \multirow{2}{*}{(1,0.5)} & MLE & 0.045 & 0.081 & 0.086 & 0.093 & 95.2\\
		&  & PMLE & 0.012 & 0.081 & 0.087 & 0.081 & 94.6\\
		\bottomrule 
	\end{tabular}
	\vspace{0.4cm}
	\caption{Simulation results for the estimates of $(\bbeta_T,\alpha)$ under the Clayton copula with sample size $n=400$. BIAS: relative bias; ESD: relative empirical standard deviation; ASE: relative average standard error; rMSE: relative root mean square error; CP: empirical coverage percentage of 95\% confidence interval.}
	\label{tab:est_clayton400}
\end{table}

\begin{table}[htbp]
	\centering
	\begin{tabular}{ll cc l rcccc}
		\toprule
		$\tau$ & $\bbeta_{T}$ & Time & Prob. & Method & \multicolumn{1}{c}{BIAS} & ESD & ASE & rMSE & CP  \\
		\midrule
		\multirow{12}{*}{0.4} & \multirow{6}{*}{(1,1)}  & \multirow{2}{*}{0.863} & \multirow{2}{*}{0.750} & MLE & 0.004 & 0.053 & 0.053 & 0.053 & 94.7\\
		&  &  &  & PMLE & 0.003 & 0.054 & 0.054 & 0.054 & 94.4\\	
		\cmidrule{3-10} 
		&  & \multirow{2}{*}{2.079} & \multirow{2}{*}{0.500}  & MLE & 0.006 & 0.124 & 0.121 & 0.124 & 94.4\\
		&  &  &  & PMLE & 0.004 & 0.126 & 0.123 & 0.126 & 94.3\\
		\cmidrule{3-10} 
		&  & \multirow{2}{*}{4.159} & \multirow{2}{*}{0.250} & MLE & 0.065 & 0.317 & 0.265 & 0.324 & 84.2\\
		&  &  &  & PMLE & 0.068 & 0.320 & 0.267 & 0.327 & 83.9\\ 
		\cmidrule{3-10} 
		& \multirow{6}{*}{(1,0.5)} & \multirow{2}{*}{0.863} & \multirow{2}{*}{0.750} & MLE & 0.004 & 0.054 & 0.053 & 0.054 & 93.9\\
		&  &  &  & PMLE & 0.003 & 0.055 & 0.053 & 0.055 & 93.3\\
		\cmidrule{3-10} 
		&  & \multirow{2}{*}{2.079} & \multirow{2}{*}{0.500} & MLE & 0.007 & 0.123 & 0.120 & 0.123 & 93.4\\
		&  &  &  & PMLE & 0.005 & 0.125 & 0.121 & 0.125 & 93.6\\
		\cmidrule{3-10} 
		&  & \multirow{2}{*}{4.159} & \multirow{2}{*}{0.250} & MLE & 0.020 & 0.291 & 0.260 & 0.291 & 89.2\\
		&  &  &  & PMLE & 0.024 & 0.294 & 0.262 & 0.295 & 88.7\\
		\midrule
		\multirow{12}{*}{0.6} & \multirow{6}{*}{(1,1)} & \multirow{2}{*}{0.863} & \multirow{2}{*}{0.750} & MLE & 0.005 & 0.051 & 0.052 & 0.052 & 95.0\\
		&  &  &  & PMLE & 0.003 & 0.054 & 0.053 & 0.054 & 93.9\\
		\cmidrule{3-10} 
		& & \multirow{2}{*}{2.079} & \multirow{2}{*}{0.500} & MLE & 0.008 & 0.120 & 0.117 & 0.120 & 94.4\\
		&  &  &  & PMLE & 0.005 & 0.124 & 0.120 & 0.124 & 94.3\\
		\cmidrule{3-10} 
		&  & \multirow{2}{*}{4.159} & \multirow{2}{*}{0.250} & MLE & 0.046 & 0.295 & 0.255 & 0.298 & 88.1\\
		&  &  &  & PMLE & 0.052 & 0.301 & 0.260 & 0.305 & 87.2\\
		\cmidrule{3-10}  
		& \multirow{6}{*}{(1,0.5)} & \multirow{2}{*}{0.863} & \multirow{2}{*}{0.750}& MLE & 0.005 & 0.052 & 0.051 & 0.052 & 94.8\\
		&  &  &  & PMLE & 0.003 & 0.054 & 0.052 & 0.054 & 93.5\\
		\cmidrule{3-10} 
		&  & \multirow{2}{*}{2.079} & \multirow{2}{*}{0.500} & MLE & 0.008 & 0.118 & 0.115 & 0.119 & 94.2\\
		&  &  &  & PMLE & 0.005 & 0.123 & 0.119 & 0.123 & 93.9\\
		\cmidrule{3-10} 
		&  & \multirow{2}{*}{4.159} & \multirow{2}{*}{0.250} & MLE & 0.013 & 0.270 & 0.246 & 0.270 & 92.2\\
		&  &  &  & PMLE & 0.019 & 0.277 & 0.252 & 0.277 & 91.6\\
		\midrule
		\multirow{12}{*}{0.8} & \multirow{6}{*}{(1,1)} & \multirow{2}{*}{0.863} & \multirow{2}{*}{0.750} & MLE & 0.006 & 0.050 & 0.051 & 0.051 & 95.2\\
		&  &  &  & PMLE & 0.003 & 0.054 & 0.053 & 0.054 & 93.4\\
		\cmidrule{3-10}
		&  & \multirow{2}{*}{2.079} & \multirow{2}{*}{0.500}& MLE & 0.006 & 0.116 & 0.115 & 0.116 & 94.7\\
		&  &  &  & PMLE & 0.002 & 0.123 & 0.121 & 0.123 & 93.7\\
		\cmidrule{3-10} 
		&  & \multirow{2}{*}{4.159} & \multirow{2}{*}{0.250}& MLE & 0.046 & 0.275 & 0.245 & 0.279 & 88.4\\
		&  &  &  & PMLE & 0.055 & 0.285 & 0.256 & 0.291 & 88.9\\
		\cmidrule{3-10} 
		& \multirow{6}{*}{(1,0.5)} & \multirow{2}{*}{0.863} & \multirow{2}{*}{0.750} & MLE & 0.006 & 0.051 & 0.050 & 0.051 & 94.3\\
		&  &  &  & PMLE & 0.002 & 0.054 & 0.052 & 0.054 & 92.9\\
		\cmidrule{3-10}
		&  & \multirow{2}{*}{2.079} & \multirow{2}{*}{0.500} & MLE & 0.006 & 0.116 & 0.114 & 0.116 & 94.6\\
		&  &  &  & PMLE & 0.003 & 0.123 & 0.120 & 0.123 & 94.1\\
		\cmidrule{3-10}
		&  & \multirow{2}{*}{4.159} & \multirow{2}{*}{0.250}& MLE & 0.017 & 0.256 & 0.238 & 0.256 & 92.0\\
		&  &  &  & PMLE & 0.023 & 0.268 & 0.249 & 0.269 & 91.7\\
		\bottomrule 
	\end{tabular}
	\vspace{0.4cm}
	\caption{Simulation results for the baseline survival function of $T$ under the Gumbel copula with sample size $n=400$. BIAS: relative bias; ESD: relative empirical standard deviation; ASE: relative average standard error; rMSE: relative root mean square error; CP: empirical coverage percentage of 95\% confidence interval.}
	\label{tab:BHaz_gumbel400}
\end{table}

\begin{table}[htbp]
	\centering
	\begin{tabular}{ll cc l rcccc}
		\toprule
		$\tau$ & $\bbeta_{T}$ & Time & Prob. & Method & \multicolumn{1}{c}{BIAS} & ESD & ASE & rMSE & CP  \\
		\midrule
		\multirow{12}{*}{0.4} & \multirow{6}{*}{(1,1)}  & \multirow{2}{*}{0.863} & \multirow{2}{*}{0.750} & MLE & 0.000 & 0.053 & 0.053 & 0.053 & 94.2\\
		&  &  &  & PMLE & 0.001 & 0.054 & 0.054 & 0.054 & 94.2\\
		\cmidrule{3-10} 
		&  & \multirow{2}{*}{2.079} & \multirow{2}{*}{0.500} & MLE & 0.000 & 0.121 & 0.116 & 0.121 & 94.4\\
		&  &  &  & PMLE & 0.003 & 0.123 & 0.120 & 0.123 & 93.7\\
		\cmidrule{3-10} 
		&  & \multirow{2}{*}{4.159} & \multirow{2}{*}{0.250}& MLE & 0.052 & 0.289 & 0.241 & 0.293 & 86.9\\
		&  &  &  & PMLE & 0.061 & 0.290 & 0.249 & 0.296 & 87.0\\ 
		\cmidrule{3-10} 
		& \multirow{6}{*}{(1,0.5)} & \multirow{2}{*}{0.863} & \multirow{2}{*}{0.750} & MLE & 0.001 & 0.053 & 0.052 & 0.053 & 94.1\\
		&  &  &  & PMLE & 0.000 & 0.055 & 0.054 & 0.055 & 93.4\\
		\cmidrule{3-10} 
		&  & \multirow{2}{*}{2.079} & \multirow{2}{*}{0.500}& MLE & 0.000 & 0.121 & 0.116 & 0.121 & 93.9\\
		&  &  &  & PMLE & 0.002 & 0.123 & 0.120 & 0.123 & 94.4\\
		\cmidrule{3-10} 
		&  & \multirow{2}{*}{4.159} & \multirow{2}{*}{0.250} & MLE & 0.021 & 0.266 & 0.236 & 0.267 & 90.9\\
		&  &  &  & PMLE & 0.030 & 0.268 & 0.245 & 0.270 & 91.1\\
		\midrule
		\multirow{12}{*}{0.6} & \multirow{6}{*}{(1,1)} & \multirow{2}{*}{0.863} & \multirow{2}{*}{0.750} & MLE & 0.001 & 0.048 & 0.047 & 0.048 & 94.1\\
		&  &  &  & PMLE & 0.003 & 0.053 & 0.053 & 0.053 & 93.3\\
		\cmidrule{3-10} 
		& & \multirow{2}{*}{2.079} & \multirow{2}{*}{0.500} & MLE & 0.001 & 0.107 & 0.103 & 0.107 & 94.6\\
		&  &  &  & PMLE & 0.006 & 0.118 & 0.117 & 0.118 & 94.1\\
		\cmidrule{3-10} 
		&  & \multirow{2}{*}{4.159} & \multirow{2}{*}{0.250} & MLE & 0.048 & 0.257 & 0.215 & 0.261 & 86.1\\
		&  &  &  & PMLE & 0.066 & 0.270 & 0.241 & 0.278 & 87.9\\
		\cmidrule{3-10}  
		& \multirow{6}{*}{(1,0.5)} & \multirow{2}{*}{0.863} & \multirow{2}{*}{0.750}& MLE & 0.001 & 0.048 & 0.047 & 0.048 & 93.8\\
		&  &  &  & PMLE & 0.002 & 0.053 & 0.053 & 0.053 & 94.2\\
		\cmidrule{3-10} 
		&  & \multirow{2}{*}{2.079} & \multirow{2}{*}{0.500} & MLE & 0.001 & 0.107 & 0.102 & 0.107 & 94.5\\
		&  &  &  & PMLE & 0.003 & 0.118 & 0.116 & 0.118 & 94.5\\
		\cmidrule{3-10} 
		&  & \multirow{2}{*}{4.159} & \multirow{2}{*}{0.250} & MLE & 0.014 & 0.233 & 0.210 & 0.233 & 90.9\\
		&  &  &  & PMLE & 0.029 & 0.251 & 0.237 & 0.253 & 92.7\\
		\midrule
		\multirow{12}{*}{0.8} & \multirow{6}{*}{(1,1)} & \multirow{2}{*}{0.863} & \multirow{2}{*}{0.750} & MLE & 0.002 & 0.043 & 0.042 & 0.043 & 94.6\\
		&  &  &  & PMLE & 0.005 & 0.054 & 0.054 & 0.054 & 92.9\\
		\cmidrule{3-10}
		&  & \multirow{2}{*}{2.079} & \multirow{2}{*}{0.500} & MLE & 0.001 & 0.095 & 0.094 & 0.095 & 94.8\\
		&  &  &  & PMLE & 0.006 & 0.120 & 0.119 & 0.120 & 94.2\\
		\cmidrule{3-10} 
		&  & \multirow{2}{*}{4.159} & \multirow{2}{*}{0.250} & MLE & 0.056 & 0.234 & 0.193 & 0.241 & 86.0\\
		&  &  &  & PMLE & 0.080 & 0.267 & 0.242 & 0.278 & 88.3\\
		\cmidrule{3-10} 
		& \multirow{6}{*}{(1,0.5)} & \multirow{2}{*}{0.863} & \multirow{2}{*}{0.750} & MLE & 0.004 & 0.042 & 0.042 & 0.042 & 93.7\\
		&  &  &  & PMLE & 0.004 & 0.055 & 0.054 & 0.055 & 92.6\\
		\cmidrule{3-10}
		&  & \multirow{2}{*}{2.079} & \multirow{2}{*}{0.500} & MLE & 0.000 & 0.095 & 0.093 & 0.095 & 94.9\\
		&  &  &  & PMLE & 0.005 & 0.120 & 0.118 & 0.120 & 95.2\\
		\cmidrule{3-10}
		&  & \multirow{2}{*}{4.159} & \multirow{2}{*}{0.250} & MLE & 0.013 & 0.210 & 0.192 & 0.210 & 92.0\\
		&  &  &  & PMLE & 0.030 & 0.254 & 0.244 & 0.256 & 92.6\\
		\bottomrule 
	\end{tabular}
	\vspace{0.4cm}
	\caption{Simulation results for the baseline survival function of $T$ under the Clayton copula with sample size $n=400$. BIAS: relative bias; ESD: relative empirical standard deviation; ASE: relative average standard error; rMSE: relative root mean square error; CP: empirical coverage percentage of 95\% confidence interval.}
	\label{tab:BHaz_claytonl400}
\end{table}

\begin{table}[htbp]
	\centering
		\begin{tabular}{cl ccc ccc}
			\toprule
			\multicolumn{2}{c}{} & \multicolumn{3}{c}{BIAS} & \multicolumn{3}{c}{ESD} \\
            \cmidrule{3-8}
			Parameter & Method & $n=200$ & $n=400$ & $n=1,000$ & $n=200$ & $n=400$ & $n=1,000$ \\
			\midrule
			\multicolumn{8}{c}{\textbf{Gumbel}} \\
            \cmidrule{1-8}
			\multirow{2}{*}{$\beta_{T,1}$} & MLE & 0.007 & 0.006 & 0.000 & 0.156 & 0.111 & 0.070\\
			 & PMLE & 0.005 & 0.002 & 0.003 & 0.168 & 0.121 & 0.075\\
			\cmidrule{3-8}
			\multirow{2}{*}{$\beta_{T,2}$} & MLE & 0.024 & 0.011 & 0.005 & 0.196 & 0.139 & 0.086\\
			 & PMLE & 0.006 & 0.003 & 0.002 & 0.211 & 0.151 & 0.090\\
			\cmidrule{3-8}
			\multirow{2}{*}{$\alpha$} & MLE & 0.096 & 0.049 & 0.021 & 0.105 & 0.065 & 0.039\\
		    & PMLE & 0.020 & 0.010 & 0.004 & 0.093 & 0.062 & 0.040\\
			\midrule
			\multicolumn{8}{c}{\textbf{Clayton}}\\
            \cmidrule{1-8}
			\multirow{2}{*}{$\beta_{T,1}$}& MLE & 0.034 & 0.016 & 0.008 & 0.133 & 0.091 & 0.056\\
			& PMLE & 0.027 & 0.016 & 0.010 & 0.173 & 0.122 & 0.077\\
			\cmidrule{3-8}
			\multirow{2}{*}{$\beta_{T,2}$} & MLE & 0.001 & 0.002 & 0.000 & 0.161 & 0.111 & 0.068\\
			& PMLE & 0.025 & 0.018 & 0.010 & 0.223 & 0.156 & 0.094\\
			\cmidrule{3-8}
			\multirow{2}{*}{$\alpha$} & MLE & 0.101 & 0.046 & 0.019 & 0.138 & 0.080 & 0.051\\
		    & PMLE & 0.011 & 0.011 & 0.006 & 0.125 & 0.080 & 0.053\\
			\bottomrule	
			
		\end{tabular}%
    \vspace{0.4cm}
	\caption{Simulation results for the scenario of Kendall's $\tau=0.8$ and $\bbeta_T=(1,1)$ with the Gumbel and Clayton copula under sample sizes $n=200$, $400$ and $1,000$. BIAS: relative bias; ESD: relative empirical standard deviation.}
	\label{tab:nsamp_comp}
\end{table}

\begin{table}[htbp]
	\centering
	\begin{tabular}{lll rcccc}
		\toprule
		$\tau$ & $\bbeta_{T}$ & Method & BIAS & ESD & ASE & rMSE & CP  \\
		\midrule
		\multicolumn{3}{c}{} & \multicolumn{5}{c}{$\widehat\beta_{D,1}$} \\
		\cmidrule{4-8}
		\multirow{4}{*}{0.4} & \multirow{2}{*}{(1,1)}   & MLE & 0.007 & 0.164 & 0.159 & 0.164 & 94.0\\
		&  & PMLE & 0.006 & 0.170 & 0.165 & 0.170 & 93.9\\     
		\cmidrule{4-8} 
		& \multirow{2}{*}{(1,0.5)}  & MLE & 0.007 & 0.164 & 0.159 & 0.164 & 94.1\\
		&  & PMLE & 0.006 & 0.170 & 0.165 & 0.170 & 93.9\\
		\midrule
		\multirow{4}{*}{0.6} & \multirow{2}{*}{(1,1)}  & MLE & 0.005 & 0.160 & 0.154 & 0.160 & 94.2\\
		&  & PMLE & 0.003 & 0.168 & 0.164 & 0.169 & 94.0\\
		\cmidrule{4-8} 
		& \multirow{2}{*}{(1,0.5)}& MLE & 0.006 & 0.160 & 0.155 & 0.160 & 94.2\\
		&  & PMLE & 0.003 & 0.168 & 0.164 & 0.169 & 94.0\\
		\midrule
		\multirow{4}{*}{0.8} & \multirow{2}{*}{(1,1)}  & MLE & 0.004 & 0.154 & 0.151 & 0.154 & 94.6\\
		&  & PMLE & 0.006 & 0.166 & 0.164 & 0.166 & 95.5\\
		\cmidrule{4-8} 
		& \multirow{2}{*}{(1,0.5)}  & MLE & 0.005 & 0.155 & 0.152 & 0.156 & 94.9\\
		&  & PMLE & 0.006 & 0.166 & 0.164 & 0.166 & 95.5\\
		\midrule 
		\multicolumn{3}{c}{} & \multicolumn{5}{c}{$\widehat\beta_{D,2}$}\\
		\cline{4-8}
		\multirow{4}{*}{0.4} & \multirow{2}{*}{(1,1)}   & MLE & 0.002 & 0.193 & 0.189 & 0.193 & 94.8\\
		&  & PMLE & 0.001 & 0.201 & 0.196 & 0.201 & 94.3\\
		\cmidrule{4-8} 
		& \multirow{2}{*}{(1,0.5)} & MLE & 0.001 & 0.193 & 0.189 & 0.193 & 94.8\\
		&  & PMLE & 0.001 & 0.201 & 0.196 & 0.201 & 94.3\\
		
		\midrule
		\multirow{4}{*}{0.6} & \multirow{2}{*}{(1,1)}   & MLE & 0.002 & 0.191 & 0.186 & 0.191 & 93.8\\
		&  & PMLE & 0.004 & 0.203 & 0.197 & 0.203 & 93.9\\
		\cmidrule{4-8} 
		& \multirow{2}{*}{(1,0.5)} & MLE & 0.002 & 0.191 & 0.185 & 0.191 & 93.6\\
		&  & PMLE & 0.004 & 0.203 & 0.197 & 0.203 & 93.9\\
		\midrule 
		\multirow{4}{*}{0.8} & \multirow{2}{*}{(1,1)} & MLE & 0.005 & 0.191 & 0.183 & 0.191 & 93.3\\
		&  & PMLE & 0.001 & 0.207 & 0.197 & 0.207 & 93.8\\
		\cmidrule{4-8} 
		& \multirow{2}{*}{(1,0.5)} & MLE & 0.005 & 0.191 & 0.182 & 0.191 & 93.6\\
		&  & PMLE & 0.001 & 0.207 & 0.197 & 0.207 & 93.8\\
		\bottomrule 
	\end{tabular}
	\vspace{0.4cm}
	\caption{Simulation results for the estimates of $\bbeta_D$ under the Gumbel copula with sample size $n=200$. BIAS: absolute bias; ESD: empirical standard deviation; ASE: average standard error; rMSE: root mean square error; CP: empirical coverage percentage of 95\% confidence interval.}
	\label{tab:D.est_gumbel200}
\end{table}

\begin{table}[htbp]
	\centering
	\begin{tabular}{lll rcccc}
		\toprule
		$\tau$ & $\bbeta_{T}$ & Method & BIAS & ESD & ASE & rMSE & CP  \\
		\midrule
		\multicolumn{3}{c}{} & \multicolumn{5}{c}{$\widehat\beta_{D,1}$} \\
		\cmidrule{4-8}
		\multirow{4}{*}{0.4} & \multirow{2}{*}{(1,1)} & MLE & 0.005 & 0.112 & 0.112 & 0.112 & 95.0\\
		&  & PMLE & 0.004 & 0.116 & 0.115 & 0.116 & 94.4\\    
		\cmidrule{4-8} 
		& \multirow{2}{*}{(1,0.5)} & MLE & 0.005 & 0.112 & 0.112 & 0.112 & 94.5\\
		&  & PMLE & 0.004 & 0.116 & 0.115 & 0.116 & 94.4\\
		\midrule
		\multirow{4}{*}{0.6} & \multirow{2}{*}{(1,1)} & MLE & 0.003 & 0.109 & 0.109 & 0.110 & 94.9\\
		&  & PMLE & 0.002 & 0.118 & 0.116 & 0.118 & 93.5\\
		\cmidrule{4-8} 
		& \multirow{2}{*}{(1,0.5)}& MLE & 0.003 & 0.111 & 0.110 & 0.111 & 95.1\\
		&  & PMLE & 0.002 & 0.118 & 0.116 & 0.118 & 93.5\\
		\midrule
		\multirow{4}{*}{0.8} & \multirow{2}{*}{(1,1)} & MLE & 0.003 & 0.107 & 0.107 & 0.107 & 94.3\\
		&  & PMLE & 0.000 & 0.117 & 0.116 & 0.117 & 94.7\\
		\cmidrule{4-8} 
		& \multirow{2}{*}{(1,0.5)} & MLE & 0.002 & 0.109 & 0.108 & 0.109 & 94.6\\
		&  & PMLE & 0.000 & 0.117 & 0.116 & 0.117 & 94.7\\
		\midrule 
		\multicolumn{3}{c}{} & \multicolumn{5}{c}{$\widehat\beta_{D,2}$}\\
		\cline{4-8}
		\multirow{4}{*}{0.4} & \multirow{2}{*}{(1,1)} & MLE & 0.002 & 0.136 & 0.134 & 0.136 & 94.4\\
		&  & PMLE & 0.001 & 0.141 & 0.138 & 0.141 & 94.5\\
		\cmidrule{4-8} 
		& \multirow{2}{*}{(1,0.5)} & MLE & 0.002 & 0.135 & 0.133 & 0.135 & 94.1\\
		&  & PMLE & 0.001 & 0.141 & 0.138 & 0.141 & 94.5\\
		
		\midrule
		\multirow{4}{*}{0.6} & \multirow{2}{*}{(1,1)} & MLE & 0.001 & 0.135 & 0.131 & 0.135 & 94.0\\
		&  & PMLE & 0.001 & 0.144 & 0.138 & 0.144 & 93.7\\
		\cmidrule{4-8} 
		& \multirow{2}{*}{(1,0.5)} & MLE & 0.002 & 0.134 & 0.131 & 0.134 & 93.9\\
		&  & PMLE & 0.001 & 0.144 & 0.138 & 0.144 & 93.7\\
		\midrule 
		\multirow{4}{*}{0.8} & \multirow{2}{*}{(1,1)} & MLE & 0.004 & 0.133 & 0.129 & 0.133 & 93.8\\
		&  & PMLE & 0.003 & 0.145 & 0.138 & 0.145 & 93.7\\
		\cmidrule{4-8} 
		& \multirow{2}{*}{(1,0.5)} & MLE & 0.004 & 0.133 & 0.129 & 0.133 & 93.4\\
		&  & PMLE & 0.003 & 0.145 & 0.138 & 0.145 & 93.7\\
		\bottomrule 
	\end{tabular}
	\vspace{0.4cm}
	\caption{Simulation results for the estimates of $\bbeta_D$ under the Gumbel copula with sample size $n=400$. BIAS: absolute bias; ESD: empirical standard deviation; ASE: average standard error; rMSE: root mean square error; CP: empirical coverage percentage of 95\% confidence interval.}
	\label{tab:D.est_gumbel400}
\end{table}

\begin{table}[htbp]
	\centering
	\begin{tabular}{lll rcccc}
		\toprule
		$\tau$ & $\bbeta_{T}$ & Method & BIAS & ESD & ASE & rMSE & CP  \\
		\midrule
		\multicolumn{3}{c}{} & \multicolumn{5}{c}{$\widehat\beta_{D,1}$} \\
		\cmidrule{4-8}
		\multirow{4}{*}{0.4} & \multirow{2}{*}{(1,1)}   & MLE & 0.004 & 0.149 & 0.145 & 0.149 & 94.7\\
		&  & PMLE & 0.005 & 0.171 & 0.165 & 0.171 & 94.3\\     
		\cmidrule{4-8} 
		& \multirow{2}{*}{(1,0.5)}  & MLE & 0.004 & 0.150 & 0.147 & 0.150 & 95.2\\
		&  & PMLE & 0.005 & 0.171 & 0.165 & 0.171 & 94.3\\
		\midrule
		\multirow{4}{*}{0.6} & \multirow{2}{*}{(1,1)}  & MLE & 0.002 & 0.135 & 0.130 & 0.135 & 93.3\\
		&  & PMLE & 0.003 & 0.167 & 0.165 & 0.168 & 95.2\\
		\cmidrule{4-8} 
		& \multirow{2}{*}{(1,0.5)} & MLE & 0.002 & 0.136 & 0.132 & 0.136 & 93.8\\
		&  & PMLE & 0.003 & 0.167 & 0.165 & 0.168 & 95.2\\
		\midrule
		\multirow{4}{*}{0.8} & \multirow{2}{*}{(1,1)} & MLE & 0.005 & 0.121 & 0.122 & 0.121 & 94.8\\
		&  & PMLE & 0.005 & 0.167 & 0.165 & 0.167 & 95.1\\
		\cmidrule{4-8} 
		& \multirow{2}{*}{(1,0.5)} & MLE & 0.005 & 0.122 & 0.123 & 0.122 & 95.0\\
		&  & PMLE & 0.005 & 0.166 & 0.165 & 0.166 & 95.1\\
		\midrule 
		\multicolumn{3}{c}{} & \multicolumn{5}{c}{$\widehat\beta_{D,2}$}\\
		\cline{4-8}
		\multirow{4}{*}{0.4} & \multirow{2}{*}{(1,1)} & MLE & 0.014 & 0.180 & 0.174 & 0.181 & 93.2\\
		&  & PMLE & 0.004 & 0.202 & 0.197 & 0.202 & 94.3\\
		\cmidrule{4-8} 
		& \multirow{2}{*}{(1,0.5)}& MLE & 0.010 & 0.180 & 0.174 & 0.180 & 93.0\\
		&  & PMLE & 0.004 & 0.202 & 0.197 & 0.202 & 94.3\\
		\midrule
		\multirow{4}{*}{0.6} & \multirow{2}{*}{(1,1)} & MLE & 0.026 & 0.164 & 0.156 & 0.166 & 93.5\\
		&  & PMLE & 0.001 & 0.204 & 0.197 & 0.204 & 93.6\\
		\cmidrule{4-8} 
		& \multirow{2}{*}{(1,0.5)} & MLE & 0.017 & 0.165 & 0.156 & 0.166 & 93.1\\
		&  & PMLE & 0.001 & 0.204 & 0.197 & 0.204 & 93.6\\
		\midrule 
		\multirow{4}{*}{0.8} & \multirow{2}{*}{(1,1)} & MLE & 0.036 & 0.152 & 0.145 & 0.156 & 93.0\\
		&  & PMLE & 0.005 & 0.209 & 0.197 & 0.209 & 93.7\\
		\cmidrule{4-8} 
		& \multirow{2}{*}{(1,0.5)} & MLE & 0.022 & 0.152 & 0.145 & 0.154 & 92.8\\
		&  & PMLE & 0.005 & 0.208 & 0.197 & 0.208 & 93.8\\
		\bottomrule 
	\end{tabular}
	\vspace{0.4cm}
	\caption{Simulation results for the estimates of $\bbeta_D$ under the Clayton copula with sample size $n=200$. BIAS: absolute bias; ESD: empirical standard deviation; ASE: average standard error; rMSE: root mean square error; CP: empirical coverage percentage of 95\% confidence interval.}
	\label{tab:D.est_clayton200}
\end{table}

\begin{table}[htbp]
	\centering
	\begin{tabular}{lll rcccc}
		\toprule
		$\tau$ & $\bbeta_{T}$ & Method & BIAS & ESD & ASE & rMSE & CP  \\
		\midrule
		\multicolumn{3}{c}{} & \multicolumn{5}{c}{$\widehat\beta_{D,1}$} \\
		\cmidrule{4-8}
		\multirow{4}{*}{0.4} & \multirow{2}{*}{(1,1)} & MLE & 0.005 & 0.105 & 0.102 & 0.105 & 93.3\\
		&  & PMLE & 0.005 & 0.117 & 0.116 & 0.117 & 95.0\\     
		\cmidrule{4-8} 
		& \multirow{2}{*}{(1,0.5)} & MLE & 0.006 & 0.107 & 0.103 & 0.107 & 92.7\\
		&  & PMLE & 0.005 & 0.117 & 0.116 & 0.117 & 95.0\\
		\midrule
		\multirow{4}{*}{0.6} & \multirow{2}{*}{(1,1)}& MLE & 0.005 & 0.094 & 0.090 & 0.094 & 93.8\\
		&  & PMLE & 0.003 & 0.117 & 0.116 & 0.117 & 94.8\\
		\cmidrule{4-8} 
		& \multirow{2}{*}{(1,0.5)} & MLE & 0.005 & 0.095 & 0.091 & 0.095 & 93.2\\
		&  & PMLE & 0.003 & 0.117 & 0.116 & 0.117 & 94.8\\
		\midrule
		\multirow{4}{*}{0.8} & \multirow{2}{*}{(1,1)}& MLE & 0.002 & 0.084 & 0.082 & 0.084 & 94.5\\
		&  & PMLE & 0.001 & 0.116 & 0.116 & 0.116 & 95.1\\
		\cmidrule{4-8} 
		& \multirow{2}{*}{(1,0.5)}& MLE & 0.003 & 0.085 & 0.082 & 0.085 & 94.6\\
		&  & PMLE & 0.000 & 0.116 & 0.116 & 0.116 & 95.1\\
		\midrule 
		\multicolumn{3}{c}{} & \multicolumn{5}{c}{$\widehat\beta_{D,2}$}\\
		\cline{4-8}
		\multirow{4}{*}{0.4} & \multirow{2}{*}{(1,1)}& MLE & 0.009 & 0.125 & 0.122 & 0.126 & 94.5\\
		&  & PMLE & 0.000 & 0.142 & 0.138 & 0.142 & 93.9\\
		\cmidrule{4-8} 
		& \multirow{2}{*}{(1,0.5)} & MLE & 0.007 & 0.125 & 0.123 & 0.126 & 94.2\\
		&  & PMLE & 0.000 & 0.142 & 0.138 & 0.142 & 93.9\\
		\midrule
		\multirow{4}{*}{0.6} & \multirow{2}{*}{(1,1)} & MLE & 0.018 & 0.112 & 0.108 & 0.113 & 93.4\\
		&  & PMLE & 0.001 & 0.145 & 0.138 & 0.145 & 94.2\\
		\cmidrule{4-8} 
		& \multirow{2}{*}{(1,0.5)} & MLE & 0.013 & 0.112 & 0.108 & 0.113 & 93.5\\
		&  & PMLE & 0.001 & 0.145 & 0.138 & 0.145 & 94.2\\
		\midrule 
		\multirow{4}{*}{0.8} & \multirow{2}{*}{(1,1)} & MLE & 0.025 & 0.101 & 0.098 & 0.104 & 93.6\\
		&  & PMLE & 0.002 & 0.144 & 0.138 & 0.144 & 94.2\\
		\cmidrule{4-8} 
		& \multirow{2}{*}{(1,0.5)}& MLE & 0.016 & 0.100 & 0.098 & 0.102 & 93.7\\
		&  & PMLE & 0.002 & 0.144 & 0.138 & 0.144 & 94.2\\
		\bottomrule 
	\end{tabular}
	\vspace{0.4cm}
	\caption{Simulation results for the estimates of $\bbeta_D$ under the Clayton copula with sample size $n=400$. BIAS: absolute bias; ESD: empirical standard deviation; ASE: average standard error; rMSE: root mean square error; CP: empirical coverage percentage of 95\% confidence interval.}
	\label{tab:D.est_clayton400}
\end{table}

\begin{table}[htbp]
	\centering
	\begin{tabular}{ll cc l rcccc}
		\toprule
		$\tau$ & $\bbeta_{T}$ & Time & Prob. & Method & \multicolumn{1}{c}{BIAS} & ESD & ASE & rMSE & CP  \\
		\midrule
		\multirow{12}{*}{0.4} & \multirow{6}{*}{(1,1)}  & \multirow{2}{*}{0.863} & \multirow{2}{*}{0.750} & MLE & 0.006 & 0.078 & 0.075 & 0.078 & 92.9\\
		 &  &  &  & PMLE & 0.002 & 0.079 & 0.076 & 0.079 & 92.3\\		
		\cmidrule{3-10} 
		&  & \multirow{2}{*}{2.079} & \multirow{2}{*}{0.500} & MLE & 0.000 & 0.169 & 0.166 & 0.169 & 94.8\\
		 &  &  &  & PMLE & 0.002 & 0.173 & 0.171 & 0.173 & 93.9\\
		\cmidrule{3-10} 
		&  & \multirow{2}{*}{4.159} & \multirow{2}{*}{0.250}& MLE & 0.037 & 0.332 & 0.328 & 0.334 & 93.9\\
		 &  &  &  & PMLE & 0.035 & 0.340 & 0.336 & 0.342 & 94.0\\ 
		\cmidrule{3-10} 
		& \multirow{6}{*}{(1,0.5)} & \multirow{2}{*}{0.863} & \multirow{2}{*}{0.750} & MLE & 0.006 & 0.078 & 0.075 & 0.078 & 93.0\\
		 &  &  &  & PMLE & 0.002 & 0.079 & 0.076 & 0.079 & 92.3\\
		\cmidrule{3-10} 
		&  & \multirow{2}{*}{2.079} & \multirow{2}{*}{0.500}& MLE & 0.000 & 0.169 & 0.166 & 0.169 & 94.6\\
		 &  &  &  & PMLE & 0.002 & 0.173 & 0.171 & 0.173 & 93.9\\
		\cmidrule{3-10} 
		&  & \multirow{2}{*}{4.159} & \multirow{2}{*}{0.250}& MLE & 0.036 & 0.332 & 0.327 & 0.334 & 94.0\\
		 &  &  &  & PMLE & 0.035 & 0.340 & 0.336 & 0.342 & 94.0\\
		\midrule
		\multirow{12}{*}{0.6} & \multirow{6}{*}{(1,1)} & \multirow{2}{*}{0.863} & \multirow{2}{*}{0.750} & MLE & 0.008 & 0.075 & 0.074 & 0.076 & 95.0\\
		 &  &  &  & PMLE & 0.002 & 0.077 & 0.076 & 0.077 & 93.7\\
		\cmidrule{3-10} 
		& & \multirow{2}{*}{2.079} & \multirow{2}{*}{0.500} & MLE & 0.002 & 0.163 & 0.164 & 0.163 & 94.6\\
		 &  &  &  & PMLE & 0.002 & 0.170 & 0.171 & 0.170 & 94.3\\
		\cmidrule{3-10} 
		&  & \multirow{2}{*}{4.159} & \multirow{2}{*}{0.250} & MLE & 0.039 & 0.322 & 0.324 & 0.324 & 94.1\\
		 &  &  &  & PMLE & 0.038 & 0.334 & 0.338 & 0.336 & 94.1\\
		\cmidrule{3-10}  
		& \multirow{6}{*}{(1,0.5)} & \multirow{2}{*}{0.863} & \multirow{2}{*}{0.750}& MLE & 0.008 & 0.075 & 0.074 & 0.076 & 94.8\\
		 &  &  &  & PMLE & 0.002 & 0.077 & 0.076 & 0.077 & 93.7\\
		\cmidrule{3-10} 
		&  & \multirow{2}{*}{2.079} & \multirow{2}{*}{0.500}& MLE & 0.002 & 0.163 & 0.164 & 0.163 & 94.7\\
		 &  &  &  & PMLE & 0.002 & 0.170 & 0.171 & 0.170 & 94.3\\
		\cmidrule{3-10} 
		&  & \multirow{2}{*}{4.159} & \multirow{2}{*}{0.250} & MLE & 0.040 & 0.322 & 0.324 & 0.324 & 93.3\\
		 &  &  &  & PMLE & 0.038 & 0.334 & 0.338 & 0.336 & 94.1\\
		\midrule
		\multirow{12}{*}{0.8} & \multirow{6}{*}{(1,1)} & \multirow{2}{*}{0.863} & \multirow{2}{*}{0.750} & MLE & 0.013 & 0.073 & 0.073 & 0.074 & 95.8\\
		 &  &  &  & PMLE & 0.003 & 0.076 & 0.077 & 0.076 & 93.9\\
		\cmidrule{3-10}
		&  & \multirow{2}{*}{2.079} & \multirow{2}{*}{0.500}& MLE & 0.011 & 0.162 & 0.162 & 0.162 & 94.6\\
		 &  &  &  & PMLE & 0.001 & 0.172 & 0.171 & 0.172 & 93.0\\
		\cmidrule{3-10} 
		&  & \multirow{2}{*}{4.159} & \multirow{2}{*}{0.250} & MLE & 0.030 & 0.315 & 0.318 & 0.316 & 93.7\\
		 &  &  &  & PMLE & 0.033 & 0.337 & 0.337 & 0.339 & 93.2\\
		\cmidrule{3-10} 
		& \multirow{6}{*}{(1,0.5)} & \multirow{2}{*}{0.863} & \multirow{2}{*}{0.750}& MLE & 0.013 & 0.073 & 0.073 & 0.074 & 95.9\\
		 &  &  &  & PMLE & 0.003 & 0.076 & 0.077 & 0.076 & 93.9\\
		\cmidrule{3-10}
		&  & \multirow{2}{*}{2.079} & \multirow{2}{*}{0.500} & MLE & 0.010 & 0.163 & 0.162 & 0.163 & 94.3\\
		 &  &  &  & PMLE & 0.001 & 0.172 & 0.171 & 0.172 & 93.0\\
		\cmidrule{3-10}
		&  & \multirow{2}{*}{4.159} & \multirow{2}{*}{0.250} & MLE & 0.032 & 0.316 & 0.319 & 0.317 & 93.6\\
		 &  &  &  & PMLE & 0.033 & 0.337 & 0.337 & 0.339 & 93.2\\
		\bottomrule 
	\end{tabular}
	\vspace{0.4cm}
	\caption{Simulation results for the baseline survival function of $D$ under the Gumbel copula with sample size $n=200$. BIAS: relative bias; ESD:  relative empirical standard deviation; ASE:  relative average standard error; rMSE: relative root mean square error; CP: empirical coverage percentage of 95\% confidence interval.}
	\label{tab:BHaz_D_gumbel200}
\end{table}

\begin{table}[htbp]
	\centering
	\begin{tabular}{ll cc l rcccc}
		\toprule
		$\tau$ & $\bbeta_{T}$ & Time & Prob. & Method & \multicolumn{1}{c}{BIAS} & ESD & ASE & rMSE & CP  \\
		\midrule
		\multirow{12}{*}{0.4} & \multirow{6}{*}{(1,1)}  & \multirow{2}{*}{0.863} & \multirow{2}{*}{0.750} & MLE & 0.006 & 0.074 & 0.070 & 0.074 & 93.0\\
		&  &  &  & PMLE & 0.002 & 0.079 & 0.076 & 0.079 & 92.3\\	
		\cmidrule{3-10} 
		&  & \multirow{2}{*}{2.079} & \multirow{2}{*}{0.500}& MLE & 0.003 & 0.161 & 0.155 & 0.161 & 94.4\\
		&  &  &  & PMLE & 0.005 & 0.175 & 0.170 & 0.175 & 93.7\\
		\cmidrule{3-10} 
		&  & \multirow{2}{*}{4.159} & \multirow{2}{*}{0.250} & MLE & 0.020 & 0.314 & 0.304 & 0.315 & 93.7\\
		&  &  &  & PMLE & 0.040 & 0.344 & 0.337 & 0.347 & 93.9\\ 
		\cmidrule{3-10} 
		& \multirow{6}{*}{(1,0.5)} & \multirow{2}{*}{0.863} & \multirow{2}{*}{0.750}& MLE & 0.005 & 0.073 & 0.070 & 0.073 & 92.8\\
		&  &  &  & PMLE & 0.002 & 0.079 & 0.076 & 0.079 & 92.3\\
		\cmidrule{3-10} 
		&  & \multirow{2}{*}{2.079} & \multirow{2}{*}{0.500}& MLE & 0.001 & 0.160 & 0.155 & 0.160 & 94.6\\
		&  &  &  & PMLE & 0.005 & 0.175 & 0.170 & 0.175 & 93.7\\
		\cmidrule{3-10} 
		&  & \multirow{2}{*}{4.159} & \multirow{2}{*}{0.250}  & MLE & 0.024 & 0.314 & 0.305 & 0.315 & 93.8\\
		&  &  &  & PMLE & 0.040 & 0.344 & 0.337 & 0.347 & 93.9\\
		\midrule
		\multirow{12}{*}{0.6} & \multirow{6}{*}{(1,1)} & \multirow{2}{*}{0.863} & \multirow{2}{*}{0.750} & MLE & 0.007 & 0.067 & 0.065 & 0.068 & 93.3\\
		&  &  &  & PMLE & 0.003 & 0.077 & 0.076 & 0.077 & 93.7\\
		\cmidrule{3-10} 
		& & \multirow{2}{*}{2.079} & \multirow{2}{*}{0.500} & MLE & 0.009 & 0.148 & 0.143 & 0.148 & 93.7\\
		&  &  &  & PMLE & 0.001 & 0.170 & 0.171 & 0.170 & 93.7\\
		\cmidrule{3-10} 
		&  & \multirow{2}{*}{4.159} & \multirow{2}{*}{0.250}& MLE & 0.003 & 0.282 & 0.279 & 0.282 & 94.7\\
		&  &  &  & PMLE & 0.028 & 0.333 & 0.336 & 0.334 & 94.3\\
		\cmidrule{3-10}  
		& \multirow{6}{*}{(1,0.5)} & \multirow{2}{*}{0.863} & \multirow{2}{*}{0.750}& MLE & 0.005 & 0.067 & 0.065 & 0.067 & 93.3\\
		&  &  &  & PMLE & 0.003 & 0.077 & 0.076 & 0.077 & 93.7\\
		\cmidrule{3-10} 
		&  & \multirow{2}{*}{2.079} & \multirow{2}{*}{0.500}& MLE & 0.004 & 0.147 & 0.143 & 0.147 & 94.0\\
		&  &  &  & PMLE & 0.001 & 0.170 & 0.171 & 0.170 & 93.7\\
		\cmidrule{3-10} 
		&  & \multirow{2}{*}{4.159} & \multirow{2}{*}{0.250}& MLE & 0.013 & 0.283 & 0.281 & 0.283 & 94.3\\
		&  &  &  & PMLE & 0.028 & 0.333 & 0.336 & 0.334 & 94.3\\
		\midrule
		\multirow{12}{*}{0.8} & \multirow{6}{*}{(1,1)} & \multirow{2}{*}{0.863} & \multirow{2}{*}{0.750} & MLE & 0.002 & 0.062 & 0.061 & 0.062 & 93.6\\
		&  &  &  & PMLE & 0.004 & 0.077 & 0.077 & 0.077 & 94.5\\
		\cmidrule{3-10}
		&  & \multirow{2}{*}{2.079} & \multirow{2}{*}{0.500}& MLE & 0.005 & 0.137 & 0.136 & 0.137 & 94.0\\
		&  &  &  & PMLE & 0.003 & 0.173 & 0.171 & 0.173 & 93.5\\
		\cmidrule{3-10} 
		&  & \multirow{2}{*}{4.159} & \multirow{2}{*}{0.250} & MLE & 0.008 & 0.262 & 0.262 & 0.262 & 95.1\\
		&  &  &  & PMLE & 0.026 & 0.343 & 0.336 & 0.344 & 93.1\\
		\cmidrule{3-10} 
		& \multirow{6}{*}{(1,0.5)} & \multirow{2}{*}{0.863} & \multirow{2}{*}{0.750}& MLE & 0.000 & 0.062 & 0.061 & 0.062 & 92.6\\
		&  &  &  & PMLE & 0.004 & 0.077 & 0.077 & 0.077 & 94.6\\
		\cmidrule{3-10}
		&  & \multirow{2}{*}{2.079} & \multirow{2}{*}{0.500}& MLE & 0.002 & 0.137 & 0.136 & 0.137 & 93.8\\
		&  &  &  & PMLE & 0.003 & 0.172 & 0.171 & 0.172 & 93.6\\
		\cmidrule{3-10}
		&  & \multirow{2}{*}{4.159} & \multirow{2}{*}{0.250} & MLE & 0.006 & 0.264 & 0.264 & 0.264 & 94.4\\
		&  &  &  & PMLE & 0.025 & 0.340 & 0.336 & 0.341 & 93.2\\
		\bottomrule 
	\end{tabular}
	\vspace{0.4cm}
	\caption{Simulation results for the baseline survival function of $D$ under the Clayton copula with sample size $n=200$. BIAS: relative bias; ESD:  relative empirical standard deviation; ASE:  relative average standard error; rMSE: relative root mean square error; CP: empirical coverage percentage of 95\% confidence interval.}
	\label{tab:BHaz_D_clayton200}
\end{table}


\begin{table}[htbp]
	\centering
	\begin{tabular}{ll cc l rcccc}
		\toprule
		$\tau$ & $\bbeta_{T}$ & Time & Prob. & Method & \multicolumn{1}{c}{BIAS} & ESD & ASE & rMSE & CP  \\
		\midrule
		\multirow{12}{*}{0.4} & \multirow{6}{*}{(1,1)}  & \multirow{2}{*}{0.863} & \multirow{2}{*}{0.750} & MLE & 0.006 & 0.055 & 0.053 & 0.055 & 94.0\\
		&  &  &  & PMLE & 0.004 & 0.056 & 0.054 & 0.056 & 93.2\\
		\cmidrule{3-10} 
		&  & \multirow{2}{*}{2.079} & \multirow{2}{*}{0.500} & MLE & 0.008 & 0.120 & 0.118 & 0.120 & 94.0\\
		&  &  &  & PMLE & 0.006 & 0.123 & 0.121 & 0.124 & 94.5\\
		\cmidrule{3-10} 
		&  & \multirow{2}{*}{4.159} & \multirow{2}{*}{0.250}& MLE & 0.005 & 0.239 & 0.232 & 0.239 & 93.7\\
		&  &  &  & PMLE & 0.005 & 0.246 & 0.238 & 0.246 & 93.3\\
		\cmidrule{3-10} 
		& \multirow{6}{*}{(1,0.5)} & \multirow{2}{*}{0.863} & \multirow{2}{*}{0.750} & MLE & 0.006 & 0.055 & 0.053 & 0.055 & 94.2\\
		&  &  &  & PMLE & 0.004 & 0.056 & 0.054 & 0.056 & 93.2\\
		\cmidrule{3-10} 
		&  & \multirow{2}{*}{2.079} & \multirow{2}{*}{0.500} & MLE & 0.008 & 0.120 & 0.118 & 0.120 & 94.3\\
		&  &  &  & PMLE & 0.006 & 0.123 & 0.121 & 0.124 & 94.5\\
		\cmidrule{3-10} 
		&  & \multirow{2}{*}{4.159} & \multirow{2}{*}{0.250} & MLE & 0.005 & 0.238 & 0.232 & 0.238 & 93.8\\
		&  &  &  & PMLE & 0.005 & 0.246 & 0.238 & 0.246 & 93.3\\
		\midrule
		\multirow{12}{*}{0.6} & \multirow{6}{*}{(1,1)} & \multirow{2}{*}{0.863} & \multirow{2}{*}{0.750} & MLE & 0.007 & 0.053 & 0.052 & 0.054 & 94.7\\
		&  &  &  & PMLE & 0.003 & 0.056 & 0.054 & 0.056 & 93.4\\
		\cmidrule{3-10} 
		& & \multirow{2}{*}{2.079} & \multirow{2}{*}{0.500} & MLE & 0.008 & 0.118 & 0.117 & 0.119 & 94.5\\
		&  &  &  & PMLE & 0.005 & 0.125 & 0.121 & 0.125 & 93.8\\
		\cmidrule{3-10} 
		&  & \multirow{2}{*}{4.159} & \multirow{2}{*}{0.250} & MLE & 0.003 & 0.235 & 0.229 & 0.235 & 93.0\\
		&  &  &  & PMLE & 0.004 & 0.248 & 0.238 & 0.248 & 93.3\\
		\cmidrule{3-10}  
		& \multirow{6}{*}{(1,0.5)} & \multirow{2}{*}{0.863} & \multirow{2}{*}{0.750}& MLE & 0.007 & 0.053 & 0.052 & 0.054 & 94.3\\
		&  &  &  & PMLE & 0.003 & 0.056 & 0.054 & 0.056 & 93.4\\
		\cmidrule{3-10} 
		&  & \multirow{2}{*}{2.079} & \multirow{2}{*}{0.500} & MLE & 0.008 & 0.118 & 0.116 & 0.118 & 94.7\\
		&  &  &  & PMLE & 0.005 & 0.125 & 0.121 & 0.125 & 93.8\\
		\cmidrule{3-10} 
		&  & \multirow{2}{*}{4.159} & \multirow{2}{*}{0.250} & MLE & 0.003 & 0.235 & 0.229 & 0.235 & 92.9\\
		&  &  &  & PMLE & 0.004 & 0.248 & 0.238 & 0.248 & 93.3\\
		\midrule
		\multirow{12}{*}{0.8} & \multirow{6}{*}{(1,1)} & \multirow{2}{*}{0.863} & \multirow{2}{*}{0.750} & MLE & 0.009 & 0.052 & 0.051 & 0.052 & 95.1\\
		&  &  &  & PMLE & 0.003 & 0.056 & 0.054 & 0.056 & 93.5\\
		\cmidrule{3-10}
		&  & \multirow{2}{*}{2.079} & \multirow{2}{*}{0.500} & MLE & 0.011 & 0.114 & 0.115 & 0.114 & 95.3\\
		&  &  &  & PMLE & 0.003 & 0.122 & 0.121 & 0.122 & 94.2\\
		\cmidrule{3-10} 
		&  & \multirow{2}{*}{4.159} & \multirow{2}{*}{0.250} & MLE & 0.002 & 0.230 & 0.225 & 0.230 & 94.0\\
		&  &  &  & PMLE & 0.005 & 0.245 & 0.238 & 0.245 & 94.4\\
		\cmidrule{3-10} 
		& \multirow{6}{*}{(1,0.5)} & \multirow{2}{*}{0.863} & \multirow{2}{*}{0.750} & MLE & 0.009 & 0.052 & 0.051 & 0.053 & 94.9\\
		&  &  &  & PMLE & 0.003 & 0.056 & 0.054 & 0.056 & 93.5\\
		\cmidrule{3-10}
		&  & \multirow{2}{*}{2.079} & \multirow{2}{*}{0.500}& MLE & 0.009 & 0.114 & 0.115 & 0.114 & 95.1\\
		&  &  &  & PMLE & 0.003 & 0.122 & 0.121 & 0.122 & 94.2\\
		\cmidrule{3-10}
		&  & \multirow{2}{*}{4.159} & \multirow{2}{*}{0.250} & MLE & 0.005 & 0.231 & 0.226 & 0.231 & 94.3\\
		&  &  &  & PMLE & 0.005 & 0.245 & 0.238 & 0.245 & 94.4\\
		\bottomrule 
	\end{tabular}
	\vspace{0.4cm}
	\caption{Simulation results for the baseline survival function of $D$ under the Gumbel copula with sample size $n=400$. BIAS: relative bias; ESD:  relative empirical standard deviation; ASE:  relative average standard error; rMSE: relative root mean square error; CP: empirical coverage percentage of 95\% confidence interval.}
	\label{tab:BHaz_D_gumbel400}
\end{table}

\begin{table}[htbp]
	\centering
	\begin{tabular}{ll cc l rcccc}
		\toprule
		$\tau$ & $\bbeta_{T}$ & Time & Prob. & Method & \multicolumn{1}{c}{BIAS} & ESD & ASE & rMSE & CP  \\
		\midrule
		\multirow{12}{*}{0.4} & \multirow{6}{*}{(1,1)}  & \multirow{2}{*}{0.863} & \multirow{2}{*}{0.750} & MLE & 0.007 & 0.052 & 0.050 & 0.053 & 92.9\\
		&  &  &  & PMLE & 0.005 & 0.056 & 0.054 & 0.057 & 94.1\\
		\cmidrule{3-10} 
		&  & \multirow{2}{*}{2.079} & \multirow{2}{*}{0.500} & MLE & 0.011 & 0.112 & 0.110 & 0.112 & 95.5\\
		&  &  &  & PMLE & 0.006 & 0.122 & 0.121 & 0.123 & 95.1\\
		\cmidrule{3-10} 
		&  & \multirow{2}{*}{4.159} & \multirow{2}{*}{0.250}& MLE & 0.009 & 0.220 & 0.215 & 0.220 & 94.6\\
		&  &  &  & PMLE & 0.003 & 0.243 & 0.238 & 0.243 & 93.7\\
		\cmidrule{3-10} 
		& \multirow{6}{*}{(1,0.5)} & \multirow{2}{*}{0.863} & \multirow{2}{*}{0.750} & MLE & 0.006 & 0.053 & 0.050 & 0.053 & 92.8\\
		&  &  &  & PMLE & 0.005 & 0.056 & 0.054 & 0.057 & 94.1\\
		\cmidrule{3-10} 
		&  & \multirow{2}{*}{2.079} & \multirow{2}{*}{0.500}& MLE & 0.010 & 0.113 & 0.111 & 0.113 & 94.9\\
		&  &  &  & PMLE & 0.006 & 0.122 & 0.121 & 0.123 & 95.1\\
		\cmidrule{3-10} 
		&  & \multirow{2}{*}{4.159} & \multirow{2}{*}{0.250} & MLE & 0.006 & 0.222 & 0.216 & 0.222 & 94.2\\
		&  &  &  & PMLE & 0.003 & 0.243 & 0.238 & 0.243 & 93.7\\
		\midrule
		\multirow{12}{*}{0.6} & \multirow{6}{*}{(1,1)} & \multirow{2}{*}{0.863} & \multirow{2}{*}{0.750} & MLE & 0.007 & 0.048 & 0.046 & 0.048 & 93.8\\
		&  &  &  & PMLE & 0.003 & 0.056 & 0.054 & 0.056 & 94.0\\
		\cmidrule{3-10} 
		& & \multirow{2}{*}{2.079} & \multirow{2}{*}{0.500} & MLE & 0.013 & 0.102 & 0.100 & 0.102 & 96.2\\
		&  &  &  & PMLE & 0.005 & 0.123 & 0.121 & 0.123 & 94.3\\
		\cmidrule{3-10} 
		&  & \multirow{2}{*}{4.159} & \multirow{2}{*}{0.250}& MLE & 0.017 & 0.200 & 0.195 & 0.201 & 95.3\\
		&  &  &  & PMLE & 0.004 & 0.242 & 0.238 & 0.242 & 94.7\\
		\cmidrule{3-10}  
		& \multirow{6}{*}{(1,0.5)} & \multirow{2}{*}{0.863} & \multirow{2}{*}{0.750}& MLE & 0.006 & 0.048 & 0.046 & 0.048 & 93.2\\
		&  &  &  & PMLE & 0.003 & 0.056 & 0.054 & 0.056 & 94.0\\
		\cmidrule{3-10} 
		&  & \multirow{2}{*}{2.079} & \multirow{2}{*}{0.500} & MLE & 0.011 & 0.102 & 0.101 & 0.102 & 96.0\\
		&  &  &  & PMLE & 0.005 & 0.123 & 0.121 & 0.123 & 94.3\\
		\cmidrule{3-10} 
		&  & \multirow{2}{*}{4.159} & \multirow{2}{*}{0.250} & MLE & 0.012 & 0.201 & 0.196 & 0.201 & 94.9\\
		&  &  &  & PMLE & 0.004 & 0.242 & 0.238 & 0.242 & 94.7\\
		\midrule
		\multirow{12}{*}{0.8} & \multirow{6}{*}{(1,1)} & \multirow{2}{*}{0.863} & \multirow{2}{*}{0.750} & MLE & 0.004 & 0.043 & 0.042 & 0.043 & 94.0\\
		&  &  &  & PMLE & 0.003 & 0.056 & 0.054 & 0.056 & 93.7\\
		\cmidrule{3-10}
		&  & \multirow{2}{*}{2.079} & \multirow{2}{*}{0.500} & MLE & 0.010 & 0.094 & 0.092 & 0.094 & 95.5\\
		&  &  &  & PMLE & 0.002 & 0.123 & 0.121 & 0.123 & 93.7\\
		\cmidrule{3-10} 
		&  & \multirow{2}{*}{4.159} & \multirow{2}{*}{0.250}& MLE & 0.022 & 0.181 & 0.177 & 0.182 & 95.6\\
		&  &  &  & PMLE & 0.009 & 0.243 & 0.238 & 0.243 & 94.9\\
		\cmidrule{3-10} 
		& \multirow{6}{*}{(1,0.5)} & \multirow{2}{*}{0.863} & \multirow{2}{*}{0.750} & MLE & 0.002 & 0.043 & 0.042 & 0.043 & 93.9\\
		&  &  &  & PMLE & 0.002 & 0.056 & 0.054 & 0.056 & 93.7\\
		\cmidrule{3-10}
		&  & \multirow{2}{*}{2.079} & \multirow{2}{*}{0.500}& MLE & 0.005 & 0.094 & 0.092 & 0.094 & 94.8\\
		&  &  &  & PMLE & 0.002 & 0.123 & 0.121 & 0.123 & 93.7\\
		\cmidrule{3-10}
		&  & \multirow{2}{*}{4.159} & \multirow{2}{*}{0.250} & MLE & 0.012 & 0.182 & 0.179 & 0.182 & 95.7\\
		&  &  &  & PMLE & 0.010 & 0.243 & 0.239 & 0.244 & 94.9\\
		\bottomrule 
	\end{tabular}
	\vspace{0.4cm}
	\caption{Simulation results for the baseline survival function of $D$ under the Clayton copula with sample size $n=400$. BIAS: relative bias; ESD:  relative empirical standard deviation; ASE:  relative average standard error; rMSE: relative root mean square error; CP: empirical coverage percentage of 95\% confidence interval.}
	\label{tab:BHaz_D_claytonl400}
\end{table}

\clearpage

\newpage

\begin{figure}[htbp]
    \centering
    \includegraphics{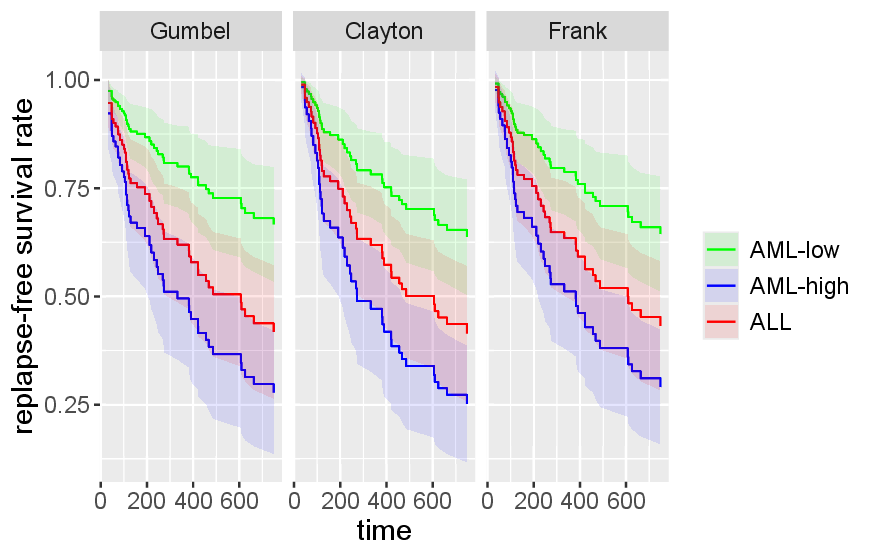}
    \caption{MLE estimates of relapse-free survival probabilities $\Pbb(T>t)$ for each disease group. The shaded areas indicate the 95\% confidence bands.}\label{fig:data_ana}
\end{figure}

\clearpage

\newpage

\bibliographystyle{apalike}
\bibliography{PMLE_SCR}